\documentclass[1p]{elsarticle}

\usepackage{amssymb}
\usepackage{amsmath}
\usepackage{verbatim} 
\usepackage{subeqn}
\usepackage[nolist]{acronym}
\usepackage{subfigure}
\usepackage[titletoc]{appendix}

\usepackage{graphicx}        

\usepackage{hyperref}
\usepackage{color}
\newcommand{\bk}{{\bf k}}                   

\begin{document}

\begin{frontmatter}

\title{One-Dimensional Optical Wave Turbulence: Experiment and Theory}

\author[lyon]{Jason Laurie\corref{cor1}}
\ead{jason.laurie@ens-lyon.fr}

\author[nice]{Umberto Bortolozzo}
\ead{umberto@gmail.com}

\author[warwick]{Sergey Nazarenko}
\ead{s.v.nazarenko@warwick.ac.uk}

\author[nice]{Stefania Residori}
\ead{stefania.residori@inln.cnrs.fr}

\cortext[cor1]{Corresponding author}
\address[lyon]{Laboratoire de Physique, Ecol\'{e} Normale Sup\'{e}reiure de Lyon, 46 all\'{e}e d'Italie, Lyon, 69007, France}

\address[nice]{INLN, Universit\'e de Nice, Sophia-Antipolis,
CNRS,
1361 route
des Lucioles, 06560, Valbonne, France}

\address[warwick]{Mathematics Institute, University of Warwick,
Coventry CV4
7AL, United Kingdom}

\begin{abstract}
We present a review of the latest developments in \ac{1D} \ac{OWT}.  Based on an original experimental setup that allows for the implementation of \ac{1D} \ac{OWT}, we are able to show that an inverse cascade occurs through the spontaneous evolution of the nonlinear field up to the point when modulational instability leads to soliton formation. After solitons are formed, further interaction of the solitons among themselves and with incoherent waves leads to a final condensate state dominated by a single strong soliton. Motivated by the observations, we develop a theoretical description, showing that the inverse cascade develops through six-wave interaction, and that this is the basic mechanism of nonlinear wave coupling for \ac{1D} \ac{OWT}. We describe theory, numerics and experimental observations while trying to incorporate all the different aspects into a consistent context. 
The experimental system is described by two coupled nonlinear equations, which we explore  within two wave limits allowing for the expression of the evolution of the complex amplitude in a single dynamical equation. The long-wave limit corresponds to waves with wave numbers smaller than the electrical coherence length of the liquid crystal, and the opposite limit, when wave numbers are larger.  We show that both of these systems are of a dual cascade type, analogous to \ac{2D} turbulence, which can be described by \ac{WT} theory, and conclude that the
cascades are induced by a six-wave resonant interaction process.  \ac{WT} predicts several stationary solutions (non-equilibrium and thermodynamic) to both the long- and short-wave systems, and we investigate the necessary conditions required for their realization.  Interestingly, the long-wave system is close to the integrable \ac{1D} \ac{NLSE} (which contains exact nonlinear soliton solutions), and as a result during the inverse cascade, nonlinearity of the system at low wave numbers becomes strong. Subsequently, due to the focusing nature of the nonlinearity, this leads to \ac{MI} of the condensate and the formation of solitons.  Finally, with the aid of the the \ac{PDF} description of \ac{WT} theory, we explain the coexistence and mutual interactions between solitons and the weakly nonlinear random wave background in the form of a \ac{WTLC}.  
\end{abstract}


\end{frontmatter}

\begin{acronym}
\acro{WT}{wave turbulence}
\acro{KE}{kinetic equation}
\acro{PDF}{probability density function}
\acro{KZ}{Kolmogorov-Zakharov}
\acro{BEC}{Bose-Einstein condensate}
\acro{WTLC}{wave turbulence life cycle}
\acro{1D}{one-dimensional}
\acro{MMT}{Majda-McLaughlin-Tabak}
\acro{OWT}{optical wave turbulence}
\acro{RPA}{random phase and amplitude}
\acro{GF}{generating functional}
\acro{CT}{canonical transformation}
\acro{DAM}{differential approximation model}
\acro{LWE}{long-wave equation}
\acro{SWE}{short-wave equation}
\acro{NLSE}{nonlinear Schr\"odinger equation}
\acro{2D}{two-dimensional}
\acro{ZT}{Zakharov transform}
\acro{IR}{infrared}
\acro{UV}{ultraviolet}
\acro{5D}{five-dimensional}
\acro{3D}{three-dimensional}
\acro{CB}{critical balance}
\acro{MHD}{magneto-hydrodynamics}
\acro{GCFR}{Gallavotti-Cohen fluctuation relation}
\acro{LC}{liquid crystal}
\acro{MI}{modulational instability}
\acro{SLM}{spatial light modulator}
\acro{CFL}{Courant-Friedrichs-Lewy}
\acro{KAM}{Kolmogorov-Arnold-Moser}
\end{acronym}

\tableofcontents

\newpage
\acresetall 
\section{Introduction}
\label{sec:intro}

\begin{figure}[ht!]
\centerline{\includegraphics[width=12 cm]{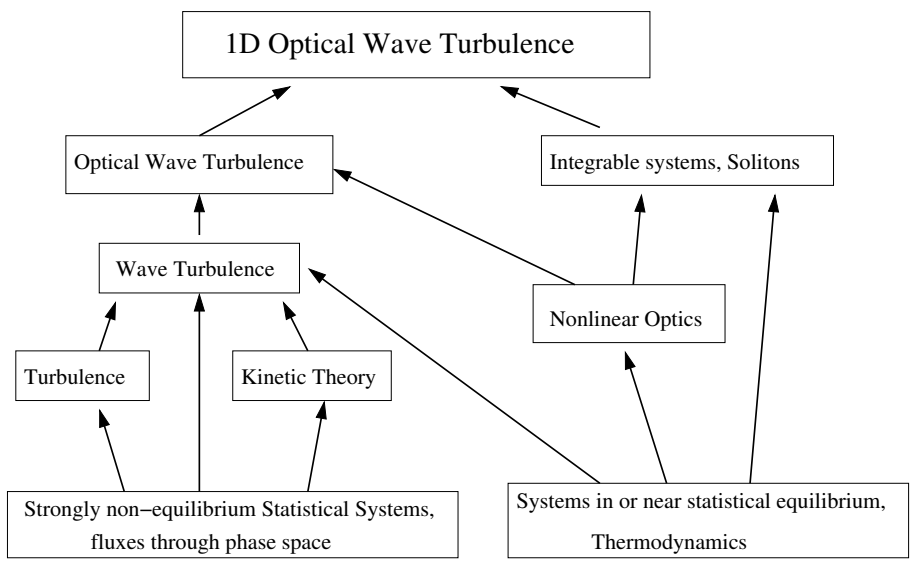}}
\caption{Interconnections of \ac{1D} \ac{OWT}  with other research areas.}
\label{fig:owt_links}
\end{figure}

One-dimensional (\acsu{1D}) \ac{OWT} is an extremely interesting physical phenomenon whose importance arises from its intrinsic overlap with several strategic research areas.  This interdisciplinary nature allows for the application of non-conventional approaches to familiar facts and routes. These areas include \ac{WT}, \ac{BEC} and lasing, integrable systems and solitons, and, on a more fundamental level, general turbulence, nonlinear optics, equilibrium and non-equilibrium statistical mechanics.  A hierarchical diagram showing these areas and their links to \ac{1D} \ac{OWT} is shown in figure~\ref{fig:owt_links}. 

The main aspects and phenomena in \ac{1D} \ac{OWT} which will be at the focus of the present review include:
\begin{itemize}
\item The inverse transfer of wave action (from short to long wave lengths), its relation to the \ac{BEC} of light, and to
\ac{WT} as an active medium for lasing;
\item The role of turbulent cascades (fluxes) versus thermodynamic potentials (temperature and chemical potential) in  \ac{1D} \ac{OWT} and in \ac{WT} in general;
\item The proximity to integrability, \ac{MI} and the generation of solitons;
\item The coexistence, interactions and mutual transformations of random waves and solitons: this constitutes the \ac{WTLC} and
the evolution  towards a final single soliton state.
\end{itemize}

 
Since \ac{WT} concepts are central for our review, we will begin by giving a brief introduction to \ac{WT}. 
\ac{WT} can be generally defined as a random set of interacting waves with a wide range of wave lengths.  \ac{WT} theory has been applied to several physical systems including water surface gravity and capillary waves in oceans \cite{ZF66,ZF67,ZLF92,OOSRPZB02,AS01,AS06,AS06-2,N06-1,DLN07,LNMD09,LNMD10,LNP06,CNP01,FLF07,FFL07,FFBF09}; internal, inertial, and Rossby waves in atmospheres and oceans \cite{CZ00,LT01,G03,L-HGK67,ZP88,MP87,BN90,BNZ90-1,BNZ91}; Alfv\'en waves in solar wind and interstellar turbulence \cite{I64,FP03,GNNP00,GNNP02,GNN01,GN08,GS95,N07,BGP08,ABPG07,SPPM02,S04}; Kelvin waves on quantized vortex lines in superfluid helium \cite{VN02,KS04,VTM03,N06,LNR07,LNR08,LLNR10,LN10,BDLLNP11}; waves in \acp{BEC} and nonlinear optics \cite{NO06,NO07,PNO09,DNPZ92,BLNR09}; waves in fusion plasmas \cite{V67,SG69,GS79,ZLF92}; and waves on vibrating, elastic plates \cite{DJR06}.  A thorough and detailed list of examples where the \ac{WT} approach has been applied, from quantum to astrophysical scales, can be found in the recent book \cite{N10}. The most developed part of the \ac{WT} theory assumes that waves have random phases and amplitudes and that their interactions are  weakly nonlinear  \cite{ZLF92,N10}, in which case a natural asymptotic closure arises for the statistical description of \ac{WT}. 
The most familiar outcome of such a closure is the \ac{KE} for the wave action spectrum and its stationary solutions describing energy and wave action cascades through scales called 
\ac{KZ} spectra \cite{ZF67,ZLF92,N10}. It is the similarity of the \ac{KZ} spectra to the Kolmogorov energy cascade spectrum in classical \ac{3D} Navier-Stokes turbulence that allows one to classify
 \ac{WT} as \em turbulence\em. On the other hand, in most physical applications, besides weakly nonlinear random waves there are also strongly nonlinear coherent structures which, even when not energetically dominant, interact with the random wave component, i.e. by random-to-coherent and coherent-to-random transformations, which may 
provide a route to the turbulence sink via wave breaking or wave collapses. In other words, together with the random weakly nonlinear waves, the strongly nonlinear coherent structures are also a fundamentally important part of the \ac{WTLC} \cite{N10}.

Let us now discuss realizations of \ac{WT} in optical systems. Very briefly, we define \ac{OWT} as the \ac{WT} of light. 
As such, \ac{OWT}  is a subject within the more general nonlinear optics field, dealing with situations involving the propagation of light in nonlinear media which is fully or partially random. \ac{OWT} is a niche area within the nonlinear optics field, and excludes a large section of the field, including systems fully dominated by strong coherent structures, such as solitons. The term  \em turbulence\em,  is even more relevant for \ac{OWT} because of the similarities between the nonlinear light behavior to fluid dynamics, such as vortex-like solutions \cite{AGRR91, SL92} and shock waves \cite{BWSF07}.  Although there have been numerous theoretical and numerical studies of \ac{OWT} \cite{DNPZ92, MRZ85, NZ05, CJPPR05, NO06}, there have been few experimental observations to date \cite{BLNR09}.  \ac{OWT} was theoretically predicted to exhibit dual cascade properties when two conserved quantities cascade to opposite regions of wave number space \cite{DNPZ92}.  This is analogous to \ac{2D} turbulence, where we observe an inverse cascade of energy and a direct cascade of enstrophy \cite{K67,K71}.  In the context of \ac{OWT}, energy cascades to high wave numbers, while wave action cascades towards low wave numbers \cite{DNPZ92, MRZ85,NZ05, CJPPR05}.  An interesting property of \ac{OWT} is the inverse cascade of wave action which in the optical context implies the condensation of photons - the optical analogue of \ac{BEC}. 

It is the \ac{BEC} processes  that make \ac{OWT} an attractive and important setup to study.
Experimental implementation of \ac{BEC} in alkali atoms was first achieved in 1995, and subsequently awarded the 2001 Nobel prize \cite{AEMWC95, PS03}.  This work involved developing a sophisticated cooling technique to micro-Kelvin temperatures, in order to make the de Broglie wave length exceed the average inter-particle distance.  This is known as the \ac{BEC} condition.  
Photons were actually the first bosons introduced by Bose in 1924 \cite{B24}, and the \ac{BEC} condition is easily satisfied by light at room temperature.  However, there was the belief that optical \ac{BEC} would be impossible, because of the fundamental difference between atoms, whose numbers are conserved, and photons which can be randomly emitted and absorbed.  However, there exist situations where light is neither emitted nor absorbed, e.g. light in an optical cavity, reflected back and forth by mirrors \cite{CB99}, or by light freely propagating through a transparent medium.  In the latter, the movement of photons to different energy states (specifically the lowest one corresponding to \ac{BEC}) can be achieved by nonlinear wave interactions.  The mechanism for these nonlinear wave interactions is provided by the Kerr effect which permits wave mixing. Moreover, the nonlinear interactions are crucial for the \ac{BEC} of light, because no condensation is possible in non-interacting \ac{1D} and \ac{2D} Bose systems. 

When the nonlinearity of the system is weak, \ac{OWT} can be described by weak \ac{WT} theory \cite{ZLF92,N10}, with the prediction of two \ac{KZ} states in a dual cascade system. One aspect of \ac{OWT} is that 
the nonlinearity of the system is predicted to grow in the inverse cascade with the progression of wave action towards large scales. This will eventually lead to a violation of the weak nonlinearity assumption of \ac{WT} theory.  The high nonlinearity at low wave numbers will lead to the formation of coherent structures \cite{DNPZ92,NZ05,NO06,MMT97,CMMT01,ZDP04}. In \ac{OWT} this corresponds to the formation of solitons and collapses for focusing nonlinearity \cite{BLNR09}, or to a quasi-uniform condensate and vortices in the de-focusing case \cite{NO06}.  

Experimentally, \ac{OWT} is produced by propagating light through a nonlinear medium \cite{Newell-book}.  However, the nonlinearity is typically very weak and it is a challenge to make it overpower the dissipation.  This is the main obstacle regarding the photon condensation setup in a \ac{2D} optical Fabry-Perot cavity, theoretically suggested in \cite{CB99} but never experimentally implemented.

This brings us to the discussion of the exceptional role played by \ac{1D} optical systems. Firstly, 
 it is in \ac{1D} that the first ever  \ac{OWT} experiment  was implemented \cite{BLNR09}.
 The key feature in our setup is to trade one spatial dimension for a time axis.
Namely, we consider a time-independent \ac{2D} light field where the principal direction of the light propagation acts as an effective time. 
This allows us to use a nematic \ac{LC}, which provides a high level of
tunable optical nonlinearity \cite{TSZ86,K95}. The slow relaxation time of
the re-orientational dynamics of the \ac{LC} molecules is not a restriction of our setup because the
system is steady in time. Similar experiments were first reported in \cite{BFL93}, where a beam propagating inside a nematic
layer undergoes a strong self-focusing effect followed by filamentation,
soliton formation and an increase in light intensity. Recently, a
renewed interest in the same setup has led to further studies on optical solitons
and the \ac{MI} regime \cite{PCALU04,PCA03,CPA06}. However, all the previous experiments  used a high input intensity, implying a strong nonlinearity of the system, and therefore the soliton condensate appears immediately, bypassing the \ac{WT} regime.
In our experiment, we carefully set up an initial condition of weakly nonlinear waves situated at high wave numbers from a laser beam.  We randomize the phase of the beam, so that we produce a wave field as close to a \ac{RPA} wave field as possible.  The nonlinearity of the system is provided by the \ac{LC}, controlled by a voltage applied across the \ac{LC} cell.  This provides the means for nonlinear wave mixing via the Kerr effect. The \ac{LC} we use is of a focusing type, causing any condensate that forms to become unstable and the formation of solitons to occur via \ac{MI}. 

Secondly, from the theoretical point of view \ac{1D} \ac{OWT} is very interesting because it represents a system close to an integrable one, namely the \ac{1D} \ac{NLSE}.  Thus it inherits many features of the integrable model, e.g. the significant role of solitons undergoing nearly elastic collisions. On the other hand, deviations from the integrability are important, because they upset the time recursions of the integrable system thereby leading to
turbulent cascades of energy and wave action through scales. We show that the process responsible for such cascades is a six-wave resonant interaction (wave mixing). Another example of a
nearly integrable \ac{1D} six-wave system can be found in superfluid turbulence - it is the \ac{WT} of Kelvin waves on quantized vortex lines \cite{KS04,KS05,LNR07,LNR08,LLNR10,BDLLNP11} 
(even though non-local interactions make the six-wave process effectively a four-wave one
in this case).  Some properties of the six-wave systems are shared with four-wave systems, particularly \ac{WT} in the \ac{MMT} model reviewed in Physical Reports by Zakharov {\em et al} \cite{ZDP04}. For example, both the four-wave and the six-wave systems are dual cascade systems, and in both systems solitons (or 
quasi-solitons) play a significant role in the \ac{WTLC}. There are also significant differences 
between these two types of systems. Notably, pure \ac{KZ} solutions appear to be much less 
important for the six-wave optical systems than for the four-wave \ac{MMT} model - instead the 
spectra have a thermal component which is  dominant over the flux component. Moreover, the number of solitons in \ac{1D} \ac{OWT} decrease in time due to soliton mergers, so that asymptotically there is only a single strong soliton left in the system.


Similar behavior was extensively theoretically studied in various
settings for non-integrable Hamiltonian systems starting with the 
paper by Zakharov {\em et al} \cite{ZPSY88}, and then subsequently in \cite{RN01,RN03,BKKMMP09,ET06,JJ00,RCK00,JTZ00,PPM08,PLJP06,PY87}.
The final state, with a single soliton and small scale noise, was interpreted as a
statistical attractor, and an analogy was pointed
out to the over-saturated vapor system, where the solitons are similar to
droplets and the random waves behave as molecules \cite{PY87}. Indeed,
small droplets evaporate whilst large droplets gain in size from free molecules, resulting in a decrease in the number of droplets.  
On the other hand, in the \ac{1D} \ac{OWT} context, the remaining strong soliton is actually a narrow coherent beam of light. This allows us to interpret the \ac{WT} evolution leading to the formation of such a beam as a lasing process. Here, the role of an active medium where the initial energy is stored is played by the weakly nonlinear random wave component, and the major mechanism for channeling this energy to the coherent beam is provided
by the \ac{WT} inverse cascade. For this reason, we can call such a system a \em \ac{WT} laser\em. It is quite possible that the described \ac{WT} 
lasing mechanism is responsible for spontaneous formation of coherent beams in stars or molecular clouds, although it would be premature to make any definite claims about this at present.

Generally, in spite of recent advances, the study of \ac{1D} \ac{OWT} is far from being complete. 
The present review provides a report on the current state of this area describing
not only what we have managed to learn and explain so far, but also the results 
which we do not know yet how to explain, discussing the existing theory and the gaps within it
that are yet to be filled. We compare the experimental observations with the predictions of \ac{WT} theory and independently juxtapose our findings with direct numerical simulation of the governing equations. In particular, we describe some puzzles related to the wave action spectra 
obtained in the experiment and in the numerical experiments.
Furthermore, we will describe the recent extensions of weak \ac{WT} theory onto the wave \ac{PDF}, which marks the beginning of developing a formalism for describing \ac{WT} intermittency  and the role of coherent structures.  On the other hand, a theory for the \ac{WTLC} incorporating interacting random waves and 
coherent structures/solitons is still to be investigated, with only of a few pioneering works reporting on the study of the interaction between coherent structures and the radiating background \cite{Rumpf-PRL}.

\section{The Experiment}
\label{sec:exp}

The \ac{1D} optical system has been designed to meet the major requirements of \ac{OWT}.
Especially important are the careful calibrations that have been taken to fulfill the 
balance between low dissipation and low nonlinearity. Indeed, the main experimental challenge in observing the \ac{WT} regime is in keeping the nonlinearity weak enough to let the \ac{WT} regime develop and, at the same time, 
high enough to make it overpower the dissipation. 
Our setup is based on a nematic liquid crystal layer in which a laminar shaped beam propagates.
\acp{LC} are particularly suitable for the observation of the \ac{WT} regime because of their well known optical properties, such as their high and tunable nonlinearity, transparency (slow absorption) over a wide range of optical wave lengths, the realization of large cells and the possibility to drive them with low voltage externally applied fields \cite{K95}.

\subsection{Description of the Experimental Setup}
\label{sec:setup}

\begin{figure}[ht!]
\centerline{\includegraphics[width=0.6 \columnwidth]{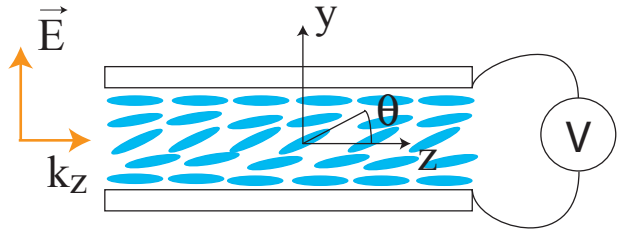}}
\caption{Schematic of the \ac{LC} cell: molecules, initially aligned parallel to the confining walls, are oriented, through the application of the external voltage $V$, at an average angle $\theta$, around which further reorientations are induced by the optical field $\vec E$.}
\label{fig:cell}
\end{figure}

The liquid crystal, \ac{LC}, cell is is schematically depicted in  figure~\ref{fig:cell}.
It is made by sandwiching a nematic layer, (E48), of
thickness $d=50$ $\mu m$, between two $20 \times 30$ $mm^2$, glass windows and on the interior, the glass walls are coated with indium-tin-oxide transparent electrodes. We have pre-treated the indium-tin-oxide surfaces with
polyvinyl-alcohol, polymerized and then rubbed, in order to align all the
molecules parallel to the confining walls. When a voltage is applied across the
cell, \ac{LC} molecules tend to orientate in such a way as to become
parallel to the direction of the electric field. By applying a $1$ $kHz$ electric
field with a rms voltage of $V_0=2.5$ $V$ we preset the molecular director to an
average tilt angle $\Theta$.

\begin{figure}[ht!]
\centerline{\includegraphics[width=\columnwidth]{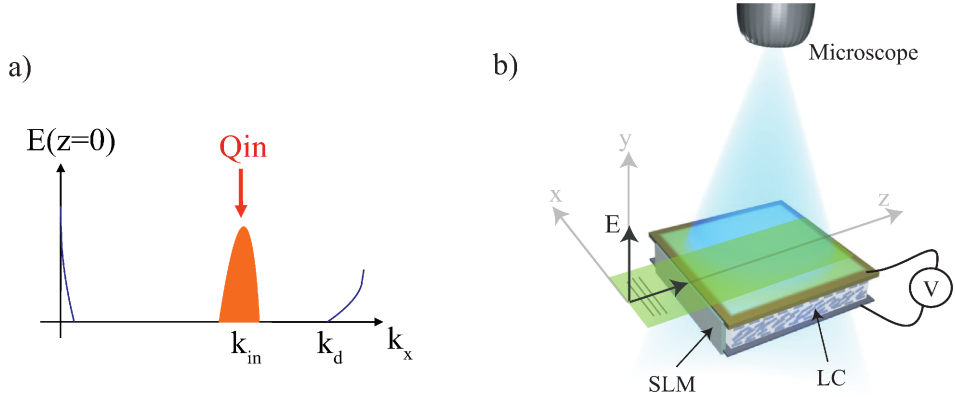}}
\caption{a) Spatial forcing realized as initial condition $Q_{in}$ by appropriate preparation of the input beam; $k_{in}$ is the chosen wave number around which random phase modulations are introduced. b) Sketch of the experimental setup: a laminar shaped
input beam propagates inside the \ac{LC} layer; random phase
modulations are imposed at the entrance of the cell by means of a spatial light modulator, \ac{SLM}. }
\label{fig:setup}
\end{figure}

The experimental apparatus is shown in figure~\ref{fig:setup}. It consists of a
\ac{LC} cell, inside which a laminar shaped beam propagates.  
An important point is that the
input beam is carefully prepared in such a way as to have an initial condition of weakly nonlinear random waves.
As depicted in figure~\ref{fig:setup}a, the preparation of the input beam
is such that the system is forced with an initial condition $Q_{in}$, which is at a intermediate spatial scale between the large scale, $k=0$, and the dissipative scale, $k_d$. Moreover, phases are randomized so that a narrow bandwidth
forcing is realized around the initial spatial modulation at the chosen wave number $k_{in}$. 

The setup is schematically represented in figure~\ref{fig:setup}b.
The input
light originates from a diode pumped, solid state laser, with $\lambda =473$ $nm$, 
polarized along $y$ and shaped as a thin laminar Gaussian beam of $30$ $\mu m$
thickness.
The input  light intensity is kept very low, with an input intensity of $I= 30$ $\mu W/ cm^2$ to
ensure the weakly nonlinear regime. A \ac{SLM}, at the
entrance of the cell is used to produce suitable intensity masks for
injecting random phased fields with large wave numbers. 
This is made by creating a random
distribution of diffusing spots with the average size $\simeq 35$ $\mu m$ through the \ac{SLM}. 

The beam evolution inside the cell is monitored with an optical microscope and a
CCD camera. 
The \ac{LC} layer behaves as a positive uni-axial medium, where $n_\parallel =n_z=1.7$
is the extraordinary refractive index and $n_{\perp} =1.5$ is the ordinary refractive index
\cite{K95}. The \ac{LC} molecules tend to align along the applied field and the
refractive index, $n( \Theta )$, follows the distribution of the tilt angle
$\theta$.  When a linearly polarized beam is injected into the cell, the \ac{LC} molecules
orientate towards the direction of the incoming beam polarization, thus, realizing 
a re-orientational optical Kerr effect. Because the refractive index increases when molecules orient themselves towards the
direction of the input beam polarization, the sign of the nonlinear index change is positive, hence, we have a focusing nonlinearity.

\begin{figure}[ht!]
\centerline{\includegraphics[width=0.85 \columnwidth]{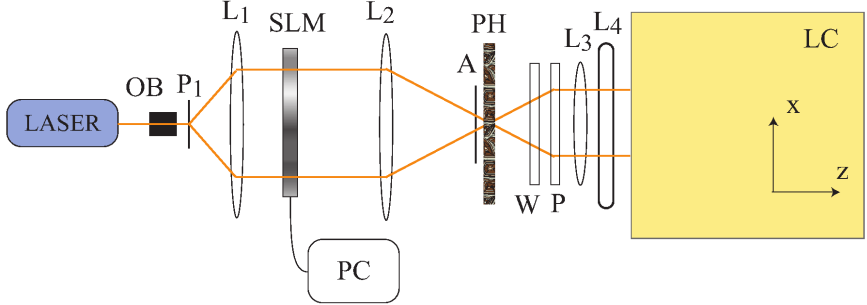}}
\caption{Detailed representation of the experimental setup,
showing the initialization of the input laser beam. OB: objective, $P_1$: pinhole, $L_1,L_2,L_3,L_4$: lenses, SLM: spatial light modulator, PC: computer, A: variable aperture, PH: random phase plate, W: half-wave plate, P: polarizer, LC: liquid crystals. }
\label{fig:setupdetail}
\end{figure}

Figure~\ref{fig:setupdetail} depicts in more detail how the input light beam is
prepared before entering the \ac{LC} cell. The beam is expanded and collimated through the spatial light filter shown in
figure~\ref{fig:setupdetail}. 
The objective, OB, focuses the light into the $20$ $\mu m$ pinhole, PH, the lens
$L_1$ collimates the beam with a waist of $18$ $mm$. After that, the light passes
through the \ac{SLM}, which is a LCD screen working in
transmission with a resolution of $800 \times 600$, with $8$ bits pixels, of size $14$ $\mu m$. Each pixel is controlled through a personal computer PC, to ensure that the outgoing light is intensity modulated. In our case we use a cosinus modulation having a colored noise envelope. 
The lenses, $L_3$ and $L_4$, are used to focus the image from the LCD screen at the
entrance of the \ac{LC} cell. The half wave-plate, W, together with the
polarizer, P, are used to control the intensity and the polarization, which is
linear along the $y$-axis. The circular aperture is inserted in the focal plane
to filter out the diffraction given by the pixelization of the \ac{SLM} and
the diffuser, PH, is inserted to spatially randomize the phase of the light. In
order to inject the light inside the \ac{LC} cell, we use a cylindrical
lens, $L_4$, close to the entrance of the \ac{LC} layer.

\subsection {The Evolution of the Light Intensity and the Inverse Cascade}

\begin{figure}[ht!]
\centerline{\includegraphics[width=\columnwidth]{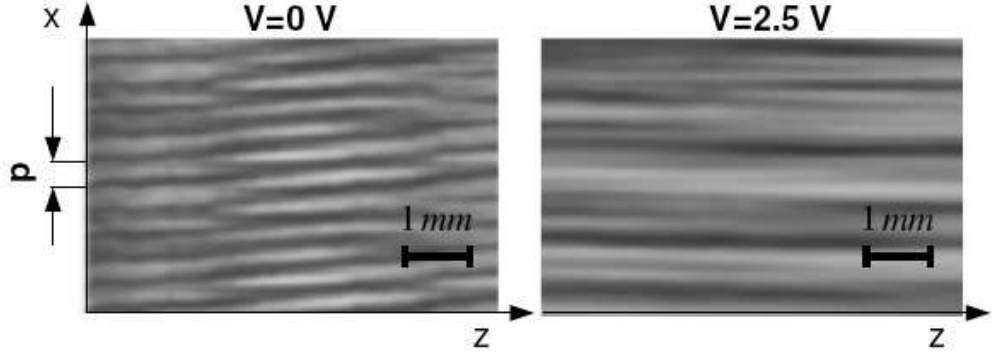}}
\caption{Intensity distribution $I(x,z)$ showing the beam evolution during
propagation; a) linear case (no voltage applied to the \ac{LC} cell), b) weakly nonlinear case (the voltage applied to the \ac{LC} cell is set to $2.5$ $V$). $d$ is the spatial period of the input beam modulation at the entrance plane of the cell.}
\label{fig:comparison}
\end{figure}

The inverse cascade can be observed directly in the experiment by inspecting the light
pattern in the $(x,z)$ plane of the \ac{LC} cell. Recall that $z$ has here the role of time. 
Two magnified images of the intensity distribution
$I(x,z)$ showing the beam evolution during propagation in the experiment are displayed in
figure~\ref{fig:comparison}. For comparison, in figure~\ref{fig:comparison}a and b, we show the beam evolution in the linear and in the weakly
nonlinear regimes, respectively. In figure~\ref{fig:comparison}a, we set a periodic initial
condition with a uniform phase and apply no voltage to the \ac{LC} cell. We see that the linear propagation is
characterized by the periodic recurrence of the pattern with the same period, a
phase slip occurring at every Talbot distance.  This is defined by
$p^2/ \lambda$, with $p$, the period of the initial condition and $\lambda$, the
laser wave length \cite{T36}. In figure~\ref{fig:comparison}b, we apply a voltage, $V=2.5$ $V$ to the \ac{LC} cell. The initial condition
is periodic with the same period as in figure~\ref{fig:comparison}a, but now with random
phases. We observe that the initial period of the pattern is becoming larger as the light beam propagates along $z$.

\begin{figure}[ht!]
\centerline{\includegraphics[width=0.8\columnwidth]{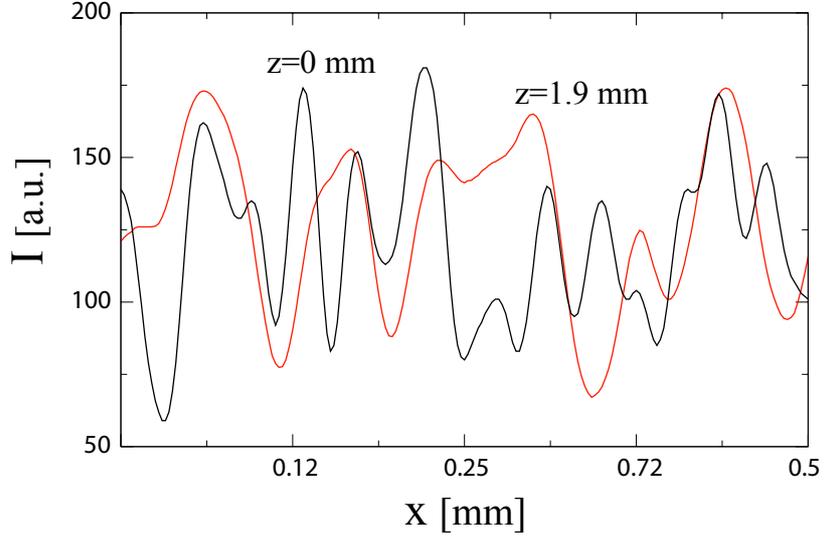}}
\caption{Two intensity profiles $I(x)$ recorded at $z=0$ and $z=1.9$ $mm$ in the
weakly nonlinear regime, with $V=2.5$ $V$, showing the smoothing associated with the
inverse cascade. }
\label{fig:profiles}
\end{figure}

While the linear propagation in figure~\ref{fig:comparison}a, forms Talbot intensity carpets \cite{BMS01},
with the initial intensity distribution reappearing periodically along the propagation direction $z$, the weak nonlinearity in figure~\ref{fig:comparison}b, leads to wave interactions, with different spatial frequencies mixing and the periodic occurrence of the Talbot carpet being broken.
In figure~\ref{fig:profiles}, we show two intensity profiles taken in the nonlinear
case at two different stages of the beam propagation. The inverse cascade is
accompanied by a smoothing of the intensity profile and the amplification of low
wave number components.

\begin{figure}[ht!]
\centerline{\includegraphics[width=0.8\columnwidth]{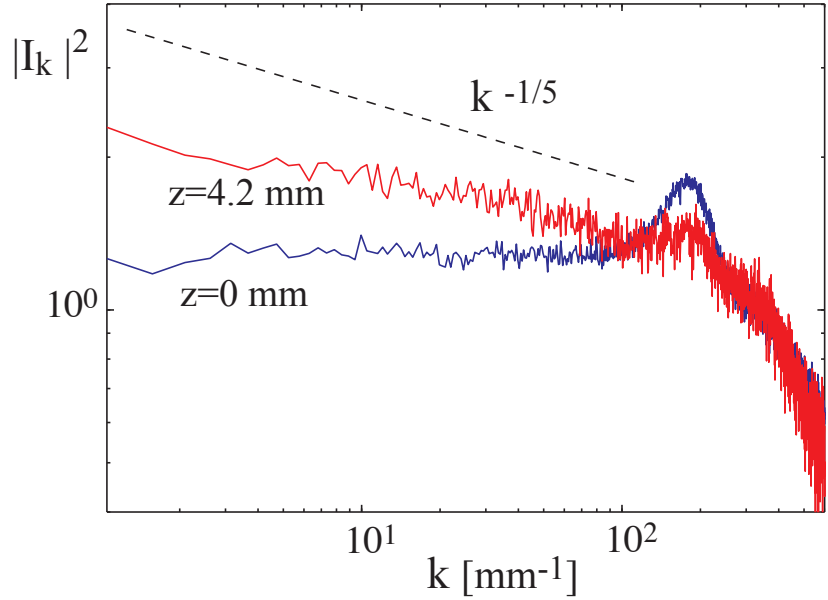}}
\caption{Experimental spectrum of the light intensity, $N_\bk=|I_\bk|^2$ at two different distances $z$. }
\label{fig:inverse-exp}
\end{figure}

The inverse cascade can be measured directly by recording the evolution of the transverse light pattern $I(x,z)$ along $z$. However, experimentally we measure the light intensity $I(x,z)$ as we do not have direct access to the phases. Therefore, we measure the spectrum of intensity $N(k,z) = |I_k(z)|^2$, for which an appropriate scaling should be derived from the theory. The experimental scaling for $N_k$ in the inverse cascade is obtained by fitting the
experimental spectrum of the light intensity, and gives
$N_k \sim |k|^{-1/5}$ as shown in figure~\ref{fig:inverse-exp}.
 One can see an inverse cascade excitation of the lower $k$ states, and a good agreement
with the WT prediction.

\subsection{The Long Distance Evolution and Soliton Formation}

\begin{figure}[ht!]
\centerline{
\includegraphics
[width=0.8\columnwidth]{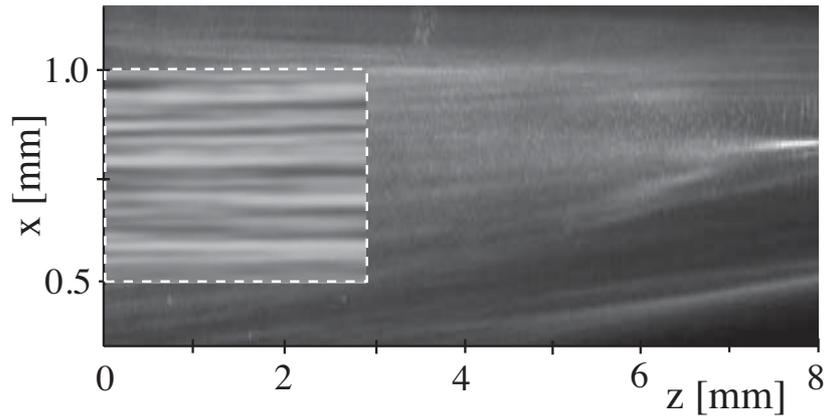}}
\caption{Experimental results for intensity distribution $I(x,z)$. The area marked
by the dashed line is shown at a higher
resolution (using a larger magnification objectif).}
\label{fig:xz-exp}
\end{figure}

The intensity distribution $I(x,z)$, showing the beam evolution
during propagation for longer distances is displayed in figures~\ref{fig:xz-exp}. 
In the high resolution inset we can observe that the typical wavelength of the waves increases along the beam, which corresponds
to an inverse cascade process. Furthermore, one can see the formation of coherent solitons out of the random initial wave field, such
that in the experiment, one strong soliton is dominant at the largest distance $z$.

\begin{figure}[ht!]
\centerline{\includegraphics[width=0.8\columnwidth]{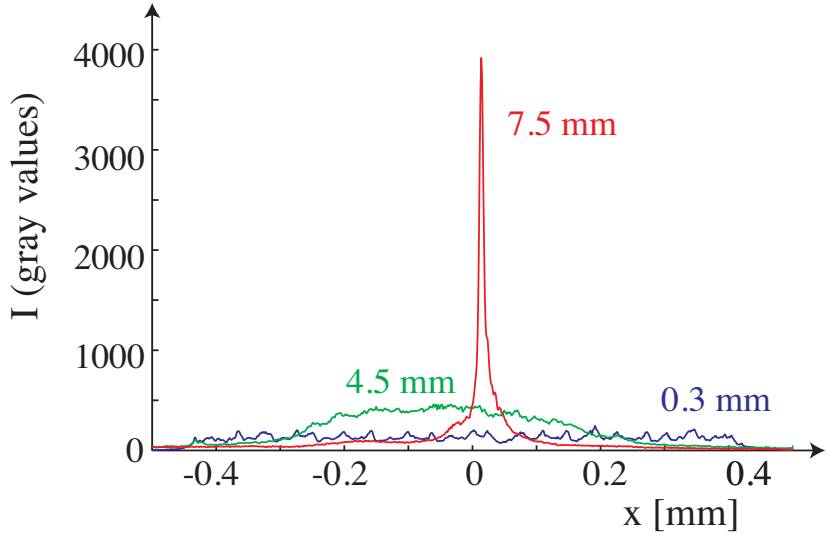}}
\caption{Linear intensity profiles $I(x)$ taken at different
propagation distances, $z=0.3$, $4.5$ and $7.5$ $mm$. }
\label{fig:condensation}
\end{figure}

The experimental evolution reported in figure~\ref{fig:xz-exp} indicates that the total number of solitons reduces.
The observed increase of the scale and formation of coherent structures
represents the condensation of light. Experimentally, the condensation into one dominant soliton is well revealed by
the intensity profiles $I(x)$ taken at different propagation distances,
as shown in figure~\ref{fig:condensation} for $z=0.3$, $4.5$ and $7.5$ $mm$. Note
that the amplitude of the final dominant soliton is three orders of magnitude
larger than the amplitude of the initial periodic modulation.

\begin{figure}[ht!]
\centerline{\includegraphics[width=0.7\columnwidth]{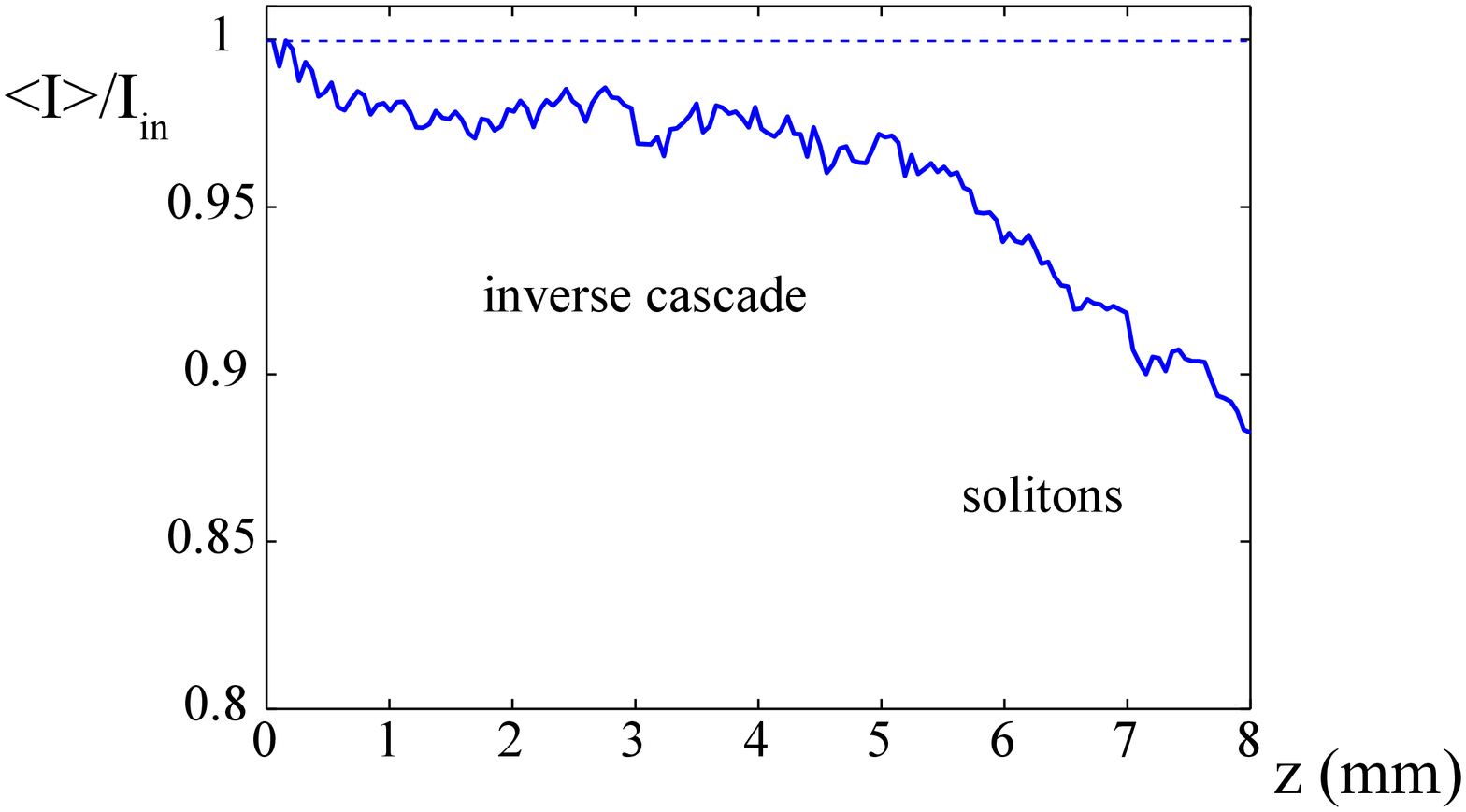}}
\caption{Evolution of the normalized $x$-averaged light intensity $<I>/I_{in}$, where $I_{in}$ is the input intensity, as a function of the propagation distance $z$. }
\label{intensity}
\end{figure}

As for energy dissipation, we should note that this is mainly due to radiation losses, whereas absorption in the liquid crystals 
is practically negligible \cite{Smyth}. 
In order to give an estimation of the losses occurred during the $z$ evolution we have averaged the light intensity $I(x,z)$ along $x$ and calculated the ratio of the $x$-averaged intensity $<I>$ to the input intensity $I_{in}$. The result is plotted in figure~\ref{intensity}, from which we observe that after $8$ $mm$ of propagation the losses are about $15 \%$. Moreover, by comparing figure~\ref{intensity} with the intensity distribution $I(x,z)$ (see figure~\ref{fig:xz-exp}), we can note that during the inverse cascade the total light intensity remains practically constant, whereas losses become more important when solitons start to appear.

\subsection{The Probability Density Function of the Intensity}

\begin{figure}[ht!]
\centerline{\includegraphics[width=0.8\columnwidth]{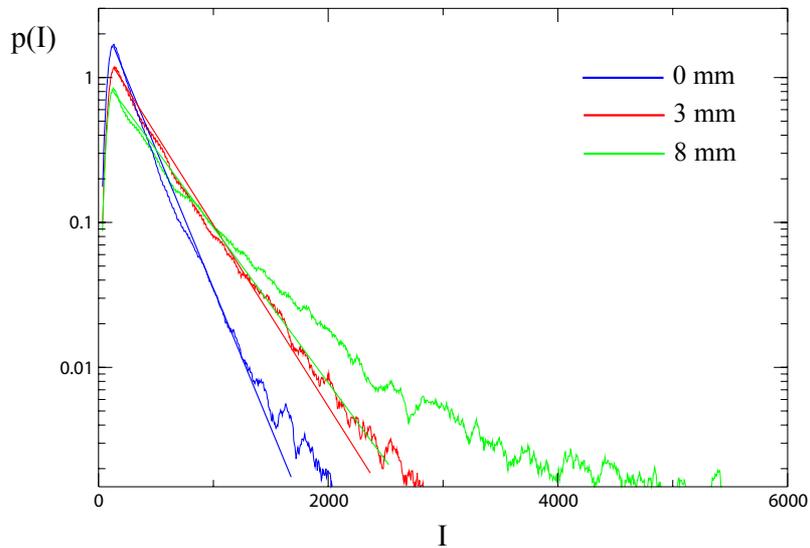}}
\caption{\acsp{PDF} of the wave intensity within the experimental cell
at three different distances along the cell, $z=0$ $mm$, $z=3$ $mm$ and $z=8$
$mm$. Straight lines correspond to the Rayleigh \acs{PDF}'s  corresponding to Gaussian wave fields (these fits have
the same mean as the respective numerical \acp{PDF}). 
}
\label{fig:pdf_exp}
\end{figure}

Figure~\ref{fig:pdf_exp} displays three experimentally obtained profiles of the \ac{PDF} of the wave
intensity along the cell at distances $z=0$ $mm$, $z=3$ $mm$ and $z=8$
$mm$ in lin-log coordinates.  In a pure Gaussian wave field, we would observe the Rayleigh distribution \eqref{eq:pdfsol1} which would correspond to a straight profile of the
\ac{PDF} in lin-log coordinates.  However, in figure~\ref{fig:pdf_exp}, non-Gaussianity is observed with the deviation from the straight lines, indicating a slower that exponential decay of the \ac{PDF} tails.  Non-Gaussianity corresponds to {\em intermittency} of \ac{WT}.  

Intermittency implies that there is a significantly higher occurrence of high intensity structures compared to that predicted by a Gaussian wave field and that the system is in the presence of an \ac{MI} process leading to solitons.  The observation of intermittency could be a sign of the development of coherent structures (solitons) in the system.

Analogies can be drawn of such high amplitude solitons with the rogue waves appearing in systems characterized by many nonlinearly interacting waves, as seen on the ocean surface \cite{Onorato}, in nonlinear optical systems \cite{Solli,Akhmediev,Arecchi}, and in superfluids \cite{McClintock}.  For all of which a description in terms of nonlinear coherent structures emerging from modulational instability in the nonlinear Schr\"{o}dinger equation can been outlined \cite{Zakharov}. In particular, possible links between rogue waves and wave turbulence in optical systems are discussed in \cite{Picozzi2011}.

\vskip 1 cm
\subsection{The Relation to Previous Studies of Optical Solitons}

Optical re-orientation of the \ac{LC} occurs under the action of the light itself, with the 
\ac{LC} molecules tending to align along the direction of the laser beam polarization, and giving rise to a Kerr effect that produces a self-focusing nonlinearity \cite{TSZ86}.
This effect was shown to lead to laser beam filamentation \cite{BFL93} and, recently, has been exploited to demonstrate the stable propagation of spatial optical solitons, also called nematicons, inside nematic \ac{LC} cells \cite{PCALU04}. For the same type of system, \ac{MI} has also been reported \cite{PCA03,CPA06}. However, the previous experiments used a high input intensity, implying a strong nonlinearity of the system, and therefore filamentation and solitons appeared immediately, bypassing the \ac{WT} regime. 

Solitons are well-known and widely studied in nonlinear optics \cite{KA03}, where they are understood as light pulses that maintain their shape, unaltered during propagation in a nonlinear medium, and where the nonlinearity implies a change in the refractive index that is induced by the intensity of the light itself. In this context solitons are classified as being either spatial or temporal, whether the self-confinement of the light beam occurs in space or in time during their propagation. The nonlinearity of the medium corresponds, respectively, to a self-focusing or a self phase modulation effect. 

While \ac{WT} implies the presence of many random waves interacting with low nonlinearity, previous work conducted in the nonlinear optics field was mainly aimed at realizing self-confined beams in a different regime, where the nonlinearity was relatively high and the medium seeded by a single pulse of Gaussian shape.
A spatial soliton is therefore obtained by imposing a tightly focused beam as an initial condition and, then by letting the beam propagate inside a medium where the self-focusing nonlinearity compensates the transverse beam widening by diffraction. 
These types of spatial solitons have been observed in a number of diverse optical media, such as, photorefractive crystals, atomic vapors, and semiconductor wave guides \cite{AA05,Springerbook}. 
On the other hand, temporal solitons correspond to cases when the nonlinearity compensates the temporal broadening of a light pulse due to the natural dispersion of the traversed medium \cite{K95}, and have been observed in optical fibers.

In both cases, the theoretical approach to spatial and the temporal solitons are based on the \ac{NLSE}. 
We therefore expect the generic behavior of the system to have an \ac{OWT} regime existing before the formation of solitons.  This entails an inverse cascade, the development of soliton turbulence and subsequently, a final single soliton acting as a statistical attractor of the system \cite{ZPSY88}. 
 In the spatial case, the light propagation direction, usually denoted as $z$, plays the role of time, therefore the wave dispersion relation is of the type 
$\omega=k^2$, whereas in the temporal case the time derivative is in the second order dispersion term, hence the dispersion relation is 
of the type $k=\omega^2$. Theoretical predictions of \ac{OWT} regimes should, therefore, be different in both cases.
Recently, experiments achieving \ac{WT} regimes in optical fibers have been devised \cite{TFMT09} and the wave thermalization phenomena has been reported for the same type of system \cite{SRJP10}. However, up to now only in the \ac{LC} experiment has there been a genuine \ac{WT} regime, with an inverse cascade and the spontaneous emergence of spatial solitons out of random waves, been demonstrated \cite{BLNR09}.

\subsection{The Theoretical Model of the \ac{OWT} Experiment}
\label{sec:theory}

Theoretically, the experimental setup can be modeled by an evolution equation for the
input beam, coupled to a relaxation equation for the \ac{LC} dynamics given by
\begin{subequations}\label{eq:evoleq}
\begin{eqnarray}
2iq \frac{\partial \psi}{\partial z}+ \frac{\partial^2 \psi }{\partial x^2}+
k_0^2n_a^2 a \psi &= 0,\label{eq:prop}\\
\frac{\partial^2 a }{ \partial x^2} - \frac{1 }{ l_\xi^2} a + \frac{\varepsilon_0 n_a^2
}{ 4 K} | \psi |^2&=0,
\label{eq:dyn}
\end{eqnarray}
\end{subequations}
where $\psi(x,z)$ is the complex amplitude of the input beam propagating along the
time axis $z$; $x$ is the coordinate across the beam; $a(x,z)$ is the \ac{LC} reorientation angle;
$n_a=n_e-n_o$ is the birefringence of the \ac{LC}; $k_0$ is the optical wave number;
$\varepsilon_0$ is the vacuum permittivity; and $l_\xi = \sqrt {\pi K / 2 \Delta
\varepsilon} (d/ V_0)$ is the electrical coherence length of the \ac{LC}
\cite{GP93}, with $K$ being the elastic constant,
$q^2=k_0^2\left(n_o^2+n_a^2/2\right)$ and $\Delta \varepsilon$ is the dielectric
anisotropy.
Note that $l_\xi$ fixes the typical dissipation scale, limiting the extent of
the inertial range in which the \ac{OWT} cascade develops. In other contexts,   such a spatial diffusion of the molecular
deformation has been denoted as a non-local effect, see \cite{PCALU04,PCA03,CPA06}. In our experiment, when
$V_0=2.5$ $V$, we have that $l_\xi =9$ $\mu m$. By considering that a typical value of
$K$ is of the order $\sim 10$ $pN$, we can derive a typical dissipation
length scale of the order $\sim 10$  $\mu m$.


The evolution equations, (\ref{eq:evoleq}) for the complex amplitude of the input beam, $\psi(x,z)$, to the \ac{LC} reorientation angle, $a(x,z)$
can be  re-written as a single equation for field $\psi(x,z)$.
This can be achieved by formally inverting the operator on $a(x,z)$ in equation (\ref{eq:dyn}).
Equation (\ref{eq:dyn}) implies that
\begin{equation}\label{eq:dyninvert}
a(x,z) =  \frac{\varepsilon_0 n_a^2 }{ 4 K}\left(\frac{1}{l_\xi^2} -\frac{\partial^2}{
\partial x^2}\right)^{-1}  | \psi |^2.
\end{equation}
Substituting this expression 
into equation (\ref{eq:prop}), we eliminate the dependence on variable $a(x,z)$. This gives
\begin{equation}\label{eq:psi}
2iq \frac{\partial \psi }{ \partial z}+ \frac{\partial^2 \psi}{\partial x^2}+ 
\frac{k_0^2n_a^4 \varepsilon_0 }{ 4 K} \psi  \left(\frac{1 }{ l_\xi^2} -\frac{\partial^2
}{ \partial x^2}\right)^{-1} | \psi |^2 = 0.
\end{equation}
Equation (\ref{eq:psi}) is a single equation modeling the evolution of the complex amplitude, $\psi(x,z)$.  We can further simplify equation (\ref{eq:psi}) by considering the system in two limits of wave number $k$: $kl_\xi \ll 1$ and $1 \ll kl_\xi$, that we call the long- and short-wave limits respectively.  These limits enable the expansion of the nonlinear operator in power of $l_\xi \partial /\partial x$. Our experimental system is well described by the long-wave limit. The limitations imposed by the dissipation of the \ac{LC} in the current experimental setup prevents the implementation in the short-wave regime.  However, for  the completeness of our description and for the possibilities in the modification of the experimental setup  in the future for the short-wave regime, we will continue to investigate this limit theoretically and numerically.

\subsubsection{The Long-Wave Regime}
\label{sec:model}

The long-wave approximation to equation (\ref{eq:psi}) corresponds to the wavelength of the spatial light distribution, $\lambda\propto 1/k$, being
greater than the electrical coherence length of the \ac{LC}, $l_\xi$. In physical space, this limit corresponds to $l_\xi \partial/\partial x \ll 1$, which permits the expansion of the nonlinear operator of equation (\ref{eq:psi}) as
\begin{equation}\label{eq:nlexpand}
\left(\frac{1 }{ l_\xi^2} -\frac{\partial^2 }{ \partial x^2}\right)^{-1} =l_\xi^2\left(1+l_\xi^2\frac{\partial^2}{\partial x^2} +l_\xi^4\frac{\partial^4}{\partial x^4} + \cdots  \right).
\end{equation}
Taking the leading order of this expansion yields
\begin{equation}\label{eq:1dnls}
 2iq \frac{\partial \psi }{ \partial z}= -\frac{\partial^2 \psi }{ \partial x^2}-
\frac 1{2 l_\xi^2 \tilde{\psi}^2} \psi |\psi |^2
\end{equation}
where, for clarity, we have introduced a reference light intensity:
\begin{equation}\label{eq:psic}
\tilde{\psi}^2 = \frac{ 2K} {\varepsilon_0 n_a^4 l_\xi^4 k_0^2 }. 
\end{equation}
Equation~\eqref{eq:1dnls} is the \ac{1D} focusing \ac{NLSE}. As is well-known, the \ac{1D} \ac{NLSE} is an integrable system, solvable with the aid of the inverse scattering transform \cite{ZM74}, and  characterized by  solitons \footnote{The term soliton is sometimes reserved for solitary waves with special properties arising from integrability, such as the ability to pass through one another without change in shape or velocity, as is the case for the \ac{1D} \ac{NLSE}.  Hereafter, we will use the term soliton more broadly, including solitary waves in non-integrable systems  which can change their states upon mutual collisions.}. Unfortunately, this would be a poor model for \ac{OWT}, as the integrability of the \ac{1D} \ac{NLSE} implies that wave turbulent interactions are not possible.  To overcome this, we must consider the sub-leading contribution in expansion (\ref{eq:nlexpand}).  This extra nonlinear term acts as a correction breaking the integrability of the system. The resulting equation is given as
\begin{equation}
2iq \frac{\partial \psi }{ \partial z}= -\frac{\partial^2 \psi }{\partial x^2}-
\frac 1{2 l_\xi^2 \tilde{\psi}^2}
\left ( \psi |\psi |^2 + l_\xi^2 
\psi \frac{\partial^2 | \psi |^2 }{ \partial x^2} \right ).
\label{eq:eq-long}
\end{equation}
We refer to equation \eqref{eq:eq-long} as the \ac{LWE}. For the expansion \eqref{eq:nlexpand} to be valid, the additional nonlinear term must be considered smaller than the leading nonlinear term. Moreover for \ac{OWT} to be in the weakly nonlinear regime, both of the nonlinear contributions should be smaller than the linear term.  Therefore, although integrability is lost, the system will remain close to the integrable one described by (\ref{eq:1dnls}). As a result, we expect  soliton-like solutions close to the exact solutions of the \ac{1D} \ac{NLSE} (\ref{eq:1dnls}).  On the other hand, exact soliton solutions of equation (\ref{eq:1dnls}), do not change shape, and have the ability to pass through one another unchanged. We can expect that in the \ac{LWE} (\ref{eq:eq-long}), we will observe similar soliton solutions, but the non-integrability will allow solitons to interact with one another, and with the weakly nonlinear random wave background.  

\subsubsection{The Short-Wave Regime}

In the opposite limit of equation (\ref{eq:psi}), when $1\ll l^2_\xi\partial^2/\partial x^2$,  the nonlinear operator of equation (\ref{eq:psi}) can be represented in terms of a Taylor expansion of negative powers of the spatial derivative:
\begin{equation}\label{eq:nlexpand2}
\left(\frac{1 }{ l_\xi^2} -\frac{\partial^2 }{ \partial x^2}\right)^{-1}
= -\left(\frac{\partial^{2}}{\partial x^{2}}\right)^{-1} + \frac{1}{l^2_\xi}\left(\frac{\partial^{2}}{\partial x^{2}}\right)^{-2} - \cdots.  
\end{equation}
It is sufficient for us to approximate the nonlinear operator of equation (\ref{eq:psi}) with just the leading order term in expansion (\ref{eq:nlexpand2}), as integrability of the equation is not an issue. Therefore, we get an equation of the form:
\begin{equation}
2iq \frac{\partial \psi }{ \partial z} = -\frac{\partial^2 \psi }{ \partial
x^2}+ 
\frac 1{2 l_\xi^4 \tilde{\psi}^2}
\psi \left(\frac{\partial^{2} 
}{ \partial x^2}\right)^{-1}| \psi |^{2}.
\label{eq:eq-short}
\end{equation}
We call equation~\eqref{eq:eq-short}  the \ac{SWE}.  Ultimately, we have presented two dynamical equations for the complex wave amplitude $\psi(x,z)$ for \ac{1D} \ac{OWT} in two limits of wave number space. Both of these systems can be expressed in a Hamiltonian formulation, that will be utilized by \ac{WT} theory in the weakly nonlinear regime. 

\subsection{The Nonlinearity Parameter}

It is essential for the development of \ac{OWT}  that the system operates in a weakly nonlinear regime. We can quantify the linearity and nonlinearity within the system with the introduction of a nonlinear parameter, $J$, which is determined by the ratio of the linear term to the nonlinear term within the dynamical equations. 

For instance, the nonlinear parameter from the \ac{LWE} (\ref{eq:eq-long}) is defined as
\begin{equation}\label{eq:JL}
 J^{L} = \frac{2\tilde{\psi}^2k^2l_\xi^2 }{{I}}.
\end{equation}
This is derived from the ratio of the linear term and the first of the two nonlinear terms.  Here, $I = \left\langle|\psi(x,z)|^2\right\rangle$ is the average value of the light intensity.  Similarly, the \ac{SWE}, \eqref{eq:eq-short} yields a nonlinearity parameter of
\begin{equation}\label{eq:JS}
J^S = \frac{2\tilde{\psi}^2 k^4 l_\xi^4}{{I}}.
\end{equation}

Calculation of $J^L$ and $J^S$ act as a verification of the weak nonlinear assumption of \ac{WT}. This is especially important in the context of  experimental implementations of \ac{OWT}, where initially unknown quantities are often difficult to measure.

\subsection{The Hamiltonian Formulation}

Both equations (\ref{eq:eq-long}) and (\ref{eq:eq-short}) can be written in terms of a Hamiltonian system of the form
\begin{equation}\label{eq:hamsysot}
2iq\frac{\partial\psi}{\partial z} = \frac{\delta \mathcal{H}}{\delta
\psi^*}.
\end{equation}
For the \ac{LWE}, the  Hamiltonian is given as 
\begin{subequations}\label{eq:hams}
\begin{eqnarray}\label{eq:Ham_long}
\mathcal{H}^L&=& \mathcal{H}_2 + \mathcal{H}^L_4,\nonumber\\
&=& \int \left\{ {\left |\frac{\partial \psi }{ \partial x} \right |^2-
\frac 1{4  \tilde{\psi}^2}
\left [ \frac{|\psi |^4}{l_\xi^2} -
 \left ( \frac{\partial |\psi |^2 }{ \partial x}
\right )^2  \right ] } \right\} \; dx.
\end{eqnarray}
In the nonlinear energy term $\mathcal{H}_4$, the term quartic with respect to $\psi$, we have added a superscript $L$ to denote that this quartic term corresponds to the \ac{LWE}, (\ref{eq:eq-long}).  This is because the Hamiltonian of the \ac{LWE} and \ac{SWE} only differ in the expression $\mathcal{H}_4$.  For the \ac{SWE}, the Hamiltonian is given by
\begin{eqnarray}\label{eq:ham_short}
\mathcal{H}^S&=& \mathcal{H}_2 + \mathcal{H}^S_4,\nonumber\\
&=&\int \left[ \left |\frac{\partial \psi }{ \partial x} \right|^2 - 
\frac 1{4 l_\xi^4 \tilde{\psi}^2}
\left ( \frac{\partial^{-1} |\psi |^2 }{ \partial x^{-1}}
\right )^2 \right] \; dx.
\end{eqnarray}
\end{subequations}

In both the \ac{LWE} and the \ac{SWE}, the linear, (quadratic), energy $\mathcal{H}_2$ is identical.  The Hamiltonians \eqref{eq:hams} coincide with the total energy of the systems and are conserved by their respective dynamics ($\mathcal{H} =$ const).  Moreover, both the \ac{LWE} and \ac{SWE} contain an additional invariant, the wave action $\mathcal{N}$ defined as 
\begin{equation}\label{eq:waveaction}
\mathcal{N}=\int |\psi|^2 dx.
\end{equation}
Conservation of  $\mathcal{N}$ is a consequence of the $U(1)$ gauge symmetry or invariance of equations \eqref{eq:eq-long} and \eqref{eq:eq-short} with respect to a phase shift: $\psi(x,z) \to \psi(x,z)\exp\left(i\phi\right)$.

By expressing the Hamiltonian in terms of its Fourier representation
\begin{equation}
\psi(x,z) = \sum_\bk a(\bk,z) e^{i\bk x},
\end{equation}
here $\bk \in \mathbb{R}$, the general Hamiltonian structure for Hamiltonians \eqref{eq:hams} can be represented in terms of the wave amplitude variable:
\begin{equation}\label{eq:4wham-ot}
 \mathcal{H} = \sum_{\bk} \omega_k\; a_\bk a_{\bk}^* + \frac{1}{4}\sum_{1,2,3,4} T^{1,2}_{3,4}\ a_1 a_2 a_3^* a_4^*\ \delta^{1,2}_{3,4}, 
 \end{equation}
where $\omega_k = k^2$ is the linear frequency\footnote{Indeed, $\omega_k$ is the frequency with respect to the time variable which is related to the distance $z$ 
 as $t
= z/2q$. }
of non-interacting waves, $\delta^{1,2}_{3,4} = \delta(\bk_1 + \bk_2 - \bk_3 - \bk_4)$ is a Kronecker delta function, $T^{1,2}_{3,4} = T(\bk_1,\bk_2,\bk_3,\bk_4)$ is the nonlinear interaction coefficient, and the subscripts in the summation correspond to the summation over the associated wave numbers. Note that we use bold symbol $\bk$ for the wave number to emphasize that it can be either positive or negative, while $k$ is reserved specifically for the wave vector length, $k=|\bk|$.

By symmetry arguments, the interaction coefficient should not change under the permutations $\bk_1 \leftrightarrow \bk_2$, $\bk_3 \leftrightarrow \bk_4$.
Furthermore, the Hamiltonian \eqref{eq:4wham-ot} represents the total energy of the system and is therefore a real quantity.  This property implies extra symmetries of the interaction coefficient:
\begin{equation}\label{eq:sym4}
 T^{1,2}_{3,4} = T^{2,1}_{3,4} = T^{1.2}_{4,3} = (T^{3,4}_{1,2})^*.
\end{equation}

For the  \ac{LWE} Hamiltonian (\ref{eq:Ham_long}),  the interaction coefficient is defined as follows,
\begin{subequations}\label{eq:longcoefficients}
 \begin{eqnarray}
  {}^{\mathrm{L}}T^{1,2}_{3,4} &=& {}^1T^{1,2}_{3,4} + {}^2T^{1,2}_{3,4} \nonumber \\
 &=&-\frac 1{ l_\xi^2 \tilde{\psi}^2}
+\frac 1{2  \tilde{\psi}^2}
\left(\bk_1\bk_4+\bk_2\bk_3+\bk_2\bk_4+\bk_1\bk_3
-2\bk_3\bk_4-2\bk_1\bk_2\right)\label{eq:longcoefficients2}.
 \end{eqnarray}
\end{subequations}
We have denoted the two contributions to ${}^{\mathrm{L}}T^{1,2}_{3,4}$, from both nonlinear terms in (\ref{eq:eq-long}), as  ${}^1T^{1,2}_{3,4}$ and ${}^2T^{1,2}_{3,4}$, where the first arises from the usual cubic nonlinearity seen in the \ac{1D} focusing \ac{NLSE}, and the second from the sub-leading  correction. 

Similarly, the \ac{SWE} yields the following interaction coefficient,
\begin{subequations}\label{eq:shortcoefficients}
 \begin{eqnarray}
  {}^{\mathrm{S}}T^{1,2}_{3,4} &=& 
\frac 1{ 2 l_\xi^4 \tilde{\psi}^2}
\left(
\frac{1}{\bk_1\bk_3}+\frac{1}{\bk_2\bk_3}+\frac{1}{\bk_1\bk_4}+\frac{1}{\bk_2\bk_4}-\frac{2}{
\bk_1\bk_2}-\frac{2}{\bk_3\bk_4}\right).
\label{eq:shortcoefficients2}
 \end{eqnarray}
\end{subequations}
In terms of the wave amplitude variables $a(\bk)$, the Hamiltonian system \eqref{eq:4wham-ot} satisfies the Fourier space analogue of equation \eqref{eq:hamsysot}:
\begin{equation}\label{eq:hamsys}
 2iq\frac{\partial a(\bk,z)}{\partial z} = \frac{\delta \mathcal{H}}{\delta a^*(\bk,z)}.
\end{equation}

It is with equation~\eqref{eq:hamsys} that the formulation of \ac{WT} theory is applied. In the next Section, we will give a brief mathematical description of \ac{WT} theory, and outline the assumptions on the wave field that is required to apply such an approach.

\subsection{The Canonical Transformation}
\label{sec:canonical}

Nonlinear wave interactions can be classified by the lowest order of resonance interactions they undergo. For an $N \leftrightarrow M$ wave scattering process, these resonance conditions are defined as 
\begin{subequations}\label{eq:resmanfull} 
 \begin{equation}\label{eq:resk}
\bk_1 + \cdots + \bk_N = \bk_{N+1} + \cdots + \bk_{N+M},
 \end{equation}
\begin{equation}\label{eq:resomega}
\omega_1 + \cdots +\omega_N =  \omega_{N+1}+ \cdots +\omega_{N+M},
\end{equation}
\end{subequations}
where $\bk_i$ is the wave number and $\omega_i = \omega(\bk_i)$ is the frequency of wave $i$.  

The lower orders of nonlinearity can be eliminated using a quasi-identity \ac{CT} which is similar to the Poincar\'e-type  algorithm used in, e.g. the construction of the corrected wave action for the perturbed integrable systems in \ac{KAM} theory. The latter represents a recursive procedure eliminating the lower-order interaction terms one by one at each of the recursive steps, which is only possible when there are no resonances at that respective order. In \ac{WT} theory, such a recursion is ``incomplete'' - it contains a finite number of recursive steps until the later steps are prevented by the lowest order wave resonances. The \ac{CT} procedure for eliminating the non-resonant interactions in \ac{WT} theory is explained in \cite{ZLF92}, where the most prominent example given was for the system of gravity water waves, where it was used to eliminate the non-resonant cubic Hamiltonian (see also \cite{K94} where some minor mistakes made for gravity waves were corrected).

\begin{figure}
\begin{center}
\includegraphics[width=0.8\columnwidth]{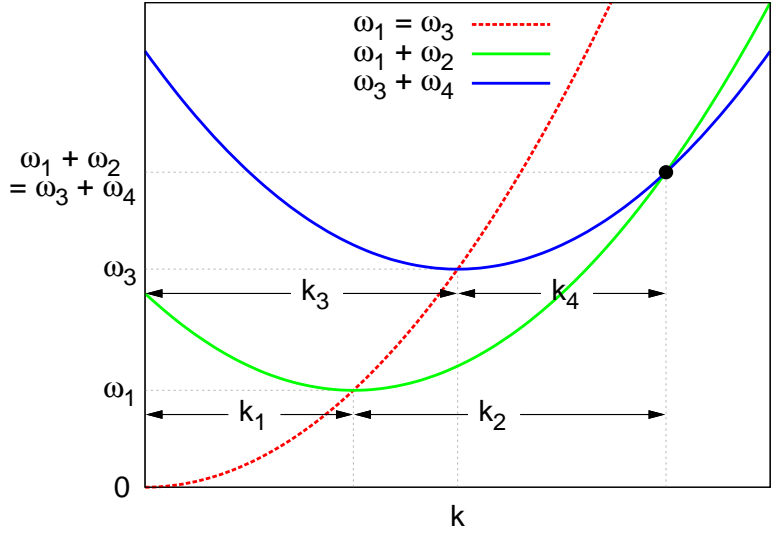}
\caption{We plot a graphical representation of the four-wave resonance condition \cite{N10}.  The four-wave resonance condition is satisfied at points where the green and blue lines intersect, (shown by the black dot).  However, for dispersion relations $\omega_k\propto\bk^\alpha$, with $\alpha > 1$, there can only be one intersection, corresponding to the trivial wave resonance: $\bk_1=\bk_4$ and $\bk_2=\bk_3$.\label{fig:wkrel}}
\end{center}
\end{figure}

Of course, apart for satisfying the resonant conditions, the respective type of the nonlinear coupling must be present.
For example, the \ac{2D} and \ac{3D} \ac{NLSE} have the dispersion relation $\omega_k = k^2$ which can satisfy the three-wave $1 \leftrightarrow 2$ resonance conditions, but the three-wave nonlinear coupling is zero. On the other hand, for gravity water waves there is a $1 \leftrightarrow 2$ wave interaction Hamiltonian (when written in terms of the natural variables - height and velocity potential), but the wave linear frequency $\omega_k = \sqrt{gk}$ does not allow for $1 \leftrightarrow 2$ resonances \cite{V67}. As a consequence, the lowest order resonant processes in all of these cases are four-wave ($2 \leftrightarrow 2$). 
For \ac{1D} \ac{OWT}, like in the \ac{NLSE},
the frequency of the linear propagating waves is given by
\begin{equation}
\omega(\bk) = k^2,
\end{equation}
In \ac{1D}, dispersion relations  of the form $\omega(\bk)\propto k^\alpha$ with $\alpha > 1$ cannot satisfy the four-wave resonance condition:
\begin{subequations}\label{eq:4wrc}
 \begin{equation}
  \bk_1 +\bk_2 = \bk_3+\bk_4,
 \end{equation}
\begin{equation}
 \omega(\bk_1) +\omega(\bk_2) =\omega(\bk_3) +\omega(\bk_4).
\end{equation}
\end{subequations}
 This can be understood by a simple graphical proof presented in figure~\ref{fig:wkrel} \cite{N10}.  In figure~\ref{fig:wkrel} we  observe the red dashed curve representing the dispersion relation $\omega_k = Ck^{\alpha}$ with $\alpha > 1$.  At two locations along this curve (at $\bk = \bk_1$ and $\bk = \bk_3$), two further dispersion curves (the  green and blue solid lines) are produced: with their minima at points $(\bk,\omega) = (\bk_1,\omega_1)$ and at $(\bk_3,\omega_3)$ respectively.  These subsequent lines represent the wave frequencies of $\omega_1 + \omega_2$ and $\omega_3 + \omega_4$, where $\bk_1$ and $\bk_3$ are now fixed, with $\bk_2$ varying  along the green solid line and $\bk_4$ varying along the blue line.  If the green and blue lines intersect, it will be when the four-wave resonance condition~\eqref{eq:4wrc} is satisfied and will occur at the point $(\bk,\omega) = (\bk_1 + \bk_2,\omega_1 + \omega_2) =(\bk_3 + \bk_4,\omega_3 + \omega_4)$. 
In figure~\ref{fig:wkrel}, we observe that such an intersection occurs only once, and it can be clearly seen that $\bk_1 = \bk_4$ and $\bk_2 = \bk_3$ must hold.  This corresponds to a trivial pairing of wave numbers, which will not provide any nonlinear energy exchange between modes.
As a consequence, resonant four-wave interactions are absent in the system.  There are no five-wave interactions either because the $U(1)$ symmetry prohibits the presence of odd orders in the wave amplitude in the interaction Hamiltonian. In situations such as these, there exists a weakly nonlinear \ac{CT} that allows us to change to new canonical variables such that the leading interaction Hamiltonian is of order six.  

\begin{figure}
\begin{center}
\includegraphics[width=0.8\columnwidth]{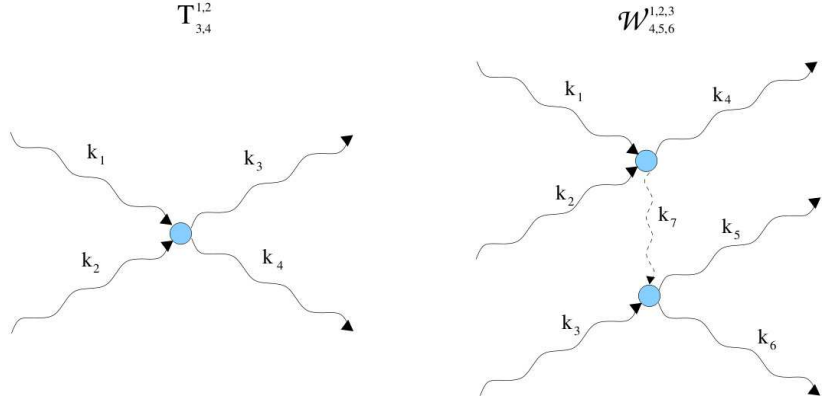}
\caption{An illustration to show the non-resonant four-wave interaction, $T^{1,2}_{3,4}$, and the resonant six-wave interaction, $\mathcal{W}^{1,2,3}_{4,5,6}$, after the \acs{CT}.  The six-wave (sextet) interaction term  is a contribution arising from two coupled four-wave (quartet) interactions via a virtual wave (dashed line). \label{fig:interaction}}
\end{center}
\end{figure}

A similar strategy was recently applied to eliminate non-resonant fourth-order interactions in the context of Kelvin waves in superfluid turbulence \cite{BCDLN09, LLNR10} and in nonlinear optics \cite{BLNR09}.  
The details of the \ac{CT} for our optical system can be found in Appendix~\ref{Appendix:ct}.  The result of the \ac{CT} is the representation of our system in a new canonical variable $c_\bk$ with the elimination of the quartic contribution $\mathcal{H}_4$.  This however results in the appearance of a sextic contribution $\mathcal{H}_6$ :
\begin{equation}\label{eq:6wham}
  \mathcal{H} = \sum_\bk \omega_k c_\bk c_\bk^* + \frac{1}{36}\sum_{1,2,3,4,5,6} \mathcal{W}^{1,2,3}_{4,5,6}\ \delta^{1,2,3}_{4,5,6}\ c_1c_2c_3c^*_4c^*_5c^*_6,
\end{equation}
where the explicit formula for $\mathcal{W}^{1,2,3}_{4,5,6}$ stemming from the \ac{CT} is given by
\begin{equation}\label{eq:6wintcoef}
\mathcal{W}^{1,2,3}_{4,5,6}=-\frac{1}{8}\displaystyle\sum^{3}_{\substack{i,j,m=1\\ i\neq j\neq
m\neq i}}\displaystyle\sum^{6}_{\substack{p,q,r=4\\ p\neq q \neq r\neq
p}}\frac{T^{p+q-i,i}_{p,q}\ T^{j+m-r,r}_{j,m}}{\omega^{j+m-r,r}_{j,m}}+\frac{T^{i+j-p,p}_{i,j}\ T^{q+r-m,m}_{q,r}}{\omega^{q+r-m,m}_{q,r}},
\end{equation}
where we have use the notation $\omega^{1,2}_{3,4} = \omega_1+\omega_2-\omega_3-\omega_4$. Note that analogous to the symmetries of the four-wave interaction coefficient $T^{1,2}_{3,4}$, we must similarly impose the following symmetry conditions on $W^{1,2,3}_{4,5,6}$ to ensure the Hamiltonian is real:
\begin{equation}\label{eq:sym6}
\mathcal{W}^{1,2,3}_{4,5,6} = \mathcal{W}^{2,1,3}_{4,5,6}=\mathcal{W}^{3,2,1}_{4,5,6} = \mathcal{W}^{1,3,2}_{4,5,6} =  \mathcal{W}^{2,3,1}_{4,5,6}=\mathcal{W}^{3,1,2}_{4,5,6}=\left(\mathcal{W}_{1,2,3}^{4,5,6}\right)^*.
\end{equation}
Hamiltonian~\eqref{eq:6wham} represents the the original Hamiltonian system \eqref{eq:4wham-ot}, but now in the new canonical variable $c_\bk$.  The interaction Hamiltonian has now been transformed from having a leading non-resonant fourth-order interaction term into one with a leading resonant six-wave interaction.  From the formula of the new six-wave interaction coefficient \eqref{eq:6wintcoef}, the six-wave interaction stems from the coupling of two non-resonant four-wave interactions connected by a {\em virtual} wave (an illustration is presented in figure~\ref{fig:interaction}).

By substituting Hamiltonian~\eqref{eq:6wham} into equation \eqref{eq:hamsys}, we derive an evolution equation for  the wave action variable $c_\bk$, 
\begin{equation}\label{eq:6_wave_evol}
 i\dot{c}_\bk = \omega_\bk c_\bk + \frac{1}{12} \sum_{2,3,4,5,6}\; \mathcal{W}^{\bk,2,3}_{4,5,6}\; c^*_2c^*_3c_4c_5c_6\; \delta^{\bk,2,3}_{4,5,6},
\end{equation}
where we denote the ``time'' derivative of $c_\bk$ as $\dot{c}_{\bk} = \partial c_k / \partial z$.
This equation is the starting point for \ac{WT} theory.  This is the six-wave analogue of the well-known Zakharov equation describing four-wave interactions of water surface waves \cite{Z68}.

\section{Wave Turbulence Theory}


General formulation of \ac{WT} theory can be found in a recent book \cite{N10}, reviews \cite{BNN01,ZDP04,RN11} and the older classical book \cite{ZLF92}.
In our review will will only outline the basic ideas and steps of \ac{WT}, following mostly the approach of \cite{N10}. More details
will be given on the parts not covered in these sources, namely the six-wave systems arising in \ac{OWT} and respective solutions
and their analysis.


\label{sec:stat}


Let us consider a \ac{1D} wave field, $a(x,z)$, in a domain which is periodic in the $x$-direction with period $L$, and let the Fourier transform of this field be represented by the Fourier amplitudes $a_l(t)=a(\bk_l,z)$, with wave number, $\bk_l = 2\pi l/L$, $l \in \mathbb{Z}$. Recall that the propagating distance $z$ in our system plays a role of ``time'', and consider the amplitude-phase  decomposition  $a_l(t)=A_l(z)\psi_l(z)$, such that $A_l$ is a real positive amplitude and $\psi_l$ is a phase factor that takes values on the unit circle in the complex plane.
Following \cite{CLNP05, LN04, CLN04,CLN05}, we  say a wave field $a(x,z)$ is an \ac{RPA} field, if  all the amplitudes $A_l(z)$ and the phase factors $\psi_l(z)$
are independent random variables and
all $\psi_l$s are uniformly distributed on the unit circle on the complex plane.  
We remark, that the  \ac{RPA} property does not require us to fix the shape of the amplitude \ac{PDF}, and therefore, we can deal with strongly non-Gaussian wave fields. This will be useful for the description of \ac{WT} intermittency.

Construction of the \ac{WT} theory for  a particular wave system consists of three main steps: {\it i}) expansion of the Hamiltonian dynamical equation (equation \eqref{eq:6_wave_evol} in our case) in powers of a small nonlinearity  parameter, {\it ii}) making the assumption that the initial wave field is \ac{RPA} and statistical averaging over the initial data, and finally {\it iii}) taking the large box limit followed by the weak nonlinearity limit.
Since the derivations for our \ac{OWT} example are rather technical and methodologically quite similar to the standard procedure described in \cite{N10}, we move these derivations  to Appendix~\ref{Appendix:wwt}.  
This approach leads to an evolution equation for the one-mode amplitude \ac{PDF} $\mathcal{P}_\bk(s_\bk)$ which is the probability of observing the wave intensity $J_\bk=|A_\bk|^2$ of the mode $\bk$ in the range $(s_\bk,s_\bk+ds_\bk)$:
\begin{equation}\label{eq:evol_pdf}
 \dot{\mathcal{P}}_\bk = -\frac{\partial \mathcal{F}_\bk}{\partial s_\bk},\qquad \mathcal{F}_\bk = -\left(s_\bk \gamma_\bk\mathcal{P}_\bk+s_\bk\eta_\bk\frac{\partial \mathcal{P}_\bk}{\partial s_\bk}\right),
\end{equation}
where we have introduced a probability space flux $\mathcal{F}_\bk$ and where
\begin{subequations}
\begin{equation}\label{eq:eta}
 \eta_\bk=\frac{\epsilon^8\pi}{6}\int|\mathcal{W}^{\bk,2,3}_{4,5,6}|^2\  \delta^{\bk,2,3}_{4,5,6}\ \delta({\omega}^{k,2,3}_{4,5,6})\ n_2 n_3 n_4 n_5 n_6\ d\bk_2  d\bk_3 d\bk_4 d\bk_5 d\bk_6,
\end{equation}
and
\begin{eqnarray}
 \gamma_\bk &=& \frac{\epsilon^8\pi}{6} \int |\mathcal{W}^{\bk,2,3}_{4,5,6}|^2\  \delta^{\bk,2,3}_{4,5,6}\ \delta({\omega}^{k,2,3}_{4,5,6}) \left[\left(n_2+n_3\right)n_4n_5n_6\right.\nonumber\\
&&\left.-n_2n_3\left(n_4n_5+n_4n_6+n_5n_6\right)\right]\   d\bk_2 d\bk_3 d\bk_4 d\bk_5 d\bk_6.\label{eq:gamma}
\end{eqnarray}
\end{subequations}
 Multiplying equation  \eqref{eq:evol_pdf} by $s_\bk$ and integrating over $s_\bk$, we then obtain the \ac{KE}, an evolution equation for the wave action density $n_\bk = \langle |c_\bk|^2 \rangle$:
\begin{equation}\label{eq:evol_n}
\dot{n}_\bk=\eta_\bk - \gamma_\bk n_\bk,
\end{equation}
Our main focus is on the non-equilibrium steady state solutions of  equations \eqref{eq:evol_pdf} and \eqref{eq:evol_n}.

\subsection{Solutions for the One-Mode PDF: Intermittency}
\label{sec:pdf}

The simplest steady state solution corresponds to the zero flux scenario, $\mathcal{F}_\bk=0$:
\begin{equation}\label{eq:pdfsol1}
\mathcal{P}_{\mathrm{hom}}=  \frac{1}{n_\bk} e^{-\frac{s_\bk}{n_\bk}},
\end{equation}
where $n_\bk$ corresponds to any stationary state of the \ac{KE}.  Solution $\mathcal{P}_{\mathrm{hom}}$, is the Rayleigh distribution.  
Subscript \em hom \em refers to the fact that this is the solution to the homogeneous part of a more general
solution for a steady state with
a constant non-zero flux, $\mathcal{F}_\bk= \mathcal{F} \neq 0$. The general solution in this case is \cite{CLNP05}
\begin{equation}\label{eq:pdfsol2}
 \mathcal{P}_\bk = \mathcal{P}_{\mathrm{hom}} + \mathcal{P}_{\mathrm{part}},
\end{equation}
where $\mathcal{P}_{\mathrm{part}}$ is the particular solution to equation \eqref{eq:evol_pdf}.
The particular solution is a correction due to the presence of a non-zero flux.  In the region of the \ac{PDF} tail, where $s_\bk \gg n_\bk $, we can expand $\mathcal{P}_{\mathrm{part}}$ in powers of $n_\bk/s_\bk$:
\begin{equation}\label{eq:Ppart}
 \mathcal{P}_{\mathrm{part}} = -\frac{\mathcal{F}}{s_\bk \gamma_\bk}-\frac{\eta_\bk \mathcal{F}}{\left(s_\bk\gamma_\bk\right)^2} - \cdots.
\end{equation}
Thus, at leading order, the \ac{PDF} tail has algebraic decay $\sim 1/s_\bk$, which corresponds to the presence of strong intermittency of \ac{WT} \cite{CLNP05}.  
From equation~\eqref{eq:Ppart}, we observe that a negative $\mathcal{F}$ implies an enhanced probability of high intensity events. Subsequently, a positive flux, $\mathcal{F}$, would imply that there is less probability in observing high intensity structures than what is expected by a Gaussian wave field. In \ac{WT} systems, we expect to observe both kinds of behavior each realized in  its own part of the $\bk$-space forming a \ac{WTLC}, which will be discussed later.

\subsection{Solutions of the Kinetic Equation}

The \ac{KE} \eqref{eq:evol_n} is the main equation in \ac{WT} theory, it describes the evolution of the wave action spectrum $n_\bk$.
It can be written a more compact form as
\begin{eqnarray}\label{eq:kinetic}
\dot{n}_\bk &=& \frac{\epsilon^8\pi}{6}\int  |\mathcal{W}^{\bk,2,3}_{4,5,6}|^2\ \delta^{\bk,2,3}_{4,5,6}\ \delta({\omega}^{k,2,3}_{4,5,6})\ n_\bk n_2n_3n_4n_5n_6\nonumber\\
&&\times\left(n_\bk^{-1}+n_2^{-1}+n_3^{-1}-n_4^{-1}-n_5^{-1}-n_6^{-1}\right) d\bk_2 d\bk_3d\bk_4d\bk_5d\bk_6.
\end{eqnarray}

The integral on the right hand side of the \ac{KE}, \eqref{eq:kinetic}, is known as the {\em collision integral}. Stationary solutions of the \ac{KE} are solutions that make the collision integral zero. 
There exist two types of stationary solutions to the \ac{KE}.  The first type are referred to as the thermodynamic equilibrium solutions.  
 The thermodynamic solutions correspond to an equilibrated system and thus refer to an absence of a $\bk$-space flux for the conserved quantities, (in our case, linear energy, $\mathcal{H}_2$, and total wave action, $\mathcal{N}$). The second type are known as the \ac{KZ} solutions.  They correspond to non-equilibrium stationary states determined by the transfer of a constant non-zero $\bk$-space flux.  They arise when the system is in the presence of forcing (source) and dissipation (sink), where there exists some intermediate range of scales, known as the {\em inertial range}, where neither forcing of dissipation influences the transfer of the cascading invariant. The discovery of the \ac{KZ} solutions for the \ac{KE} has been one of the major achievements of \ac{WT} theory, and as such these solutions have received a large amount of attention within the community.  In systems that possess more than one invariant, the \ac{KZ} solutions describe the transfer of invariants to distinct regions of $\bk$-space \cite{ZLF92}.  For many systems, these regions are usually the low and high wave number areas of $\bk$-space, however this is not necessarily the case for anisotropic wave systems \cite{NQ09}.  The directions in which the invariants cascade can be discovered by considering a Fj\o{}rtoft argument.   

\subsection{Dual Cascade Behavior}
\label{sec:fjortoft}

As \ac{1D} \ac{OWT} has two invariants, there are two \ac{KZ} solutions of the \ac{KE}, each defined by a constant flux transfer of either invariant.  This is analogous to \ac{2D} turbulence, where the enstrophy, (the integrated  squared  vorticity), cascades towards small scales and energy towards large scales \cite{K67,K71}. When a non-equilibrium statistical steady state is achieved in a weak nonlinear regime, the total energy (which is conserved) is dominated by the linear energy ($\mathcal{H}\approx \mathcal{H}_2$).  Hence, we can make the assumption that the linear energy is almost conserved.  This is important as the linear energy is a quadratic quantity in $\psi(x,z)$ and allows for the application of the Fj\o{}rtoft argument \cite{F53}.  This argument was originally derived in the context of \ac{2D} turbulence, and does not require any assumptions on the locality of wave interactions.  To begin, let us define the  energy flux $P(\bk,t)=P_\bk$ and wave action flux $Q(\bk,t)=Q_\bk$ by
\begin{equation}\label{eq:fluxes}
\frac{\partial \epsilon_\bk}{\partial z} = -\frac{\partial P_\bk}{\partial \bk}, \qquad \frac{\partial n_\bk}{\partial z} = -\frac{\partial Q_\bk}{\partial \bk},
\end{equation} 
where the energy density in Fourier space is defined as $\epsilon_\bk = \omega_k n_\bk$, such that $\mathcal{H}_2 = \int \epsilon_\bk\; d\bk$. Below, we will outline the Fj\o{}rtoft argument in the context of the six-wave \ac{OWT} system.

\begin{figure}[ht!]
\centerline{\includegraphics[width=0.8\columnwidth]{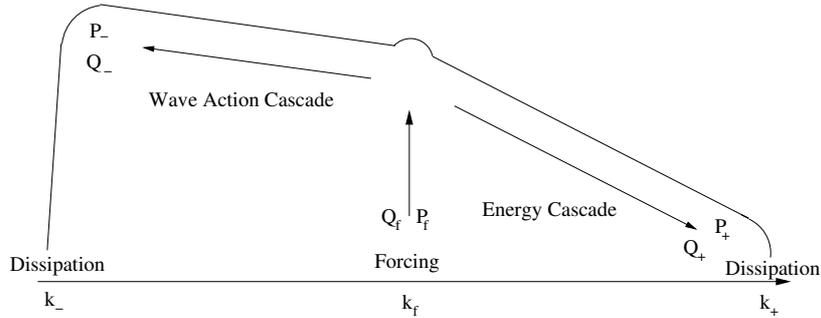}}
\caption{A Graphic to show the development of the the dual cascade regime for \ac{1D} \ac{OWT}.}
\label{fig:dualcascade}
\end{figure}

We should assume that the system has reached a non-equilibrium statistical steady state, therefore the total amount of energy flux, $P_\bk$, and wave action flux, $Q_\bk$, contained within the system must be zero, i.e. $\int P_\bk\; d\bk =0$ and $\int Q_\bk\; d\bk =0$ respectively - this corresponds to  the flux input equaling the flux output. Then, let the system be forced at a specific intermediate scale, say $\bk_f$, with both energy and wave action fluxes being generated into the system at rates $P_f$ and $Q_f$.  Moreover, let there exist two sinks, one towards small scales, say at $k_+ \gg k_f$, with energy and wave action being dissipated at rates $P_+$ and $Q_+$, and one at the large scales, say at $k_- \ll k_f$, dissipated at rates $P_-$ and $Q_-$.  Further assume that in between the forcing and dissipation, there exist two distinct inertial ranges where neither forcing or dissipation takes effect.  In the weakly nonlinear regime, the energy flux is related to the wave action flux by $P_\bk \approx \omega_k Q_\bk = k^2Q_\bk$. In a non-equilibrium statistical steady state, the energy and wave action balance implies that
\begin{equation}\label{eq:PQbal}
P_f = P_- + P_+,\qquad Q_f = Q_- + Q_+,
\end{equation}
and therefore, we approximately have 
\begin{equation}
P_f \approx k_f^2 Q_f,\qquad P_-\approx k^2_- Q_-,\qquad P_+ \approx k^2_+ Q_+.
\end{equation}
Subsequently,  the balance equations \eqref{eq:PQbal} imply
\begin{equation}\label{eq:PQbal2}
k^2_fQ_f \approx k^2_-Q_- + k^2_+Q_+,\qquad Q_f = Q_- + Q_+.
\end{equation}
Re-arranging equations \eqref{eq:PQbal2} enables us to predict at which rates the energy and wave action fluxes are dissipated at the two sinks.  From equations \eqref{eq:PQbal2} we obtain
\begin{equation}\label{eq:n_kd}
Q_+\approx\frac{k^2_f-k^2_{-}}{k^2_{+}-k^2_{-}}Q_f,\qquad Q_-\approx\frac{k^2_f-k^2_{+}}{k^2_{-}-k^2_{+}}Q_f.
\end{equation}
If we consider the region around large scales, $k_- \ll k_f < k_+$, then the first equation in \eqref{eq:n_kd} implies $k^2_fQ_f\approx k^2_{+}Q_+$,
 i.e. that energy is
mostly absorbed at the region around $k_{+}$.  Furthermore, considering the region around small scales, $k_{-} < k_f \ll k_{+}$, the second equation in \eqref{eq:n_kd} implies that $Q_f\approx Q_-$, i.e. that wave action is mostly absorbed at regions around $k_{-}$. Ultimately, if we force the system at an intermediate scale, where there exists two inertial ranges either side of $k_f$, we should expect to have that the majority of the energy flowing towards small scales and the majority of the wave action flowing towards large scales.  This determines the dual cascade picture of the six-wave system illustrated in figure~\ref{fig:dualcascade}.

\subsection{The Zakharov Transform and the Power-Law Solutions}
\label{sec:zak}

To formally derive the thermodynamic and \ac{KZ} solutions of the \ac{KE} we will use the \ac{ZT}. This requires that the interaction coefficients of the system are {\em scale invariant}. Scale invariance of an interaction coefficient is reflected by its self-similar form when the length scales are multiplied by a common factor, i.e. for any real number $\lambda\in \mathbb{R}$,  we say that an interaction coefficient is scale invariant with a {\em homogeneity coefficient} $\beta\in \mathbb{R}$, if
\begin{equation}\label{eq:homogeneity}
 \mathcal{W}(\lambda \bk_1, \lambda \bk_2, \lambda \bk_3, \lambda \bk_4, \lambda \bk_5, \lambda \bk_6) =  \lambda^\beta \mathcal{W}(\bk_1,\bk_2,\bk_3,\bk_4,\bk_5,\bk_6).
\end{equation}
Moreover, the frequency $\omega_k$ must also possess the scale invariant property, i.e.
\begin{equation}
\omega(\lambda \bk ) = \lambda^\alpha\omega(\bk),
\end{equation}
with some $\alpha \in \mathbb{R}$.  For \ac{OWT} this is indeed the case.  Since for \ac{OWT} we have $\omega(\bk) \propto k^2$,  we see that $\alpha = 2$. Let us now seek solutions of the \ac{KE} with a power-law form, 
 \begin{equation}\label{eq:powerlaw}
n_\bk = Ck^{-x},
\end{equation}
where $C$ is the constant prefactor of the spectrum, whose physical dimension is determined by the dimensional quantities within the system, and $x$ is the exponent of the spectrum.    

An informal way in determining the exponent $x$ of the \ac{KZ} and thermodynamic solutions is to apply a dimensional analysis argument.  For the thermodynamic equilibrium solutions, we assume a zero flux, i.e. that both $\epsilon_\bk$ and $n_\bk$ are scale independent. Conversely, for the derivation of the \ac{KZ} solutions, we want to consider a wave action density scaling that implies a constant flux of the cascading invariant.  This is achieved by considering $P_\bk, Q_\bk \propto k^0$ in equations
\begin{subequations}
 \begin{eqnarray}
  P_\bk &=& \int^\bk \frac{\partial \epsilon_{\bk '}}{\partial z} \; d\bk',\\
Q_\bk &=& \int^\bk \frac{\partial n_{\bk '}}{\partial z} \; d\bk',
 \end{eqnarray}
\end{subequations}
(see \eqref{eq:fluxes}), using the power-law ansatz \eqref{eq:powerlaw} and the \ac{KE}, \eqref{eq:kinetic}.   However, this method does not allow for the evaluation of the prefactor of the spectrum in \eqref{eq:powerlaw}. Therefore,  we will now describe the formal way of calculating the \ac{KE} solutions by using the \ac{ZT}.  The \ac{ZT} expresses the \ac{KE} in such a way that it overlaps sub-regions of the \ac{KE}'s domain of integration, thus at each solution, the integrand of the collision integral is set exactly to zero over the whole domain \cite{ZLF92}.  The \ac{ZT} takes advantage of the symmetries possessed within the \ac{KE} by a change of variables. In our case, this results in splitting the domain of the \ac{KE} into six sub-regions.
 
Applicability of the \ac{ZT} requires the locality of wave interactions, namely that only waves with a similar magnitude of wave number will interact.  The criterion of locality is equivalent to the convergence of the collision integral.  Locality of these solutions will be checked in the following Section. 

The \ac{ZT} is a change of variables on specific sub-regions of the domain, one such sub-region is transformed by
\begin{equation}
\bk_2 = \frac{\bk^2}{\tilde{\bk}_2},\quad \bk_3 = \frac{\bk\tilde{\bk}_3}{\tilde{\bk}_2}, \quad \bk_4 =
\frac{\bk\tilde{\bk}_4}{\tilde{\bk}_2},\quad \bk_5 = \frac{\bk\tilde{\bk}_5}{\tilde{\bk}_2} \quad \mathrm{and} \quad \bk_6 =
\frac{\bk\tilde{\bk}_6}{\tilde{\bk}_2},
\label{eq:ZT}
\end{equation}
with the Jacobian of the transformation $J =-\left(\bk /\tilde{\bk}_2\right)^6$. We must apply four similar transformations, to each of the remaining sub-regions (these are presented in Appendix~\ref{Appendix:zt}).

Using the scale invariant properties of the interaction coefficients and frequency, and the fact that a Dirac delta function scales as
\begin{equation}
\delta((\lambda \bk)^\alpha) = \lambda^{-\alpha}\delta(\bk^{\alpha}),
\end{equation}
the \ac{ZT} implies that the \ac{KE} can be expressed as
\begin{eqnarray}\label{eq:kolsol}
\dot{n}_\bk&=&\frac{C^5\epsilon^8\pi}{6}\int
|\mathcal{W}^{\bk,2,3}_{4,5,6}|^2\left|\bk\bk_2\bk_3\bk_4\bk_5\bk_6\right|^{-x}\left\{k^{x}+k_2^{x}+k_3^{x}-k_4^{x}-k_5^{x}-k_6^{
x}\right\}\nonumber\\
&\times&\left[1+\left(\frac{k_2}{k}\right)^y+\left(\frac{k_3}{k}
\right)^y-\left(\frac{k_4}{k}\right)^y-\left(\frac{k_5}{k}\right)^y-\left(\frac{k_6}{k}\right)^y\right]\nonumber\\
&\times&\ \delta^{\bk,2,3}_{4,5,6}\ \delta(\omega^{k,2,3}_{4,5,6})\ d\bk_2 d\bk_3d\bk_4d\bk_5d\bk_6,
\end{eqnarray}
where we have omitted the tildes and $y = 5x-3-2\beta$. We see that if $x=0$ or $x=2$, the integrand will vanish as we have zero by cancellation in either the curly brackets or in the square brackets (taking into account the Dirac delta function involving frequencies).  In particular, when $x=0$ or $x=2$, the solutions will correspond to thermodynamic equilibria of the equipartition of the energy and the wave action respectively:
\begin{subequations}\label{eq:thermsol}
 \begin{equation}\label{eq:thermH}
  n_\bk^{\mathrm{T}} = C^\mathrm{T}_{\mathcal{H}} k^{-2}.
\end{equation}
 \begin{equation}\label{eq:thermN}
  n_\bk^{\mathrm{T}} = C^\mathrm{T}_{\mathcal{N}} k^0,
 \end{equation}
\end{subequations}
Solutions \eqref{eq:thermsol} correspond to zero flux states - in fact both energy and wave action fluxes, $P_\bk$ and $Q_\bk$ are identically equal to zero on both equilibrium solutions. Spectra \eqref{eq:thermsol} are two limiting cases, in low and high wave number regions, of the more general thermodynamic equilibrium Rayleigh-Jeans solution
\begin{equation}\label{eq:RayleighJeans}
n_\bk^{\mathrm{RJ}}=\frac{T_\mathrm{c}}{\omega_k +\mu}.  
\end{equation}
Here $T_\mathrm{c}$ is the characteristic temperature of the system, and $\mu$ is a chemical potential.

In addition to these thermodynamic solutions, our system possesses two non-equilibrium \ac{KZ} solutions. The \ac{KZ} solutions are obtained from equation~\eqref{eq:kolsol} when either $y=0$ or $y=2$.  When either condition is met, the integrand in equation \eqref{eq:kolsol} vanishes due to cancellation in the square brackets. The corresponding solution for the direct
energy cascade from low to high wave numbers is obtained when $y=2$, this gives the following wave action spectrum scaling:
\begin{subequations}{\label{eq:kzspectra}}
\begin{equation}\label{eq:directscale}
n_\bk^{\mathcal{H}} =C_\mathcal{H}k^{-\frac{5+2\beta}{5}},
\end{equation}
where $\beta$ is the homogeneity parameter, which we will specify later depending on the \ac{LWE} or \ac{SWE} regime considered.

Note that in the experiment we do not have direct access to $n_\bk$ but we measure instead the spectrum $N(\bk,z) = |I_\bk(z)|^2$ of the light intensity. 
In Appendix~\ref{Appendix:Intspec} we present how the scaling for $N_\bk$ in the inverse cascade
state is easy to obtain from the scaling of $n_\bk$ under the random phase condition.

The wave action spectrum \eqref{eq:directscale} implies that the energy flux $P_\bk$ is
$\bk$-independent and non-zero.  The solution for the inverse wave action cascade from high to low wave numbers is
obtained when $y=0$, and is of the form
\begin{equation}\label{eq:inversescale}
n_\bk^{\mathcal{N}} =C_\mathcal{N}k^{-\frac{3+2\beta}{5}}.
\end{equation}
\end{subequations}
On each \ac{KZ} solution, the respective flux is a non-zero constant - reflecting the Kolmogorov scenario, whilst the flux of the other invariant is absent.  However, we emphasize that the \ac{KZ} solutions are only valid if they correspond to local wave interactions - an assumption of the \ac{ZT}.

The main contribution to the six-wave interaction coefficient in the \ac{LWE}, after application of the \ac{CT} and expansion in small $k l_\xi$ (see 
Appendix \ref{Appendix:LWE-W}), is 
$\bk$-independent:
\begin{equation}\label{eq:WL}
  {}^{\mathrm{L}}\mathcal{W}^{1,2,3}_{4,5,6} \approx  \frac{9}{4\tilde{\psi}^4 l_\xi^2 }. 
\end{equation}
This implies that the homogeneity coefficient for the \ac{LWE} is $\beta^{L}=0$. 

Therefore, equations \eqref{eq:kzspectra} imply that the \ac{KZ} solutions for the \ac{LWE} are given by
\begin{subequations}\label{eq:longKZ}
\begin{eqnarray}\label{eq:direct}
 {}^{\mathrm{L}}n_\bk^{\mathcal{H}} &=& C_\mathcal{H}^L\;   \left( \frac{\tilde{\psi}^2 l_\xi}{2} \right)^{4/5}
P_\bk^{1/5}   \; k^{-1},\\
  {}^{\mathrm{L}}n_\bk^{\mathcal{N}} &=& C_\mathcal{N}^L\;     
\left( \frac{\tilde{\psi}^2 l_\xi}{2} \right)^{4/5}
Q_\bk^{1/5} 
\;k^{-3/5},\label{eq:inverse}
\end{eqnarray}
\end{subequations}
where ${}^{\mathrm{L}}n_\bk^{\mathcal{H}}$, is the \ac{KZ} spectrum for the direct energy cascade, and ${}^{\mathrm{L}}n_\bk^{\mathcal{N}}$ is the inverse wave action spectrum.  Here, $C_\mathcal{H}^L$ and $C_\mathcal{N}^L$ are dimensionless constant pre-factors of the spectra.


Conversely, for the \ac{SWE} we see that the four-wave interaction coefficient, ${}^{\mathrm{S}}T^{1,2}_{3,4}$, is scale invariant and scales as $\propto k^{-2}$. Formula (\ref{eq:6wintcoef}), implies that the homogeneity coefficient for the  \ac{SWE} is $\beta^{S}=-6$. For both the \ac{LWE} and \ac{SWE} we calculated the explicit expressions for the six-wave interaction coefficients using \texttt{Mathematica}, and confirmed that $\beta^{L}=0$ and $\beta^{S}=-6$.  However we must omit the explicit expression for ${}^{\mathrm{S}}\mathcal{W}^{1,2,3}_{4,5,6}$ in this review as it is extremely long.  

Therefore, the \ac{SWE} yields the following \ac{KZ} solutions:
\begin{subequations}\label{eq:shortKZ}
\begin{eqnarray}\label{eq:directshort}
 {}^{\mathrm{S}}n_\bk^{\mathcal{H}} &=& C_\mathcal{H}^S\; 	
\left( \frac{\tilde{\psi}^2 }{2} \right)^{4/5}
P_\bk^{1/5}	\;k^{7/5},\\
 {}^{\mathrm{S}}n_\bk^{\mathcal{N}} &=& C_\mathcal{N}^S\;   
 \left( \frac{\tilde{\psi}^2 }{2} \right)^{4/5}
Q_\bk^{1/5} \; k^{9/5},\label{eq:inverseshort}
\end{eqnarray}
\end{subequations}
where,  $C_\mathcal{H}^S$ and $C_\mathcal{N}^S$ are dimensionless constants.

Spectra \eqref{eq:longKZ} and \eqref{eq:shortKZ} are valid solutions only if they correspond to local wave interactions. Therefore we must check that the collision integral converges on these spectra.

\subsection{Locality of the Kolmogorov-Zakharov Solutions}

The \ac{KZ} solutions can only be realized if they correspond to local, in $\bk$-space, wave interactions.  This entails checking that the collision integral converges, when the wave action density is of \ac{KZ} type, \eqref{eq:kzspectra}.  Our strategy for determining the locality of the \ac{KZ} solutions is to check the convergence of the collision integral when one wave number vanishes, or when one wave number diverges to $\pm\infty$.  These limits correspond to the convergence of the collision integral in the \ac{IR} and \ac{UV} regions of $\bk$-space respectively.  Although the collision integral is \ac{5D}, the two Dirac delta functions, relating to the six-wave resonance condition, imply that integration is over a \ac{3D} surface within the \ac{5D} domain. For linear frequencies of the form, $\omega_k \propto k^2$, we can parametrize the six-wave resonance condition in terms of four variables (see Appendix~\ref{Appendix:ct}), which subsequently allows us to neglect two of the integrations in \eqref{eq:kinetic}.  

The details of the locality analysis is situated in Appendix~\ref{Appendix:local}.  We performed the analysis using the \texttt{Mathematica} package, which allowed us to handle and simplify the vast number terms resulting form the \ac{CT}. Assuming that the  six-wave interaction coefficient has the following scaling as $k_6\to 0$, $\lim_{k_6 \to 0} \mathcal{W}^{\bk,1,2}_{4,5,6} \propto k_6^\xi$, where $\xi \in \mathbb{R}$.  Then we find that we have \ac{IR} convergence of the collision integral if $ x < 1+2\xi$, is satisfied.  Similarly, by assuming that the six-wave interaction coefficient scales with $k_6^\eta$ as $k_6 \to \infty$: $\lim_{\bk_6\to \infty} \mathcal{W}^{\bk,2,3}_{4,5,6} \propto k_6^\eta$, where $\eta \in \mathbb{R}$, we find that the criterion for \ac{UV} convergence of the collision integral is $\eta<x$.

We now check for convergence in the \ac{OWT} models.  We begin by investigating the locality on the \ac{LWE}.  The six-wave interaction coefficient ${}^{\mathrm{L}}\mathcal{W}^{1,2,3}_{4,5,6}$ was shown (at leading order) to be constant, with the constant given in equation \eqref{eq:WL}.  This implies that in the \ac{IR} limit  ${}^{\mathrm{L}}\mathcal{W}^{1,2,3}_{4,5,6}$ remains constant, i.e. $\xi=0$.  Hence, the condition for \ac{IR} convergence becomes $x < 1$. Due to relation~\eqref{eq:WL} being constant, we also have that  $\eta=0$, and therefore, the convergence condition for \ac{UV}  is $0<x$. Therefore, the \ac{LWE} has local \ac{KZ} spectra if their exponents are within the region $0<x<1$.  For the \ac{LWE}'s \ac{KZ} solutions \eqref{eq:longKZ}, we have that the direct cascade of energy has the exponent $x=1$, which implies divergence.  However, because this exponent is at the boundary of the convergence region, we have a slow logarithmic divergence of the collision integral.  This implies that by making  a logarithmic correction to the wave action spectrum \eqref{eq:direct}, we can prevent divergence of the collision integral - we will consider this in the next Subsection. The inverse cascade of wave action, with $x=3/5$, implies convergence of the collision integral.

Consideration of the \ac{SWE} in the \ac{IR} region, by appropriate Taylor expansion around a vanishing $\bk_6$ using \texttt{Mathematica}, reveals that the six-wave interaction coefficient, ${}^{\mathrm{S}}\mathcal{W}^{1,2,3}_{4,5,6}$, behaves as $\lim_{\bk_6\to 0} {}^{\mathrm{S}}\mathcal{W}^{1,2,3}_{4,5,6} \propto k^{-1}_6$, giving $\xi=-1$.  Therefore, the condition for \ac{IR} convergence of the \ac{KZ} solutions becomes $x<-1$.  Similarly, using \texttt{Mathematica} and expanding the interaction coefficient ${}^{\mathrm{S}}\mathcal{W}^{1,2,3}_{4,5,6}$ in the limit where $\bk_6 \to \infty$, we find that $\lim_{\bk_6\to \infty} {}^{\mathrm{S}}\mathcal{W}^{1,2,3}_{4,5,6} \propto k^{0}_6$, thus $\eta=0$.  This implies that the \ac{UV} condition for convergence is the same as that for the \ac{LWE}.  Therefore, there does not exist any region of convergence for the collision integral of the \ac{SWE}.  To be specific, both \ac{KZ} solutions of the \ac{SWE} \eqref{eq:shortKZ}, where $x=-7/5$ and $x=-9/5$ for the direct and inverse spectra respectively, we have \ac{IR} convergence and \ac{UV} divergence.  Therefore, both short-wave \ac{KZ} spectra are non-local. Non-locality of the \ac{KZ} solutions implies that the local wave interaction assumption is incorrect, and thus the approach taken to predict these spectra is invalid. However, the development of a non-local theory for the \ac{SWE} may yield further insight.

\subsection{Logarithmic Correction to the Direct Energy Spectrum}

In the previous Subsection, the direct energy cascade in the \ac{LWE}, (\ref{eq:direct}), was shown to be marginally divergent in the \ac{IR} limit.  This is to say, that the collision integral diverges at a logarithmic rate in the limit of one vanishing wave number.  However, by introducing a logarithmic dependence to the \ac{KZ} solution, we can produce a convergent collision integral. Following Kraichnan's argument for the logarithmic correction associated to the direct enstrophy cascade in \ac{2D} turbulence \cite{K67,K71}, we assume  a correction of the form:
\begin{equation}
 {}^{\mathrm{L}}n^\mathcal{H}_\bk = C_\mathcal{H}^L 
\left( \frac{\tilde{\psi}^2 l_\xi}{2} \right)^{4/5}
P_\bk^{1/5}  \,   k^{-1}\ln^{-y}(k\ell),
\end{equation}
where $y$ is some constant to be found and $\ell$ is the scale of at which energy is injected.  The exponent of the logarithmic power law, $y$, is calculated by assuming that the energy flux, $P_\bk$, remains $\bk$-independent. Subsequently, the energy flux can be expressed as
\begin{equation}
 P_\bk=\int^\bk \omega_k\frac{\partial n_\bk}{\partial t}\ d\bk \propto \int^\bk k^4n_\bk^5\ d\bk \propto \int^\bk k^{-1}\ln^{-5y}(k\ell)\ d\bk, 
\end{equation}
where we have taken into account that the collision integral (\ref{eq:kinetic}), with interaction coefficients (\ref{eq:longcoefficients}), scales as $\dot{n}_\bk \propto k^2n_\bk^5$.  Therefore, $P_\bk$ remains $\bk$-independent\footnote{The energy flux is actually proportional to $ \ln(\ln(k\ell))$.} when $y=1/5$. This implies that the logarithmically corrected direct energy \ac{KZ} spectrum is given by
\begin{equation}\label{eq:directcorrected}
 {}^{\mathrm{L}}n_\bk^{\mathcal{H}} =  C_\mathcal{H}^L 
\left( \frac{\tilde{\psi}^2 l_\xi}{2} \right)^{4/5}
P_\bk^{1/5}  \;  k^{-1}\ln^{-1/5}(k\ell).
\end{equation}
Spectrum \eqref{eq:directcorrected} now produces a convergent collision integral for the \ac{LWE} and is subsequently a valid \ac{KZ} solution. 

\subsection{Linear and Nonlinear Times and The Critical Balance Regime}
\label{sec:cb}

In this Section, we estimate the nonlinear  transfer times in the \ac{KZ} cascades and discuss the \ac{CB} states of strong \ac{WT}.  
The linear timescale is defined as $T_L=2\pi/\omega_k$.  From the \ac{KE}, we can define the nonlinear timescale as $T_{\mathrm{NL}} = 1/\frac{\partial \ln (n_\bk) }{\partial t}$. Weak \ac{WT} theory is applicable when there is a large separation between the linear and nonlinear timescales, i.e. $ \frac{T_{\mathrm{L}}}{T_{\mathrm{NL}}} \ll 1$. The estimation of $T_{\mathrm{NL}}$, with respect to $k$, can be achieved using the \ac{KE} \eqref{eq:kinetic}, giving $T_{\mathrm{NL}} \propto k^{4x-2\beta-2}$. Therefore, the condition of applicability can be written as $ {T_{\mathrm{L}}} / {T_{\mathrm{NL}}} \propto k^{-4x+2\beta} \ll 1$.  This can be violated in either of the limits $k\to 0$ or $k\to \infty$, depending on the sign of $-4x+2\beta$.  When $T_{\mathrm{L}}/T_{\mathrm{NL}} \sim 1$, then the \ac{KE} approach breaks down and we are in a strong \ac{WT} regime.

To describe  strong \ac{WT} in some physical systems the concept of \ac{CB} was suggested, first in \ac{MHD} turbulence \cite{GS95} and then later in rotating and stratified turbulence \cite{NS11} and \ac{BEC} \cite{PNO09}. \ac{CB} is defined as a turbulent state where the nonlinear evolution time $T_{\mathrm{NL}}$ is of the same order as the linear wave period $T_{\mathrm{L}}$ \em over a large range of scales. \em
A \ac{CB} scaling for the wave action density, $n_\bk$, can be made by equating $T_{\mathrm{NL}}$ with $T_{\mathrm{L}}$, on a scale by scale basis, i.e ${T_{\mathrm{L}}}/{T_{\mathrm{NL}}} \propto k^{-4x+2\beta}\sim 1$, implying a \ac{CB} spectrum
\begin{equation}
 n^{\mathrm{CB}}_\bk = C_{\mathrm{CB}} k^{-\frac{\beta}{2}}, \;\;\;\;\;\;\; C_{\mathrm{CB}} =\mbox{const}.
\end{equation}

For \ac{1D} \ac{OWT}, we estimate the ratio between the linear and nonlinear timescales using the \ac{KE} of the \ac{LWE} and \ac{SWE}.  For the \ac{LWE}, with homogeneity coefficient $\beta^L = 0$, implies that for both the direct ($x = 1$) and inverse ($x = 3/5$) \ac{KZ} spectra, the \ac{WT} criterion $ \frac{T_{\mathrm{L}}}{T_{\mathrm{NL}}} \ll 1$ will gradually be violated as $k \to 0$.  For the direct cascade, the largest scale is at the forcing. Thus, if \ac{WT} is weak at the forcing scale, it will remain weak everywhere along the cascade. In the inverse cascade, even for weak forcing, the strength of  \ac{WT} is increasing along the cascade toward the lowest wave numbers, resulting in the break down of \ac{WT} theory and the formation of nonlinear coherent structures.  For the \ac{SWE}, with homogeneity coefficient $\beta^S=-6$, and with \ac{KZ} exponents $x=-7/5$ and $x=-9/5$ for the direct and inverse \ac{KZ} spectra respectively, we have a similar situation where condition~$ \frac{T_{\mathrm{L}}}{T_{\mathrm{NL}}} \ll 1$ gets violated as the inverse cascade progress to small wave numbers.

\subsection{The Differential Approximation and the Cascade Directions}
\label{sec:dam}

The \ac{DAM} is an approximation of the \ac{KE} by assuming strongly local wave interactions.  This enables the construction of a partial differential equation for the evolution of the wave action density, $n_\bk$. We stress that the \ac{DAM} is an approximation and thus can only be justified when the \ac{KZ} solutions are proved local, i.e. the \ac{DAM} is only applicable to the \ac{KE} of the \ac{LWE}.  Therefore, we find that the \ac{DAM} is only applicable to the \ac{KE} of the \ac{LWE}. The \ac{DAM} contains both the thermodynamic and non-equilibrium \ac{KZ} solutions of the \ac{KE}, and can be further simplified to {\em reduced} \acp{DAM}, which only consider a subset of these solutions. The usefulness of the \ac{DAM} can be shown with derivation of exact analytical expressions for the fluxes and finally by its computational simplicity.  The \ac{DAM} can also be adapted to classical turbulence theory, where it is known as the Leith model \cite{L67, L68,CN04}.  This has led to the \ac{DAM} being used extensively in \ac{WT} and classical turbulence \cite{HH85,I85,ZP99,LNV04,LN06,N06,LNS06,BCDLN09}.  The derivation of the \ac{DAM} for the \ac{LWE} can be found in Appendix \ref{Appendix:dam}.  The \ac{DAM} for \ac{1D} \ac{OWT} is given by
\begin{equation}\label{eq:DAM}
\dot{n}_\omega =  S_0\omega^{1/2}\frac{\partial^2}{\partial \omega^2}\left[\omega^{\frac{9}{2}}n_\omega^6\frac{\partial^2}{\partial \omega^2}\left(\frac{1}{n_\omega}\right)\right].
\end{equation}
This equation contains the same steady state solutions (thermodynamic and non-equilibrium) as the \ac{KE}.  The thermodynamic solution, $n_\omega = T_{\mathrm{c}}/(\omega+\mu)$ is found when
\begin{equation}
R_\omega =  S_0\omega^{\frac{9}{2}}n_\omega^6\frac{\partial^2}{\partial \omega^2}\left(\frac{1}{n_\omega}\right)
\end{equation}
is identically equal to zero.  The energy and wave action fluxes, $P_\omega$ and $Q_\omega$, can be derived from $R_\omega$ by the following formulae:
\begin{subequations}\label{eq:fluxDAM}
\begin{eqnarray}
P_\omega &=& R_\omega-\omega\frac{\partial R_\omega}{\partial \omega},\label{eq:fluxDAMenergy}\\
Q_\omega &=& -\frac{\partial R_\omega}{\partial \omega}.\label{eq:fluxDAMwaveaction}
\end{eqnarray}
\end{subequations}
Consequently, both $P_\omega$ and $Q_\omega$ vanish upon the thermodynamic solution. By assuming the same power-law scaling as for the \ac{KE} \eqref{eq:powerlaw}, $n_\omega = C\omega^{-x/2}$, we can calculate formulae for the behavior of the \ac{DAM} with respect to $\omega$ and the exponent $x$: 
\begin{subequations}\label{eq:fluxdam2}
\begin{eqnarray}
\dot{n}_\omega &=& S_0 C^5 \frac{x}{2} \left(\frac{x}{2}-1\right) y \left(y-1\right) \omega^{-y-\frac{1}{2}},\\
R(\omega,x,y) &=& S_0 C^5 \frac{x}{2} \left(\frac{x}{2}-1\right) \omega^{-y+1},\\
P(\omega,x,y) &=& S_0 C^5 \frac{x}{2} \left(\frac{x}{2}-1\right)y \omega^{-y+1},\label{eq:Pomega}\\
Q(\omega,x,y) &=& S_0 C^5 \frac{x}{2} \left(\frac{x}{2}-1\right) \left(y-1\right)\omega^{-y}\label{eq:Qomega},
\end{eqnarray}
\end{subequations}
where $y = \frac{5x}{2} - \frac{3}{2}$. Stationary solutions for the \ac{DAM} are seen when $\dot{n}_\omega=0$, i.e. when $x=0$, $x=2$, $y=0$ or $y=1$, yielding the thermodynamic solutions \eqref{eq:thermsol} and the \ac{KZ} solutions of the \ac{LWE} \eqref{eq:longKZ}.  Relations \eqref{eq:Pomega} and \eqref{eq:Qomega} enable for the calculation of the sign of the \ac{KZ} fluxes.  First, we observe that both fluxes vanish upon reaching the thermodynamic solutions, when $x=0$ and $x=2$, and second, that on the \ac{KZ} solutions, the flux of the non-cascading invariant is zero.

The Fj\o{}rtoft argument of Subsection~\ref{sec:fjortoft}, showed the directions of $\bk$-space in which energy and wave action are permitted to flow.  However, the direction of the invariant cascade in the \ac{KZ} solution may sometimes contradict Fj\o{}rtoft's argument in which case they cannot be matched to any physical forcing and dissipation, and hence, they are not realizable.
Therefore, it is essential that the direction of the flux agrees with the analysis of Fj\o{}rtoft's argument.

We use the formulation of the \ac{DAM} in Subsection~\ref{sec:dam}, to determine the directions of the energy and wave action fluxes $P_\bk$ and $Q_\bk$ within the \ac{LWE}. 
By plotting formulae \eqref{eq:Pomega} and \eqref{eq:Qomega} for $P_\bk$ and $Q_\bk$ against the exponent of the \ac{KZ} solution, $x$, we can determine the sign of the fluxes.  When the exponent is that of both thermodynamic solutions \eqref{eq:thermsol}, then both fluxes should vanish.  

\begin{figure}[ht!]
\centerline{\includegraphics[width=0.8\columnwidth]{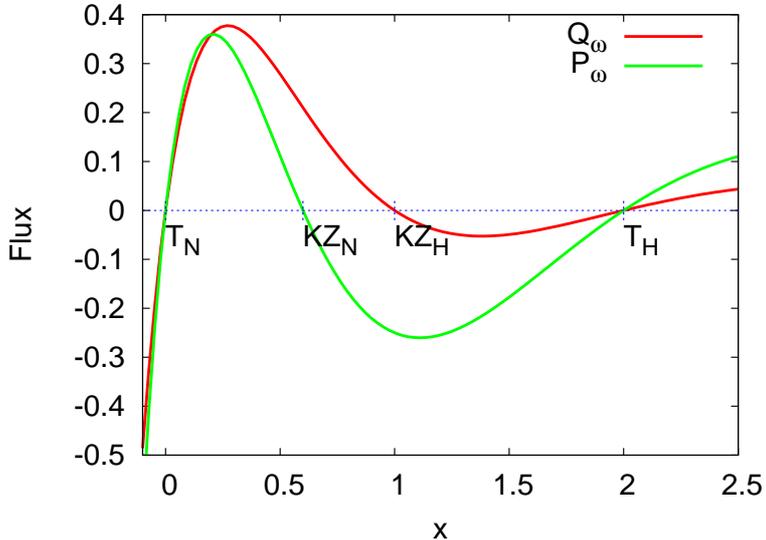}}
\caption{Plot of the energy flux $P_\omega$ and wave action flux $Q_\omega$ against the exponent $x$ of the power law distribution of $n_\omega=C\omega^{-x/2}$ in the \acs{DAM} representation of the \acs{LWE}.}
\label{fig:fluxsign}
\end{figure}

In figure~\ref{fig:fluxsign}, we plot equations (\ref{eq:fluxdam2}), for $P_\bk$ and $Q_\bk$.  we observe that at $x=0,2$ corresponding to the two thermodynamic equilibrium solutions, we have both the energy and wave action fluxes identically zero.  At $x=3/5$, corresponding to the wave action cascade, we have that the energy flux $P_\bk$ is zero, whilst the wave action flux $Q_\bk$ is {\em positive}. When $x=1$, the exponent for the energy cascade, we have the wave action flux $Q_\bk$ is zero, and the energy flux $P_\bk$ is {\em negative}.  However, to agree with Fj\o{}rtoft's argument, when $x=1$, the energy flux must be {\em positive} and the wave action flux zero.  Similarly, when the exponent agrees with the \ac{KZ} exponent of the inverse cascade, \eqref{eq:inverse}, Fj\o{}rtoft's argument implies that the wave action flux must be {\em negative}. Thus, we should not expect realizability of the pure \ac{KZ} solutions.
The authors of \cite{DNPZ92} suggest that in such situations, we might expect to observe  {\em finite temperature cascade }  solutions.  
Such ``warm cascades'' were studied in the example of the Boltzmann kinetics in \cite{PNAO11} and in the \ac{BEC} \ac{WT} context in
\cite{N10}.
These solutions are predominantly thermodynamic ones similar to (\ref{eq:RayleighJeans}), but with a correction which is small in the inertial range and which causes a sharp falloff at the dissipation scale.  A remarkable feature of the warm cascade is that its temperature is independent of the forcing rate, but dependent only on the wave numbers of the forcing and the dissipation.
 
\subsection{Modulational Instability and Solitons in the Long-Wave Equation}
\label{sec:MI}

The similarity of the \ac{LWE} (\ref{eq:eq-long}) to the integrable \ac{1D} focusing \ac{NLSE} (\ref{eq:1dnls}) means that we should expect not only random waves but also soliton-like coherent structures. During the inverse cascade, solitons appear naturally as wave action reaches the low wave numbers. Indeed, the \ac{KE} description (\ref{eq:kinetic}) becomes invalid, and the \ac{MI} of the wave field occurs resulting in the filamentation of light and its condensation into solitons.

Solitons are localized (in physical space) structures resulting from the balance between the dispersion of the wave and the nonlinear beam self-focusing.  Solitons appear in several nonlinear wave equations, with the \ac{1D} \ac{NLSE} one example.  The \ac{1D} \ac{NLSE} \eqref{eq:1dnls} can be solved exactly by the inverse scattering transform. Analytical soliton solutions of equation \eqref{eq:1dnls} have the form
\begin{eqnarray}\label{eq:soliton}
\psi_{sol}(x,z) =
2A \, l_\xi \, \tilde{\psi}\, \mathrm{sech}\left(Ax-\frac{ABz}{q}+C_1\right) 
\exp\left(i\left[Bx+\frac{\left(A^2-B^2\right)z}{2q}+C_2\right]\right),
\end{eqnarray}
where $A$, $B$, $C_1$ and $C_2$ are constants.  

Unlike
equation (\ref{eq:1dnls}), the \ac{LWE}, (\ref{eq:eq-long}), is non-integrable, and
thus, will not possess exact analytical solutions of form \eqref{eq:soliton}. 
However, the \ac{LWE}'s deviation from \ac{NLSE} is small, so we would
expect to observe coherent structures of the \ac{LWE} to be similar in shape
 to \eqref{eq:soliton}. Moreover, solitons of integrable systems possess the property of passing through one another without changing shape. However, when integrability is slightly violated, we expect to
observe weak nonlinear interactions between solitons. This may be observed as possible merging events, oscillations in profile
or interactions with weak random waves.

Formally, \ac{1D} \ac{NLSE} solitons are defined as solutions to \eqref{eq:1dnls} of the form \cite{ZDP04}
\begin{equation}
\psi_{sol}(x,z) = e^{i\Omega z /2q}\phi\left(x-\frac{Vz}{2q}\right),
\end{equation}
i.e. that the structure of the soliton is preserved as it propagates along the $z$-axis by some velocity $V$.  The Fourier transform of $\phi(x,z)$ satisfies the following integral equation:
\begin{equation}\label{eq:phik}
\phi_\bk = -\frac{1}{2\left(\Omega - \bk V + \omega_k\right)} \int T^{\bk,2}_{3,4} \phi_2^* \phi_3 \phi_4 \delta^{\bk,2}_{3,4}\; d\bk_2 d\bk_3 d\bk_4,
\end{equation}
with $T^{1,2}_{3,4}$, in the case of the \ac{1D} \ac{NLSE}, given by the first term of equation~(\ref{eq:longcoefficients}b), i.e. constant.
If the denominator of $\phi_\bk$ in equation~\eqref{eq:phik} has no zeros for $\bk \in \mathbb{R}$, then solitons may be present in the system.  If in equation~\eqref{eq:phik} the numerator and denominator contain only one zero each at $\bk = 0$ then solitons may also exist. For \ac{OWT}, when $\omega_k = k^2$, there exist soliton parameters $\Omega$ and $V$ for which the denominator is not zero. This corresponds to the cases when on the $(\bk,\omega)$-plane, the line $\omega = \bk V - \Omega$ lies below the parabola $ \omega = \omega_k =k^2$ without intersecting it. Such parameters correspond to non-radiating solitons. Let us emphasize an important difference with systems for which there exist no straight lines on the $(\bk,\omega)$-plane non-intersecting with the dispersion curve, as it is the case e.g. for the gravity waves, where $\omega_k = k^{1/2}$, and for a the \ac{MMT} model with the same dispersion relation
 \cite{ZDP04,Rumpf-PRL}. In the latter cases quasi-soliton structures are possible, but they have only a finite life time because they decay via
 radiating weak waves in a way similar to Cherenkov radiation \cite{Rumpf-PRL}. In contrast, there is no radiation of waves by solitons in our optical system, except during the soliton-soliton collisions (the latter do create wave ripples due to the weak non-integrability of the system). In fact, we will see that the dominant interactions are in the opposite direction: the waves ``condense'' into solitons with resonant parameters $\Omega$ and $V$ i.e. such that the line $\omega = \bk V - \Omega$ is tangentially touching  the parabola $\omega = \omega_k$. Subsequent evolution brings these lines further down thus decoupling the solitons from the weak waves. Hence, the solitons act as optical ``vacuum cleaners" which absorb ``optical dirt'' - the background noise consisting of weak random waves.

\begin{figure}[ht!]
\centerline{\includegraphics[width=0.8\columnwidth]{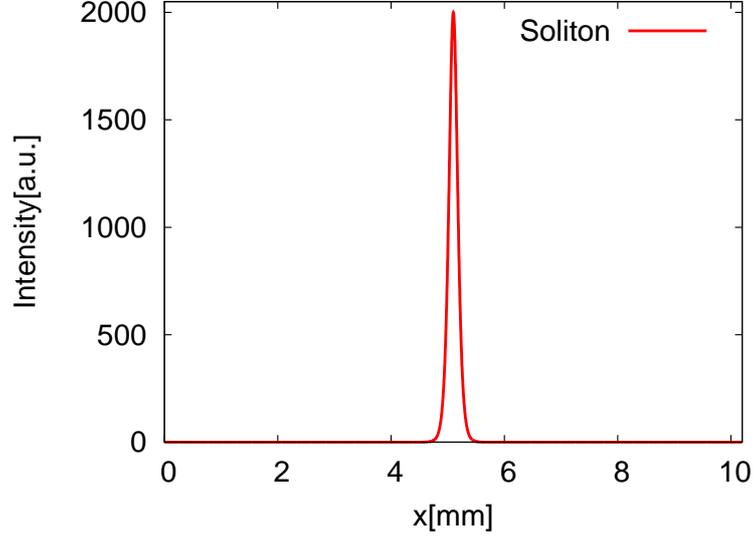}}
\caption{Intensity profile of the \acs{1D} \acs{NLSE} soliton profile given in equation (\ref{eq:soliton}).}
\label{fig:soliton_profile_32pi}
\end{figure}

\begin{figure}[ht!]
\centerline{\includegraphics[width=0.8\columnwidth]{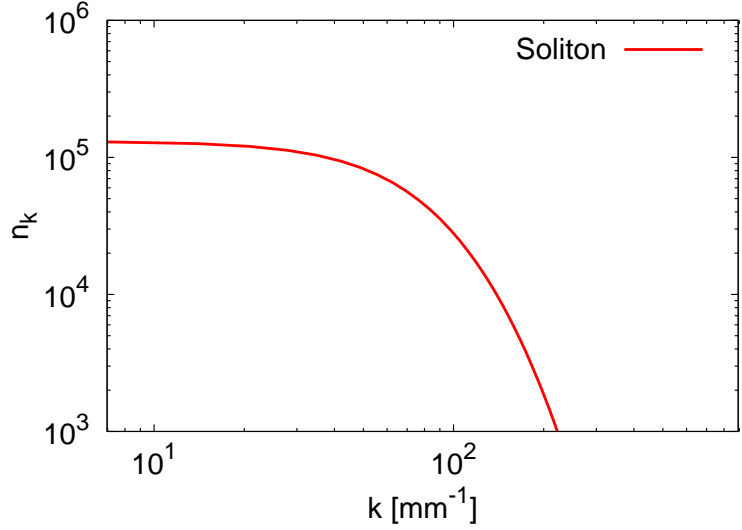}}
\caption{The corresponding wave action spectrum for the soliton profile in figure~\ref{fig:soliton_profile_32pi}}
\label{fig:soliton_spectrum_32pi}
\end{figure}

In figure~\ref{fig:soliton_profile_32pi}, we plot the profile of the \ac{NLSE} soliton \eqref{eq:soliton}, with the corresponding wave action spectrum given in figure~\ref{fig:soliton_spectrum_32pi}. We observe a scaling of $n_\bk\propto k^0$ towards low wave numbers, with a decline of the spectrum at large $k$.  However, a $n_\bk\propto k^0$ scaling is also observed from the equipartition of wave action (\ref{eq:thermsol}).  Therefore, by mere observation of the wave action spectrum, $n_\bk$, it will be difficult to determine if the wave field is comprised of equilibrated random waves or if is in the presence of solitons. To distinguish between these two states, we can numerically produce a $(\bk,\omega)$-plot, that involves an additional Fourier transform over a time window.  This method separates the random waves from the coherent component by observation of the dispersion relation, $\omega_k$ \cite{BLNR09, NO06, PNO09}. 

The inverse cascade of photons is an essential step in the creation of
solitons from a weakly nonlinear random wave field.  The cascade provides the means, via nonlinear wave interaction, for wave
action to reach large scales. As the inverse cascade develops, the nonlinearity of the system increases and a condensate forms, resulting in the dynamics of random waves deviating from the linear dispersion relation to one that is Bogoliubov modified. In the \ac{LWE}, this condensate is unstable, and undergoes \ac{MI} resulting in the formation of solitons. 
To derive an expression for the Bogoliubov dispersion relation, we must first expand the wave function $\psi(x,z)$ around a homogeneous condensate solution.   The Bogoliubov dispersion relation then comprises of the wave frequency for the disturbances upon the condensate. Its derivation can be found in Appendix \ref{Appendix:bogoliubov}.  The dispersion relation is
\begin{eqnarray}\label{eq:bogoliubov}
\omega^{\mathrm{B}}_k &=& \omega_c + \Omega_k \nonumber\\
&=& 
-\frac{I_0 }{2 \tilde{\psi}^2 l_\xi^2} 
+ \sqrt{\left(1+
\frac{I_0 }{\tilde{\psi}^2}\right)k^4 
-\frac{I_0 }{\tilde{\psi}^2 l_\xi^2} 
k^2},
\end{eqnarray}
where $I_0=|\psi_0|^2$ is the intensity of the condensate.
We observe that the Bogoliubov dispersion relation \eqref{eq:bogoliubov} is the linear wave frequency $\omega_k=k^2$ (when $I_0=0$) with a correction due to the formation of the condensate.  
 Frequency \eqref{eq:bogoliubov} becomes complex when the term inside the square root turns negative.  This gives a condition for  the \ac{MI},
\begin{equation}\label{eq:MIkcrit}
 k^2 < 
\frac{I_0}{ (\tilde{\psi}^2+ I_0) l^2_\xi   }.
\end{equation}
It is \ac{MI} that precedes and assists the formation of solitons when \ac{WT} fails to remain weak at large scales. Note that the large scale condensate is not necessarily uniform in the physical space - it may contain a range of excited large-scale modes. However, such a condensate will be felt as quasi-uniform by modes with $\bk$ significantly exceeding the lowest wave numbers. Thus, it is best to generalize the definition of the condensate intensity.  In fact, we found that the best agreement of the Bogoliubov dispersion relation to the numerical $(\bk,\omega)$-plot (figure~\ref{fig:wk.eps}) was achieved when we take $I_0$ to equal the intensity of the soliton component, i.e. the wave action contained in the region $-\omega_{\mathrm{max}}/2 < \omega < 0$.  Note that other classical examples of \ac{MI} are also found in the context of water waves, where it is known as the Benjamin-Feir instability \cite{BF67}, and in the Rossby/drift waves \cite{CNNQ10}.

Consideration of a single \ac{1D} \ac{NLSE} soliton defined by equation~\eqref{eq:soliton}, in an infinitely long domain, allows for the analytical computation of the conserved quantities of the \ac{1D} \ac{NLSE} system.  This will provide some insight into the behavior of solitons in the \ac{LWE}.    Calculation of the two energy contributions in the \ac{NLSE} Hamiltonian are given by
\begin{subequations}\label{eq:solenergy}
\begin{eqnarray}\label{eq:sollin}
\mathcal{H}_2=\int_{-\infty}^\infty \left| \frac{\partial \psi_{\mathrm{sol}}}{\partial x^2}\right|^2 \; dx &=& \frac{8A\tilde{\psi}^2l^2_\xi\left(A^2+3B^2\right)}{3},\\
\mathcal{H}_4=-\frac{1}{4\tilde{\psi}^2l_\xi^2} \int_{-\infty}^{\infty} |\psi_{\mathrm{sol}}|^4 \; dx &=& -\frac{16A^3\tilde{\psi}^2l_\xi^2}{3}.\label{eq:solnon}
\end{eqnarray}
\end{subequations} 
Although, exact \ac{NLSE} solitons do not interact, solitons in the \ac{LWE} will because of the additional non-integrable term.  Therefore, we will most probably observe behavior such as growing, shrinking, merging and collapsing. If we consider the $A$-dependence of equations \eqref{eq:solenergy}, then we observe that the nonlinear energy of the soliton increases at twice the rate (with respect to $A$) compared to the linear energy.  However, for the system to conserve energy, the linear and nonlinear energies must remain balanced for all time (with their sum equal to some initial energy, $E_0$).  From \eqref{eq:solenergy} this initial energy is given by
 \begin{equation}\label{eq:Hsol}
\mathcal{H}_{\mathrm{sol}} = E_0 = \frac{8\tilde{\psi}^2l_\xi^2A\left(3B^2-A^2\right)}{3}.
\end{equation}
For the same energy, equation~\eqref{eq:Hsol} implies  the relation
\begin{equation}
B = \sqrt{\frac{E_0}{3A}+\frac{A^2}{3}},
\end{equation}
i.e. for the soliton to grow in amplitude $A$, it must also increase its speed $\sim B/q$, in order to compensate for the balance between the linear and nonlinear energies.  However,  this would increase the wave action  $\mathcal{N}$ and the linear momentum $\mathcal{M}$ - another conserved quantity of the system.  Therefore, an important question to ask is, can two solitons merge, while wave action, energy, and linear momentum remain conserved?

\begin{figure}[ht!]
\centerline{\includegraphics[width=0.8\columnwidth]{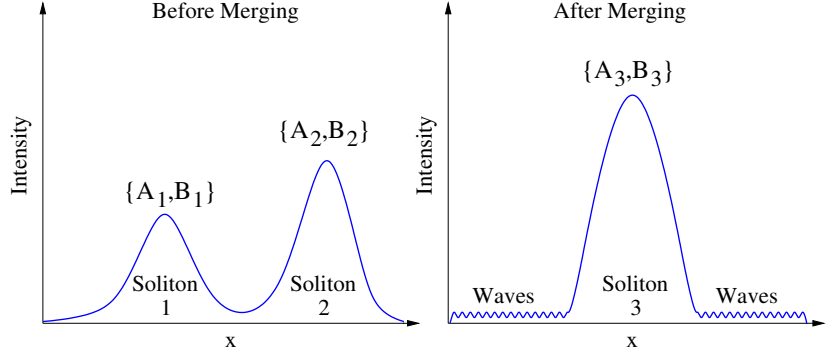}}
\caption{Graphic to show the before and after states from a soliton merger.  On the left image we see two solitons with parameters $\{A_1,B_1\}$ and $\{A_2,B_2\}$, while after we see only one soliton defined by $\{A_3,B_3\}$ but with the generation of waves, with the wave energy being determined by equation~\eqref{eq:SolEbalance}.}
\label{fig:soliton_merger}
\end{figure}

The answer to this is yes, but only if the merging of two solitons generate waves. Let us prove this fact. Consider two solitons with parameters $\{A_1, B_1\}$ and $\{A_2,B_2\}$, that defines their amplitude and speed respectively, and then assume that the two solitons merge into one soliton defined by $\{A_3,B_3\}$, see figure~\ref{fig:soliton_merger} for an illustration, 

The wave action $\mathcal{N}$ of a soliton is defined as
\begin{equation}
\mathcal{N}_{\mathrm{sol}} = \int_{-\infty}^\infty \left|\psi_{\mathrm{sol}}\right|^2 \; dx = 4A\tilde{\psi}^2l_\xi^2,
\end{equation}
and its momentum $\mathcal{M}$ is given by
\begin{equation}\label{eq:momentum}
\mathcal{M}_{\mathrm{sol}} = \frac{i}{2}\int_{-\infty}^{\infty}\left( \psi_{\mathrm{sol}}\frac{\partial \psi^*_{\mathrm{sol}}}{\partial x} - \psi_{\mathrm{sol}}^* \frac{\partial \psi_{\mathrm{sol}}}{\partial x}\right) \; dx = 8AB\tilde{\psi}^2l_\xi^2.
\end{equation}
Conservation of the total wave action and the total momentum of two solitons imply that 
\begin{subequations}\label{eq:sol_merg_cond}
\begin{eqnarray}
A_3 &=& A_1 + A_2,\\
B_3 &=& \frac{A_1B_1 + A_2B_2}{A_1+A_2},
\end{eqnarray}
\end{subequations}
must hold during the merging event.  Now if we consider the total energy conservation, but include a term that describes the generation of energy associated to waves $\mathcal{H}_{\mathrm{waves}}$ (it is reasonable to assume that low amplitude waves do not significantly contribute to the total wave action and the momentum of the system). Then the conservation of energy implies
\begin{equation}\label{eq:energybalance}
\mathcal{H}_{\mathrm{sol}1} + \mathcal{H}_{\mathrm{sol}2} = \mathcal{H}_{\mathrm{sol}3} + \mathcal{H}_{\mathrm{waves}},
\end{equation}
where $\mathcal{H}_{\mathrm{sol}i}$ is the energy defined by equation~\eqref{eq:Hsol} of soliton $i$. Then using \eqref{eq:sol_merg_cond}, \eqref{eq:energybalance} and \eqref{eq:Hsol}, we have that the energy that is generated into waves is given by
\begin{equation}\label{eq:SolEbalance}
\mathcal{H}_{\mathrm{waves}} = \frac{8\tilde{\psi}^2l_\xi^2A_1A_2\left[\left(A_1+A_2\right)^2+\left(B_1-B_2\right)^2\right]}{A_1+A_2}.
\end{equation}
Equation~\eqref{eq:SolEbalance}  states the amount of energy that is transferred to waves from the soliton merging event and is solely determined by the parameters of the initial (pre-merging) solitons.  Furthermore, equation~\eqref{eq:SolEbalance} implies that there {\em must} be waves generated from the merging of two solitons. If there were no waves created from the interaction, then the only solution to \eqref{eq:SolEbalance} would imply $A_1=0$ or $A_2 = 0$, i.e. one of the two initial solitons is absent. Therefore, we can expect the behavior of the \ac{LWE} to be as follows: As wave action reaches the largest scale by the inverse cascade, solitons begin to emerge via \ac{MI}. As these solitons evolve, they grow in size by absorbing wave action from waves and by merging with other solitons. But by merging with other solitons, the interaction emits low intensity random linear waves back into the system.  This process proceeds until there is only a single soliton remaining surrounded by a field of random linear waves.

\section{Numerical results and comparison with the experiment}
\label{sec:results}

We present the numerical results for \ac{OWT}. After presenting the numerical method, we divide this Section into two main parts, firstly, we consider the system described by the \ac{LWE} in its numerical aspect and compare with the experimental results. We begin this Subsection by considering the decaying setup with condensation at the lowest wave numbers. Then we move on to the numerical simulations of the \ac{LWE} in the forced and dissipated regime. In the final Subsection, we present the numerical results for the \ac{SWE}.

\subsection{The Numerical Method}
\label{sec:num}

For the numerical computation of the \ac{LWE} and the \ac{SWE}, we implement a pseudo-spectral method \cite{F98,CHQZ06}, with a resolution of $N=2^{12}$ Fourier modes.  The scheme utilizes the fast Fourier transform to convert physical space vectors into their Fourier representation, where differentiation of the wave field is transformed to algebraic multiplication by wave number $\bk$.  The linear terms are solved exactly in Fourier space and then converted back into physical space with the aid of integrating factors, to greatly improve the numerical stability. Nonlinear terms are more complicated. Multiplications involving $\psi(x,z)$, have to be performed in physical space.  Moreover, multiplication produces aliasing errors, due to the periodicity of the Fourier series expansion of the solution.  These errors can be removed by artificially padding the Fourier mode representation with wave number modes of zero amplitude at either ends of  $\bk$-space.  The amount of de-aliasing is subject to the degree of nonlinearity of the equation. For us, this entails that half of the wave number resolution for the solution needs to be zero - a quarter at each end of $\bk$-space \cite{CHQZ06}.

We evolve the simulations in time by using the fourth order Runge-Kutta method \cite{CHQZ06}, using a time step that is small enough so that it satisfies the \ac{CFL} condition of
\begin{equation}
 \max_{\bk}\left(\frac{\partial \omega_k}{\partial \bk}\right)\Delta t < \Delta x,
\end{equation}
where $\Delta x = L/N$ is the spacing of our spatial grid and $ L= 32\pi$, is the length of our periodic box. Numerically we compute the non-dimensionalized equation given by
\begin{equation}\label{eq:numeric}
 i \frac{\partial \psi }{ \partial z}= -\frac{\partial^2 \psi }{ \partial x^2} + N\left(\psi\right) +i\left(F_\bk - D_\bk\right),
\end{equation}
where $N(\psi)$ represents the nonlinear part of our equations. Appendix \ref{Appendix:nondim} includes the derivation of the non-dimensional versions of both the \ac{LWE} and \ac{SWE}. The non-dimensional \ac{LWE} contains an adjustable parameter $\alpha$, defined in Appendix~\ref{Appendix:nondim}.  For the decaying setup, we set $\alpha=128$, so that the numerical simulation is in the same regime as the experiment.  For the forced and dissipated cases, we set $\alpha=1024$, so that the long-wave limit is better realized.  $F_\bk=F(\bk)$ is the forcing profile, where  energy and wave action are injected into the system.  We define this in Fourier space, over a specific range of wave numbers.  For the direct cascade simulations we use a forcing profile given by
\begin{equation}\label{eq:Forcingdirect}
 F_\bk^{\mathrm{direct}} = \left\{
\begin{array}{cl}
A\exp(i\theta_\bk) & \text{if } 9\leq \frac{k}{\Delta k} \leq 11\\
0 & \text{otherwise},
\end{array}
\right.
\end{equation}
where $A$ is the amplitude of forcing and $\theta_\bk$ is a random variable chosen from a uniform distribution on $[-\pi,\pi)$ at each wave number and at each time step and $\Delta k = 2\pi / L$ is the Fourier grid spacing. For the inverse cascade simulations, we apply forcing over a small range of wave numbers situated in the high wave number region.  However, we must allow for some Fourier modes with wave numbers larger than the forcing ones, because some direct cascade range is necessary for  the development of the inverse cascade.  Our forcing profile for the inverse cascade simulations is given by
\begin{equation}\label{eq:Forcinginverse}
 F_\bk^{\mathrm{inverse}} = \left\{
\begin{array}{cl}
A\exp(i\theta_\bk) & \text{if } \frac{N}{16} -10 \leq \frac{k}{\Delta k} \leq \frac{N}{16}+10\\
0 & \text{otherwise}.
\end{array}
\right.
\end{equation}
In all simulations, we dissipate at high wave numbers by adding a hyper-viscosity term  $\propto \bk^4\; \psi_\bk$.  We use the fourth power of $\bk$, so that the dissipation profile is not too steep for the formation of a  bottleneck, or too shallow as to prevent a large enough inertial range developing.  At low wave numbers, we use two types of dissipation profile.  Firstly, we can use a hypo-viscosity term $\propto \bk^{-4}\; \psi_\bk$.  However, this type of dissipation profile in the \ac{3D} \ac{NLSE} model has led to the \ac{WT} description becoming invalid and a \ac{CB} regime to develop \cite{PNO09}.  In these situations, the \ac{CB} scenario is avoided by the use of low wave number friction.  

To summarize, our numerical dissipation profile $D_\bk=D(\bk)$, removes wave action and energy from the system at low and high wave numbers. For clarity, we split $D_\bk$ into the low and high wave number contributions $D^L_\bk$ and $D^H_\bk$.  At high wave numbers our hyper-viscosity profile is defined as
\begin{subequations}\label{eq:diss}
\begin{equation}\label{eq:Dissipation-hyper}
 D^{H}_\bk =  \nu_{\mathrm{hyper}} \; \bk^4 \; \psi_\bk,
\end{equation}
where $\nu_{\mathrm{hyper}}$ is the coefficient for the rate of dissipation.  If we apply hypo-viscosity at low wave numbers, then $D_\bk^L$ is given by
\begin{eqnarray}\label{eq:Dissipation-hypo}
 D^{L}_\bk = \left\{
\begin{array}{cl}
\nu_{\mathrm{hypo}} \; \bk^{-4}\; \psi_\bk & \text{if } \bk\neq 0\nonumber\\
\psi_\bk =0 & \text{if } \bk = 0,
\end{array}
\right. 
\end{eqnarray}
where $\nu_{\mathrm{hypo}}$ is the coefficient for the rate of dissipation at low wave numbers.  However, in situations where we use friction, then $D_\bk^L$ is defined as
\begin{equation}\label{eq:Dissipation-friction}
 D^L_\bk = \left\{
\begin{array}{cl}
\nu_{\mathrm{friction}}\; \psi_\bk & \text{if } 0\leq k \leq 6\nonumber\\
0 & \text{otherwise},
\end{array}
\right.
\end{equation}
\end{subequations}
where $\nu_{\mathrm{friction}}$ is the rate of friction dissipation.

\subsection{The Long-Wave Equation}

\subsubsection{The Decaying Inverse Cascade with Condensation}
\label{sec:decay}

The decaying simulation leading to photon condensation was originally reported by us in \cite{BLNR09}.
Both experimental and numerical setups are configured for decaying \ac{OWT}, where
an initial condition is set up and allowed to develop in the  absence of any forcing
or artificial dissipation.   We perform the numerical simulation with the same parameters as the experiment, and present results in dimensional units for comparison. 

The inverse cascade spectrum is of a finite capacity type, in a sense that
only a finite amount of the cascading invariant (wave action in this case) is
needed to fill the infinite inertial range. Indeed, this is determined as the integral of the wave action spectrum $n_\bk$ converges at $\bk=0$, i.e.
\begin{equation}
 \int_0 n_\bk \; dk \propto \int_0 k^{-3/5} \; dk < \infty.
\end{equation}
For finite capacity spectra, the turbulent systems have a
long transient (on its way to the final thermal equilibrium state) in which the
scaling is of the \ac{KZ} type (provided this spectrum is realizable, i.e. local and agrees with Fj\o{}rtoft). This is because the initial condition serves as a
huge reservoir of the cascading invariant.

Experimentally, the initial condition is setup by injecting photons at small spatial scales by modulating the
intensity of the input beam with a patterned intensity mask. We randomize the
phases by the use of a phase modulator. This is made by creating a random
distribution of diffusing spots with the average size $\simeq 35$ $\mu m$ through the \ac{SLM}.  This is done in order to create an initial condition close to
a \ac{RPA} wave field required by the theory.
The numerical initial condition is more idealized.  We restrict the initial condition to a localized  region at small scales. The initial profile is given by

\begin{eqnarray}\label{eq:numinitial}
 \psi_\bk(0) = \left\{
\begin{array}{cl}
A\exp(i\theta_\bk) & \text{if }\frac{N}{16} - 5 < \frac{k}{\Delta k} < \frac{N}{16}\\
\psi_\bk(0) =0 & \text{otherwise},
\end{array}
\right. 
\end{eqnarray}
where $A=4.608\times 10^3$ is the amplitude of forcing and $\theta_\bk$ is a random variable chosen from a uniform distribution on $[-\pi,\pi)$ at each wave number. In dimensional units, this corresponds to an initial condition in the region around $k_f\approx 1.5\times
10^2$ $mm^{-1}$.  Moreover, we
apply a Gaussian filter in physical space to achieve a beam profile comparable to that of the
experiment.

Applicability of the \ac{WT} approach is verified by the calculation of the nonlinear
parameter ${}^{\mathrm{L}}J$, equation~\eqref{eq:JL}, for the numerical simulation, which agrees with the experiment and
is of the order ${}^{\mathrm{L}}J\simeq 100$.  Experimentally, we measure the light intensity $I(x,z)=|\psi|^2$ and not the
phases of $\psi$.  Therefore the wave action spectrum $n_\bk$ is not
directly accessible. Instead, we measure the intensity spectrum, $N(\bk,z) =
|I_\bk(z)|^2$, where the $k$-scaling for $N_\bk$ in the inverse cascade
state is easily obtained from the \ac{KZ} solution \eqref{eq:inverse} and the random phase
condition, (see Appendix \ref{Appendix:Intspec}). This procedure gives an intensity spectrum of  
\begin{equation}\label{eq:N_k}
N_\bk \propto k^{-1/5}.
\end{equation}

\begin{figure}[ht!]
\centerline{\includegraphics[width=0.8\columnwidth]{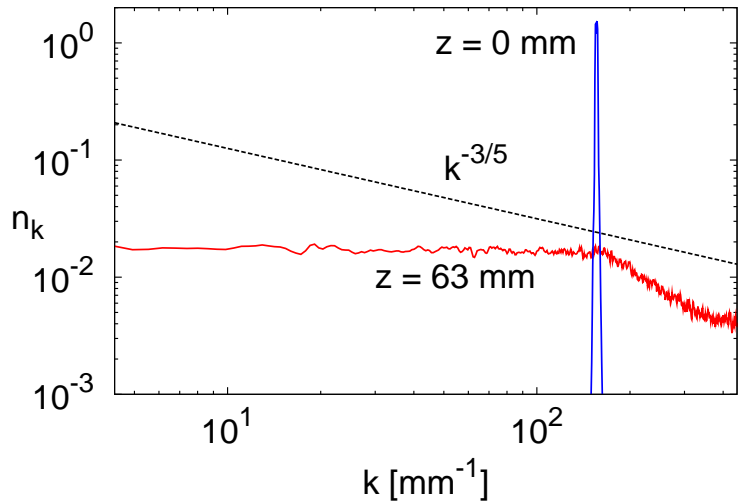}}
\caption{Numerical spectrum of the wave action, $n_\bk$, at distances $z=0$ $mm$ and $z=63$ $mm$.}
\label{inverse-num}
\end{figure}

The wave action spectrum from the numerical simulation is shown in figure~\ref{inverse-num} at two
different distances, we observe at
$z=0$ $mm$ the peak from the initial condition at high wave numbers, then at
$z=63$ $mm$ we observe evidence of an inverse cascade, as the majority of the wave action is
situated towards low wave numbers.  However, at low wave numbers we do not see the
spectrum matching our theoretical \ac{KZ} prediction \eqref{eq:inverse}. 
We showed in Section~\ref{sec:dam}, that the wave action flux, $Q_\bk$, corresponding to the \ac{KZ} solution~\eqref{eq:inverse}, has the incorrect sign for an inverse cascade. Therefore, we noted that the inverse cascade spectrum would correspond to a mixed thermal solution with a finite flux contribution.  This is the probable cause for the lack of agreement with the \ac{KZ} solution. We observe from figure~\ref{inverse-num}, that $n_\bk \propto k^0$ at low wave numbers, which may account for the mixed thermal solution (as this scaling corresponds to the equipartition of wave action \eqref{eq:thermN}) or for the presence of solitons (c.f. figure~\ref{fig:soliton_spectrum_32pi}).

\begin{figure}[ht!]
\centerline{\includegraphics[width=0.8\columnwidth]{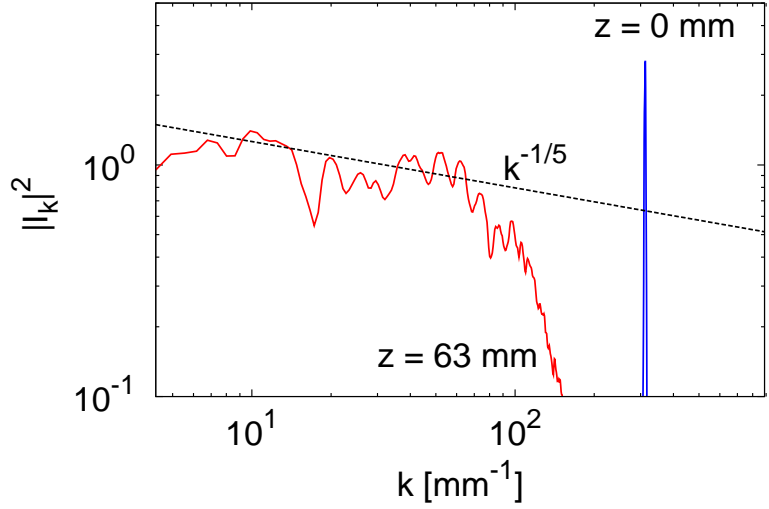}}
\caption{Numerical spectrum of the light intensity,
$N_\bk=|I_\bk|^2$ at two different distances $z$. Averaging is done over a small
finite time window and over ten realisations.}
\label{fig:inverse-num}
\end{figure}

The numerical spectrum of the light intensity is shown in
figure~\ref{fig:inverse-num}. In this plot, we observe an inverse cascade, resulting in a good agreement with the \ac{WT} prediction for the intensity spectrum (\ref{eq:N_k}).  This agreement may be coincidental, as we have shown that the flux directions are insufficient for a \ac{KZ} wave action cascade.    However, the intensity spectrum is a non-local quantity, and the system may, somehow, conspire to have better scaling properties of the \ac{KZ} type for the intensity than for the wave action density. We can only confidently say that we still do not fully understand the agreement.

Verification of the long-wave limit $kl_\xi \ll 1$ and deviation from integrability is checked by considering the ratio of the two nonlinear terms in the \ac{LWE} (\ref{eq:eq-long}), which we denote as $R$, and is estimated in Fourier space as $R \propto k^2l_\xi^2$.  We observe from the experiment and the numerical simulations, respectively, figures~\ref{fig:inverse-exp} and \ref{fig:inverse-num}, that the inverse cascade is approximately in the region $k \sim 10^4 - 10^5$  $m^{-1}$, giving an estimation of $R \sim 10^{-2} - 1$.  Note, that if $R$ is too small, then we are close to an integrable system, which would be dominated by solitons and lack cascade dynamics, and if $R$ is too high then the \ac{LWE} is a poor approximation for \ac{1D} \ac{OWT}.

\begin{figure}[ht!]
\centerline{
\includegraphics
[width=0.8\columnwidth]{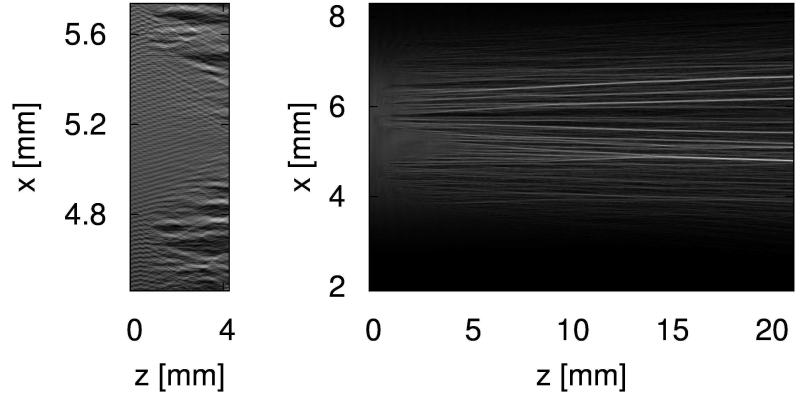}}
\caption{Numerical results for intensity distribution $I(x,z)$. The frame on the
left is a magnified section of the initial propagation of the beam.}
\label{fig:xz-num}
\end{figure}


The numerical simulation of the intensity distribution $I(x,z)$, showing the beam evolution
during propagation is displayed in figure~\ref{fig:xz-num}. Both the experimental and the numerical results, respectively figures~\ref{fig:xz-exp} and
\ref{fig:xz-num}, indicate that the total number of solitons reduces.
The observed increase of the scale and formation of coherent structures
represents the condensation of light. Experimentally, the condensation into one dominant soliton is well revealed by
the intensity profiles $I(x)$ taken at different propagation distances,
as shown in figure~\ref{fig:condensation} for $z=0.3$, $4.5$ and $7.5$ $mm$. Note
that the amplitude of the final dominant soliton is three orders of magnitude
larger than the amplitude of the initial periodic modulation.

\subsubsection{The PDF of the light intensity}

\begin{figure}[ht!]
\centerline{\includegraphics[width=0.8\columnwidth]{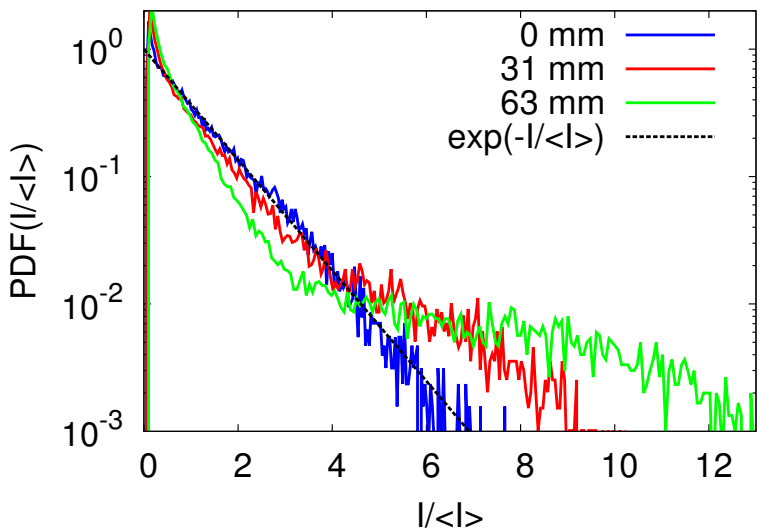}}
\caption{Normalized \acsp{PDF} of the wave intensity within the numerical cell
at three different distances along the cell, $z=0$ $mm$, $31$ $mm$ and $z=63$
$mm$. The black dashed straight line corresponds to the Rayleigh \acs{PDF} corresponding to a Gaussian wave field. 
}
\label{fig:pdf_num}
\end{figure}

Numerically, we have computed the \ac{PDF} of the light intensity and, for comparison with the experiment, we plot 
in figure~\ref{fig:pdf_num} three \acp{PDF} at distances $z=0$ $mm$, $z=31$ $mm$ and $z=63$ $mm$.
As in the experiment (see figure~\ref{fig:pdf_exp}) we observe deviations from a pure Gaussian field, with a slower that exponential decay of the \ac{PDF} tails. Again, non-Gaussianity corresponds here to {\em intermittency} of \ac{WT} and, indeed, it accompanies the development
of coherent structures (solitons) in the system.
\vskip 2cm

\subsubsection{The $k-\omega$ Plots: Solitons and Waves}

\begin{figure}[ht!]
\centerline{
\includegraphics
[width=0.8\columnwidth]{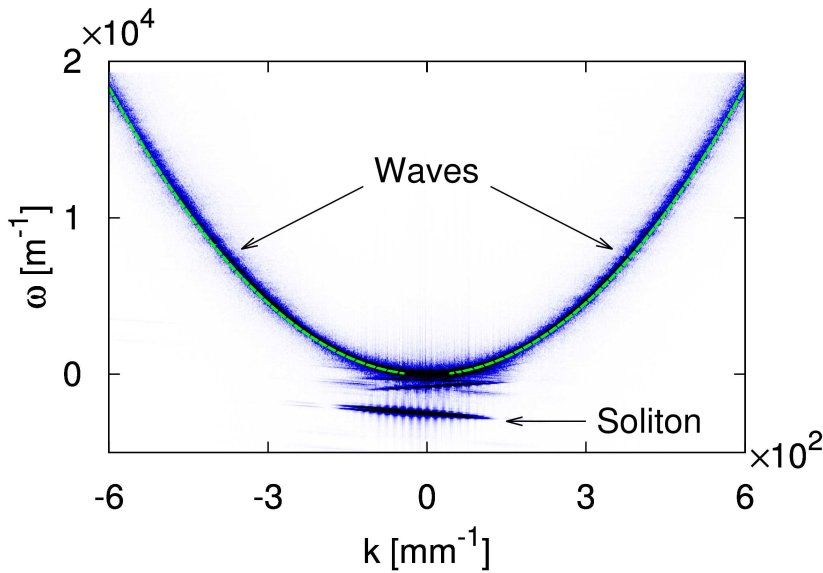}}
\caption{The $(\bk,\omega)$-plot of the wave field at $z=10$
$m$. The
Bogoliubov dispersion relation is shown by the dashed green line.}
\label{fig:wk.eps}
\end{figure}

 To allow us to capture the inverse cascade dynamics in more detail and to enable us to visualize solitons and analyze their interactions with incoherent waves, we perform an additional decaying simulation with a lower initial intensity.  However, to compensate for the lower intensity initial condition, we must run the simulation for much longer times.  Separation of the random wave and the coherent soliton components can be achieved by performing an additional Fourier transform with respect to $z$ over a finite
$z$-window which results in the \ac{2D} Fourier $(\bk,\omega)$-plots of wave action density \cite{NO06,BLNR09}. 

The numerically obtained $(\bk,\omega)$-plot enables the direct observation of the dispersion relation of random waves, and is shown in
figure~\ref{fig:wk.eps}.
Here, the incoherent wave component is distributed around the wave dispersion
relation, which is Bogoliubov-modified by
the condensate, (\ref{eq:bogoliubov}), and is shown by the solid line in figure~\ref{fig:wk.eps}. The distribution of the dispersion relation is centered around the theoretical prediction (\ref{eq:bogoliubov}).  We observe that the width of the dispersion relation is narrow for
large $\bk$, a sign of weak nonlinearity, and progressively gets wider as it approaches smaller wave numbers.  This broadening, is an indication that the nonlinearity of the system increases towards smaller $\bk$, as was theoretically predicted in Section~\ref{sec:cb}. At wave numbers around zero, we see that the theoretical Bogoliubov curve vanishes, corresponding to the region defined by equation~\eqref{eq:MIkcrit}, where the frequency becomes complex.  For such wave numbers $\bk$, \ac{MI} of the wave packets occurs and the \ac{WT} description breaks down. Below the region where the Bogoliubov relation becomes complex, we observe slanted lines. Each of these lines corresponds to a coherent soliton, whose speed is equal to the
 slope of the line.  To examine the time evolution of the system, we have put together successive $(\bk,\omega)$-plots into a movie. We observe that the formation of solitons is seen in the movie as straight lines peeling with a gradient tangential to
the dispersion curve.  Moreover, we observe the gradual migration of these lines to higher negative frequencies, as the solitons begin to grow in size by the absorption of energy from surrounding waves or by merging with other solitons.

\begin{figure}[ht!]
\centerline{
\includegraphics
[width=0.8\columnwidth]{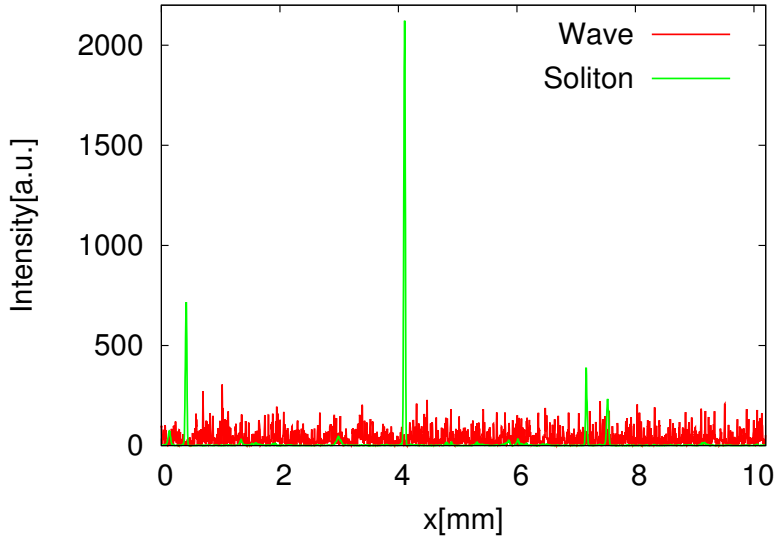}}
\caption{Plot of the intensity profile at $z=10$ $m$ of both the soliton and wave components}
\label{fig:sw-Intensity}
\end{figure}

\begin{figure}[ht!]
\centerline{
\includegraphics
[width=0.8\columnwidth]
{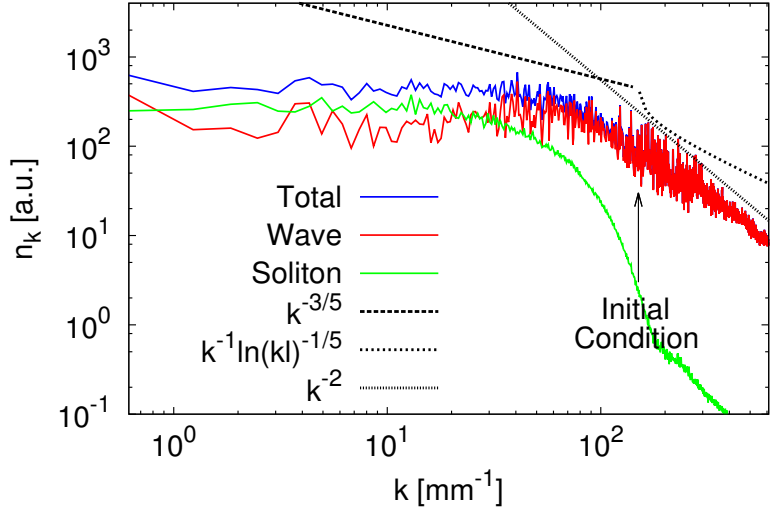}}
\caption{The wave action spectrum of both the soliton and wave components of the spectrum, with the \acs{KZ} and thermalized energy predictions for comparison.}
\label{fig:sw-Spec}
\end{figure}

Further analysis can be done by separating the wave and the soliton components of the $(\bk,\omega)$-plot.  We define the soliton region of the $(\bk,\omega)$-plot as the region defined with $-\omega_{\mathrm{max}}/2 < \omega < 0$, (where $\omega_k$ ranges from $[-\omega_{\mathrm{max}},\omega_{\mathrm{max}})$ and $\omega_{\mathrm{max}} = k_{\mathrm{max}}^2$ in the $(\bk,\omega)$-plot).  We chose this region to represent the soliton component because we found that some wave component of the dispersion relation appears in the extreme negative $\omega$-region due to the periodicity of the Fourier transform and possibly from an insufficient time window for the Fourier transform in $z$.  Consequently, the wave component is defined as the region of the $(\bk,\omega)$-plot outside the soliton region.  By inverting the Fourier transforms of each of the soliton and wave regions separately, we can recover the wave field $\psi(x,z)$ for each component.  This enables us to compare the soliton and wave components of the intensity profile, figure~\ref{fig:sw-Intensity}, and the wave action spectrum, figure~\ref{fig:sw-Spec}.  In figure~\ref{fig:sw-Intensity}, we plot the intensity profile of light in real space for the soliton and wave components.  We see that 
our procedure does indeed isolate the coherent solitons from the random wave background and we observe that the main soliton is a least an order of magnitude greater than the random wave field.  

In figure~\ref{fig:sw-Spec}, we observe that the soliton component of the wave action spectrum, is  situated towards the low wave number region and scales as $\propto k^0$.  This profile is qualitatively similar to the wave action spectrum of the exact soliton of the \ac{1D} \ac{NLSE}, seen in figure~\ref{fig:soliton_profile_32pi}.  The wave component of the wave action spectrum is more widely distributed in $\bk$-space.  Taking into account that the initial spectrum was concentrated near $k=1.5 \times 10^2 mm^{-1}$, we see that, roughly, the inverse cascade range is soliton dominated and the direct cascade range is random-wave dominated in this run. We also put a slope corresponding to the inverse cascade \ac{KZ} spectrum to show that it strongly disagrees with the spectrum in the soliton dominated range. In the direct cascade range, the spectrum is consistent with the thermalized spectrum for energy \eqref{eq:thermH}.  For comparison, we also plot the direct cascade \ac{KZ} scaling in figure~\ref{fig:sw-Spec}. We see that the thermalized spectrum is in a much better agreement with the data than \ac{KZ} .  This is not surprising, as the decaying simulation is void of any dissipation mechanism, so energy would tend to thermalize.  We would expect to see a better test of the \ac{KZ} predictions in the forced and dissipated simulations where a non-equilibrium stationary state can be achieved.  

Note that in other wave systems, the separation of the coherent and wave components, may be a useful technique in observing the \ac{KZ} scaling otherwise masked by the presence of coherent structures. 

\begin{figure}[ht!]
\centerline{
\includegraphics
[width=0.8\columnwidth]
{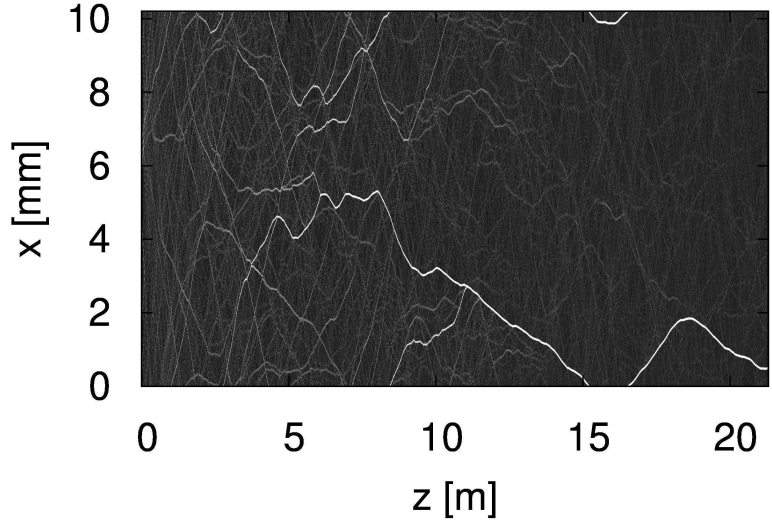}}
\caption{Numerically obtained $I(x,z)$-plot for a long time simulation with a low intensity initial condition to see the inverse cascade and soliton merging.}
\label{fig:sw-xz}
\end{figure}

\begin{figure}[ht!]
\centerline{\includegraphics[width=0.9\columnwidth]{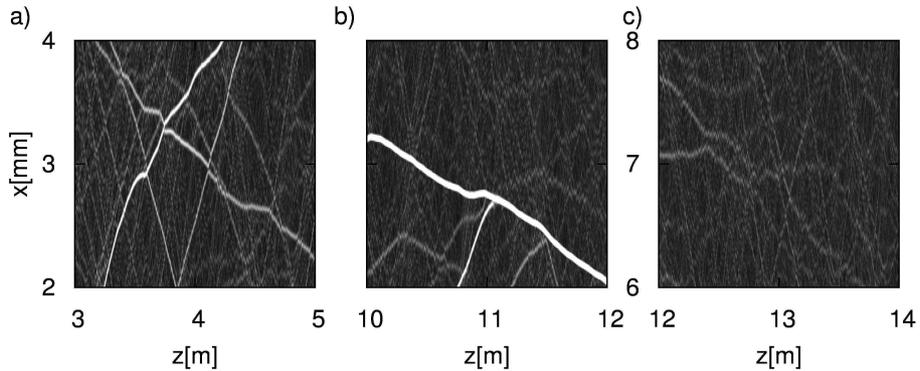}}
\caption{We present three close-up images of the (x,z)-plot of figure~\ref{fig:sw-xz}, highlighting three types of soliton behavior: a) passing through each other, b) merging  and c) dissipating.}
\label{fig:sw-xz-closeup}
\end{figure}

From the same simulation, we plot the $(x,z)$-plot for the intensity distribution, $I(x,z)$, in figure~\ref{fig:sw-xz}.  With the longer propagation distance, we observe the formation and evolution of solitons from the weakly nonlinear background.   In the initial stages of the simulation, we observe enlargement of the characteristic wavelength of the random waves followed by the formation of numerous solitons out of the random wave background.
As the simulation progresses, the solitons become more pronounced and begin to behave independently from the waves, with an almost random movement through the numerical box.  A large number of solitons are produced at the beginning of the simulation, but over long evolution times, the number of solitons reduces, with the remaining solitons being of increasing amplitude.  Indeed, there are many soliton merging events. 
In figure~\ref{fig:sw-xz-closeup}b we see a magnified section of a merging event from figure~\ref{fig:sw-xz}. At late times, figure~\ref{fig:sw-xz} shows a single dominant soliton in the system that has subsequently grown by absorption of energy from fellow solitons and the background wave field.  Such a  final single-soliton state is also observed in the experiment (figure~\ref{fig:xz-exp}), albeit at a much shorter propagation distance, with the multi-soliton range missing. 
In the future, it would be interesting to set-up a larger experiment which could capture the multi-soliton transient state.  In figure~\ref{fig:sw-xz-closeup}, we plot three zoomed sections of figure~\ref{fig:xz-exp} to highlight aspects of the soliton behavior.  In figure~\ref{fig:sw-xz-closeup}a, we observe several solitons passing through one another with little deviation in their trajectories.  Moreover, the weaker solitons (light gray in color), have almost straight trajectories, like freely propagating linear waves, whilst the stronger solitons' movements are more erratic.  

In figure~\ref{fig:sw-xz-closeup}b, we magnify a merging event between two solitons.  The larger soliton engulfs the smaller soliton almost without any deviation in its trajectory.  Therefore, in the numerical simulation, large solitons absorb energy from smaller solitons, reducing the total number within the system.  In addition, we also observe at $z=11.5$ $m$, a soliton bouncing off the larger soliton.  As the weaker soliton approaches the larger soliton, it slows, before moving away at a fast speed.  When two solitons merge and when they repel is a key question that still remains to be answered. Finally, in figure~\ref{fig:sw-xz-closeup}c, we observe two weak solitons propagating together, until around $z=13$ $m$, when they both  repel each other and disperse back into the random wave field.  This shows the break up of the coherent structures and the subsequent re-injection of wave action into high frequency waves.  All these events are parts of the \ac{WTLC}, i.e. the cycle of coexistence and  interactions of the random waves and the solitons and their mutual transformations into one another.

\begin{figure}[ht!]
\centerline{
\includegraphics
[width=0.8\columnwidth]{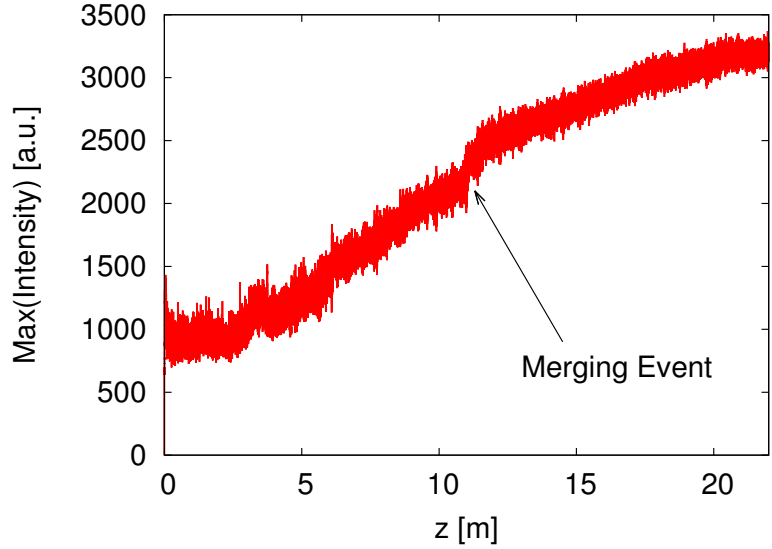}}
\caption{Plot of the maximum of the wave intensity in physical space versus $z$.}
\label{fig:sw-max}
\end{figure}

\begin{figure}[ht!]
\centerline{
\includegraphics
[width=0.8\columnwidth]{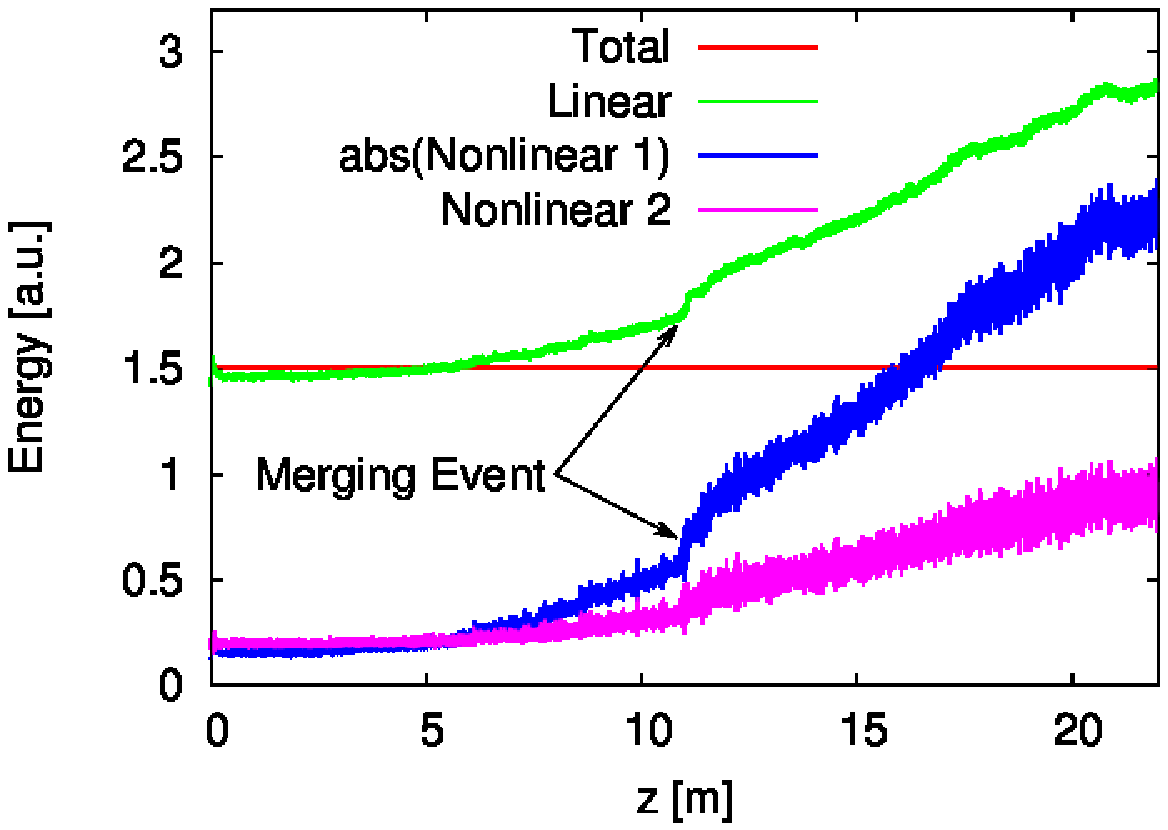}}
\caption{Plot of the energies for the long time simulation.  We see the conservation of the total energy, the growth of the linear and both nonlinear energies (corresponding to the two nonlinear terms in equation (\ref{eq:Ham_long})).}
\label{fig:sw-energy}
\end{figure}

In figures~\ref{fig:sw-max} and \ref{fig:sw-energy}, we plot the maximum of the intensity and the energy of the system with $z$.  To begin, we note that the maximum of the intensity of the wave field is always growing, and moreover, we observe that there is a sharp increase in the maximum at around $z=11$ $m$.  This jump corresponds to the merging event seen in figure~\ref{fig:sw-xz-closeup}b.  Thus, the dominant soliton instantly grows in size once it absorbs the other soliton. This merging event is also noticed in figure~\ref{fig:sw-energy}, where we observe a similar sharp increase to the linear and nonlinear energies at $z=11$ $m$, showing that as the dominant soliton's amplitude increases, there is a significant increase in the nonlinearity of the system. It is interesting that the amplitude of the final dominant soliton continues to grow at the late stages of this simulation, e.g. in the range of $z$ between $15$ $m$ and $20$ $m$. { This soliton acts as a kind of ``optical vacuum cleaner'' which, as it moves, sucks in energy of the random waves and remaining soliton debris (i.e.  soliton-like structures which are so weak that they cannot maintain their coherence for long).}

From the beginning, the nonlinear energy, associated mostly with the first nonlinear term in the evolution equation \eqref{eq:eq-long}, grows almost by an order of magnitude in size, compared to just under double in size for the linear contribution. This indicates that the inverse cascade and the subsequent soliton development is associated with an increase in the nonlinearity of the system, and hence the breakdown of the \ac{WT} description. Moreover, we note that the total energy remains conserved, verifying that our simulation is well-resolved.


\begin{figure}[ht!]
\subfigure[Evolution of the total energy.]{\includegraphics[width = 0.48\columnwidth]{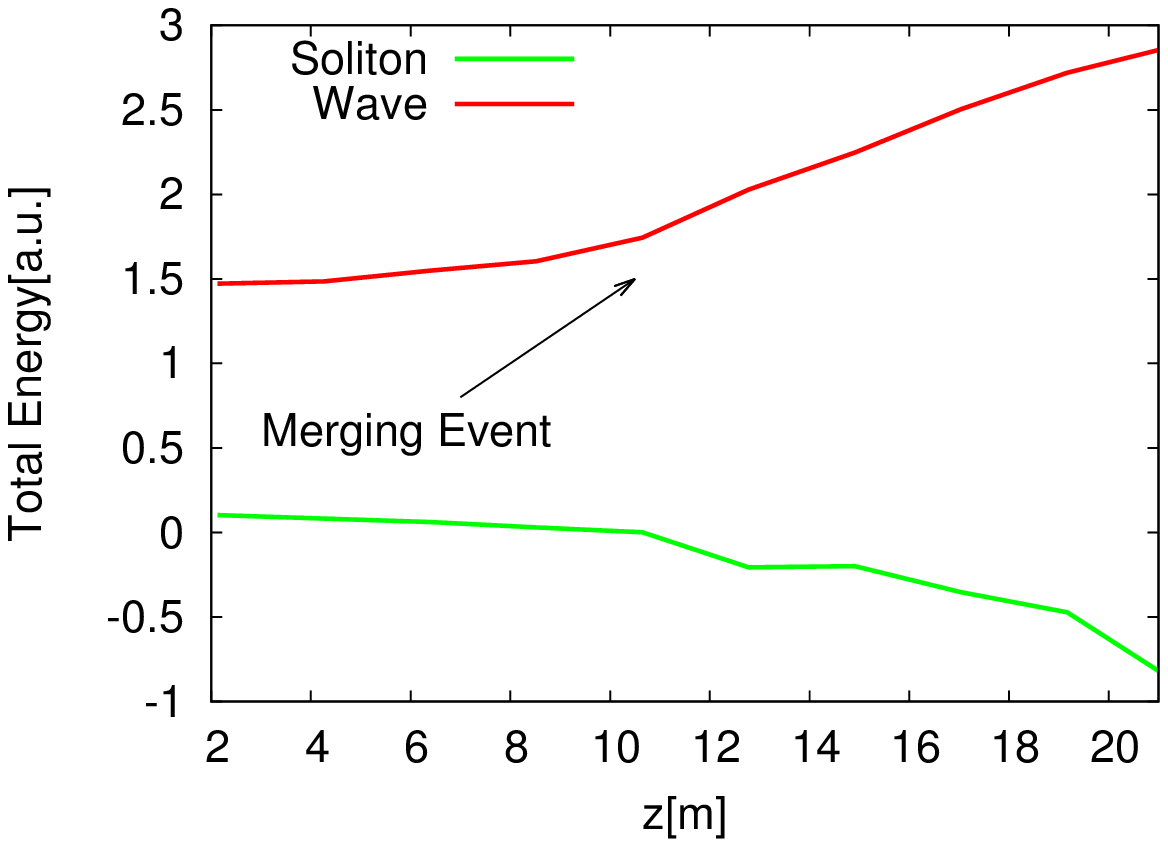}}
\hfill
\subfigure[Evolution of the linear energy.]{\includegraphics[width = 0.48\columnwidth]{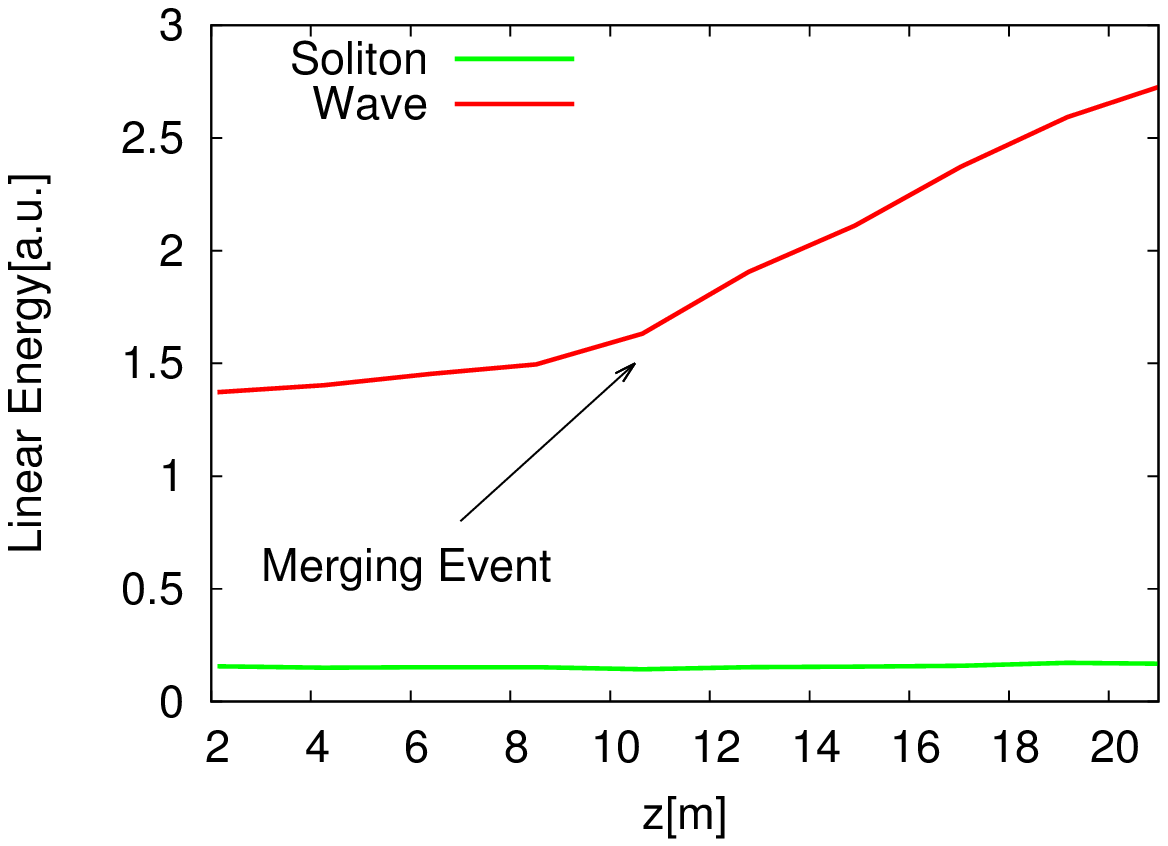}}

\subfigure[Evolution of the first nonlinear energy term.]{\includegraphics[width = 0.48\columnwidth]{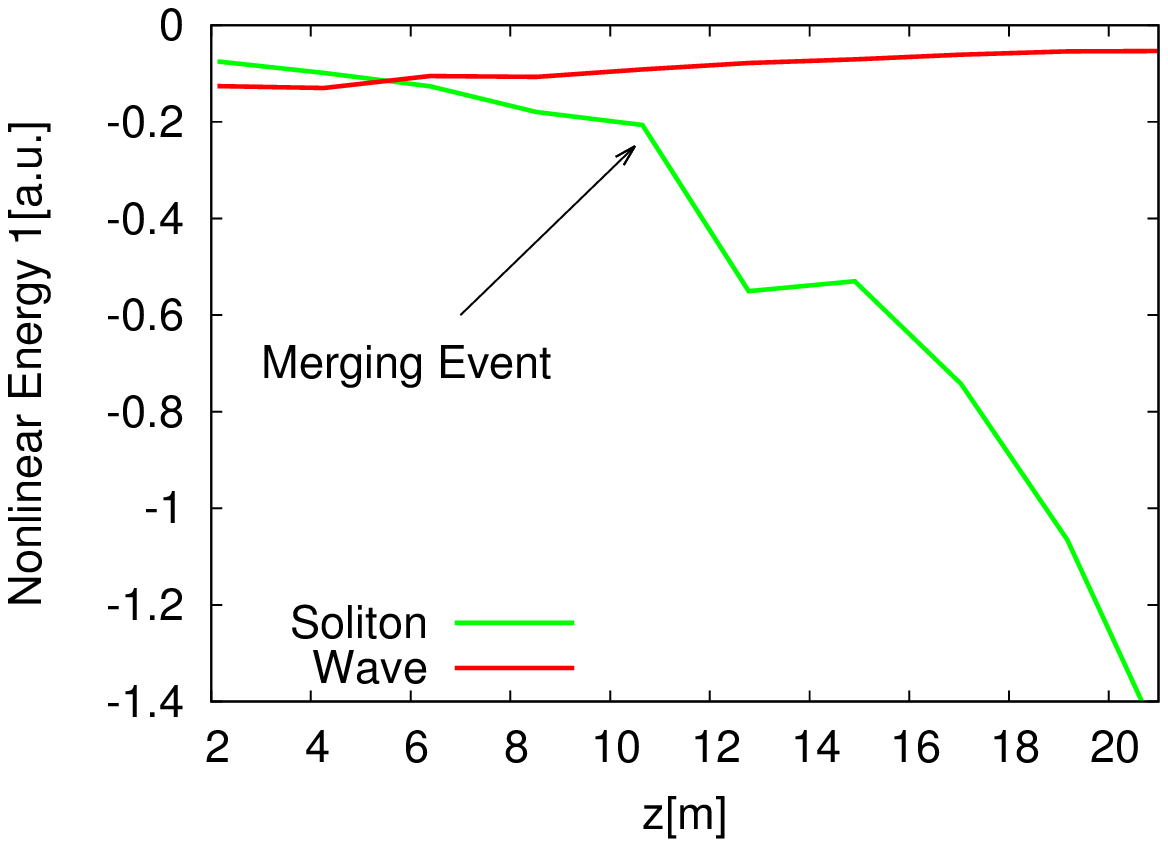}}
\hfill
\subfigure[Evolution of the second nonlinear energy term.]{\includegraphics[width = 0.48\columnwidth]{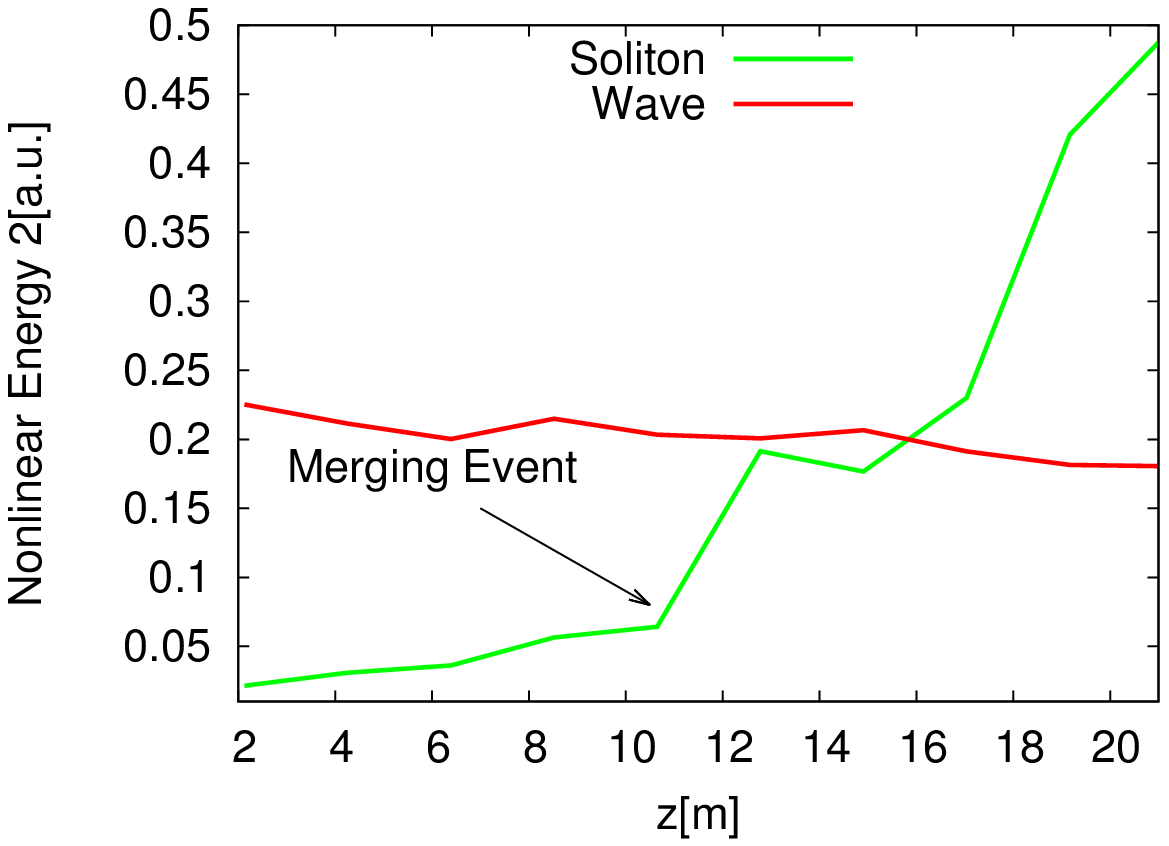}}

\caption{Plot of the evolution of the (a) total energy, (b) linear energy, (c) first nonlinear energy term, and (d) second nonlinear energy term, of the soliton  and wave components versus $z$.\label{fig:soliton-wave-t}}
\end{figure}

In figures~\ref{fig:soliton-wave-t}, we present  the soliton  and wave  components of the energies  evolving with $z$.
The data was made by using the same soliton-wave separation technique as the one we described previously, now using a series of (\bk, $\omega$)-plots and computing the energy values at each time.  In figure~\ref{fig:soliton-wave-t}a, we observe that over time the total energy of the wave component increases, whilst the energy of the soliton component decreases, crosses zero and becomes negative. However, this is not indicating a reduction in the soliton strength but the opposite, noting that the first nonlinear energy term in the long-wave Hamiltonian is negative, this indicates that the soliton component is becoming more nonlinear (an increase in the negative energy contribution). This fact  is supported by figures~\ref{fig:soliton-wave-t}b and \ref{fig:soliton-wave-t}c, where the linear and nonlinear energies of the soliton component are growing in magnitude and becoming of similar values. In figures~\ref{fig:soliton-wave-t}c and \ref{fig:soliton-wave-t}d, we have labeled the point in time where two large solitons merge (see figure \ref{fig:sw-xz-closeup}b).  At this point, we observe a sharp acceleration in the growth of both the nonlinear energies, which coincides with figures~\ref{fig:sw-max} and \ref{fig:sw-energy} showing a sharp increase in the maximum intensity and energies at the same point in $z$. Thus figures~\ref{fig:soliton-wave-t} point towards a system that is becoming more nonlinear in time, with energy flowing from waves into solitons.  In figure~\ref{fig:soliton-wave-t}b we see almost no growth with respect to the linear energy of the solitons.  This coincides with the analysis of Subsection~\ref{sec:MI}, where the consideration of the conserved quantities led us to expect the generation of waves rather than the increase in the momentum of the solitons.  This is indeed the case, as figure~\ref{fig:soliton-wave-t}b shows, and is further supported by figure~\ref{fig:sw-xz}.  If we observe closely the paths of the solitons in figure~\ref{fig:sw-xz}, we observe that the solitons at late times are not moving `faster' than those from earlier times.  Moreover, what is interesting is that in figure~\ref{fig:sw-max}, we observe an increase to the maximum intensity of the system when there is only a single soliton present ($z > 11$ $m$), i.e. the single soliton is still growing in size,  but now by the absorption of wave action from the linear wave background and not by merging with other solitons.  This can be viewed as a vacuum cleaner effect, with the soliton `hoovering' up wave action from waves.  However, we note that this process does not remove waves from the system, as clearly in figure~\ref{fig:soliton-wave-t}b we observe the wave component of the linear energy is still increasing.  Note that the growth of the wave energy is consistent with the loss of their wave action when the mean wavelength is decreasing.  It seems that the hoovering effect is more efficient when the soliton is large in size, as opposed to earlier times when solitons are small in amplitude.   The evidence for this scenario is observed in  figures~\ref{fig:sw-max} and \ref{fig:soliton-wave-t}, where, at early times, sharp soliton growth (and subsequently the increase of the linear and nonlinear energies) is more determined and dominated by discrete mergers between solitons (as seen by the sudden spike in growth during the large merger event).  However, at late times, when no more merging events occur, we still observe a growth in the maximum intensity, linear and nonlinear energies but in a more smooth and continuous fashion.  At asymptotically large times, we expect that the equilibrium state will consist of a single stationary soliton, containing  the majority of the wave action in the system, and a sea of small amplitude short random  waves, containing the majority of the linear energy of the system.  This scenario was originally predicted in \cite{ZPSY88,DZPSY89} and shown using statistical mechanics arguments in \cite{JJ00, JTZ00}.

\begin{figure}[ht!]
\centerline{
\includegraphics
[width=0.8\columnwidth]{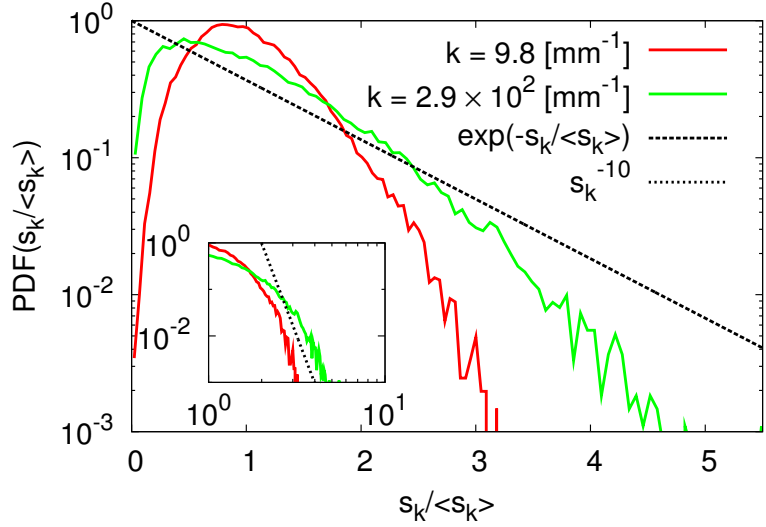}}
\caption{\acs{PDF} of intensity $ J_\bk$,  ${\mathcal P} = \langle \delta(s_\bk - J_\bk ) \rangle$, at two wave numbers, $k = 9.8$ $mm^{-1}$, and $k=2.9\times 10^2$ $mm^{-1}$. The black dashed  line is the Rayleigh distribution that corresponds to a Gaussian wave field. The inset shows the power-law behavior of the \ac{PDF} tails. The black dashed line in the inset is $s_\bk^{-10}$.}
\label{fig:PDF-cond}
\end{figure}

In figure~\ref{fig:PDF-cond}, we plot the intensity \ac{PDF} in $\bk$-space for two wave numbers, $k = 9.8$ $mm^{-1}$, and $k=2.9\times 10^2$ $mm^{-1}$ at $z\approx 20$ $m$.
 The two wave numbers are situated in the low wave number (soliton dominated) and high wave number (weak wave dominated) regions respectively. We observe at both large and small values of $J_\bk$ both \acp{PDF} show signs of depletion of their probability with respect to the Rayleigh distribution. However, there is far more depletion observed in the soliton dominated \ac{PDF} at  $k = 9.8$ $mm^{-1}$.  Probability depletion in the wave dominated scale, $k=2.9\times 10^2$, at the large (small) $J_\bk$ corresponds to a positive (negative) probability flux $\mathcal{F}_\bk$.   We believe that the depletion at both large and small $J_\bk$'s is because of the hoovering effect of the soliton - as solitons grow, their hoovering effect (absorption of wave action from waves) becomes more apparent and thus more depletion is seen.  As the soliton is composed of a wide range of wave modes, this effect will occur at a large range of wave numbers (even high wave numbers).  However, as the solitons merge, grow, or dissipate, they excite high frequency waves.  This behavior would encourage the transportation of wave action from the soliton to small $J_\bk$s.  Indeed, this process would generally only occur at high wave numbers and technically would produce an enhancement (negative probability flux) of the \ac{PDF} tail in figure~\ref{fig:PDF-cond}.  The balance between the hoovering effect and excitation of waves is probably what is occurring at high wave numbers, giving rise to an almost (slightly depleted) Rayleigh \ac{PDF} at $k=2.9\times 10^2$ $mm^{-1}$.  Over time, as solitons merge and dissipate, until only one remain, we would observe a reduction in the re-injection of wave action at high wave numbers because the mechanisms that allow this (merging and dissipation of solitons) would cease.  At low wave numbers, at $k = 9.8$ $mm^{-1}$,  the \ac{PDF} would  gradually become more and more depleted (deviated from Rayleigh) as time progresses. As this scale is soliton dominated, we cannot apply our theory of the \ac{PDF} depletion in terms of the probability flux $\mathcal{F}_\bk$ which was developed for weak waves.
In fact, such a \ac{PDF} corresponds to a certain distribution of solitons in the soliton parameter space $(A,B)$ for which a theory has not yet been developed.

\begin{figure}[ht!]
\centerline{
\includegraphics
[width=0.8\columnwidth]{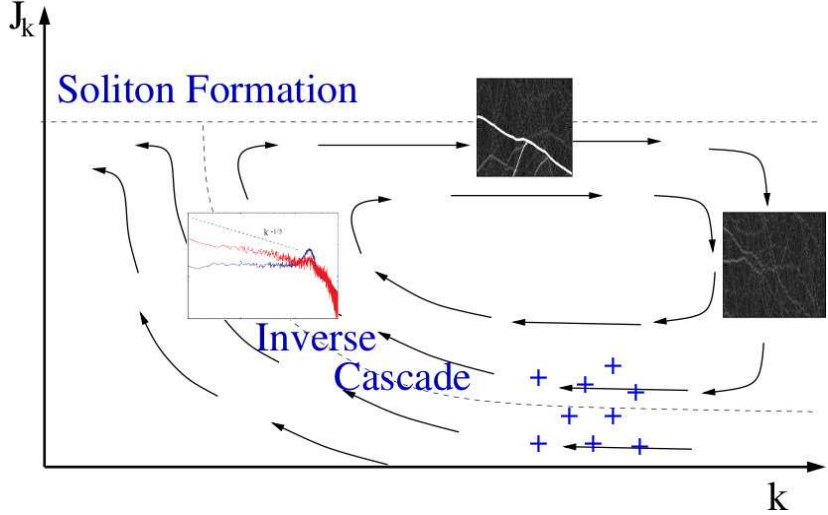}}
\caption{Diagram of the \acs{WTLC}. The high wave number initial condition subsequently gets transferred to low $k$ and high $J_\bk$ by the inverse cascade.  At low $k$, the weak \ac{WT} fails (near the $T$-junction of the dashed lines) and \ac{MI} occurs, solitons form resulting in the wave action getting spread along a $n_\bk \propto k^0$ spectrum (horizontal dashed line). The emission of waves from merging or deteriorating solitons re-injects wave action at low intensities.}
\label{fig:WTLC}
\end{figure}

The observed directions of the probability flux directions $\mathcal{F}_\bk$ - toward large $J_\bk$s at low wave numbers and to low  $J_\bk$s at high wave numbers - can be combined with the picture of the inverse cascade energy flux in the direction of low wave numbers by considering a \ac{2D} flux pattern on the $(k,J_\bk)$-plane. This yields a  diagram for the \ac{WTLC}  presented in figure~\ref{fig:WTLC}. 
Here, the initial condition is at high wave numbers and low $J_\bk$s (weak small-scale random waves). The inverse cascade transports wave action, via six-wave mixing, to small wave numbers and to higher $J_\bk$s, as the inverse cascade is associated with an increase in nonlinearity of the system.  As wave action accumulates at large scales, \ac{MI} kicks in and solitons form, with wave action spreading along the soliton spectrum $n_\bk \propto k^0$. Note that the wave modes within the soliton are coherent, i.e. correlated with each other. The non-integrability of the system allows for solitons to interact with surrounding structures, that enable energy exchange. In particular, solitons will collide, occasionally merge or deteriorate via random interactions - each of these processes allowing for the emission of energy and wave action to incoherent waves, resulting in a reversal of the probability flux $\mathcal{F}$ at high wave numbers. The wave action re-injected back into the random wave component fuels the continual process of the inverse cascade. This continuous transport of wave action can be seen as a vortex-like flux structure in the $(k,J_\bk)$-plane which we call the \ac{WTLC}  in our \ac{1D} \ac{OWT} system \cite{LNMD10,N10}.  Note that from figure~\ref{fig:PDF-cond} we see a positive probability flux even at high wave numbers!  Thus, the \ac{WTLC} picture with flux lines not emerging and not terminating anywhere, as in figure~\ref{fig:WTLC}, is only valid for a
steady state, whereas figure~\ref{fig:PDF-cond}  corresponds to an evolving transient where the net flux (produced by the hoovering effect) is still directed
from waves to (growing) solitons.
 For the \ac{LWE}, the   asymptotical final state is a single soliton containing most of wave action and surrounded by almost linear waves containing finite energy but almost no wave action.   

\subsubsection{Forced and Dissipated Simulations}

In this Subsection, we explore numerical simulations of the \ac{LWE} and \ac{SWE} with forcing and dissipation.  This setup is intended to produce more idealized conditions for non-equilibrium stationary states assumed for the realizability of pure  the \ac{KZ} solutions, (\ref{eq:longKZ}).  Thus, we can test to see if the \ac{KZ} spectra are realizable in principle, or if they can not be established at all, even in the most idealized of settings. In an infinite sized system, both the direct- and the inverse-cascade ranges can be realized in the same simulation, with forcing acting at an intermediate scale and two inertial ranges either side (see figure~\ref{fig:dualcascade}). However, computational restrictions make it impractical to perform such a simulation.  Therefore, we simulate each cascade separately, by performing a simulation for each with forcing at either ends of $\bk$-space, allowing for just one inertial range.  We numerically solve the non-dimensionalized equation (see Appendix \ref{Appendix:nondim}) with $\alpha=1024$.  We run the simulations with a time step of $\Delta t= 1\times 10^{-4}$ in a box of length $L=32\pi$, with spatial resolution of $N=2^{12}$.  The simulations are run so that a non-equilibrium steady state is achieved.  This is checked by observing stationarity of the total energy $\mathcal{H}$ and wave action $\mathcal{N}$ in the system.

\begin{figure}[ht!]
\centerline{\includegraphics[width=0.8\columnwidth]{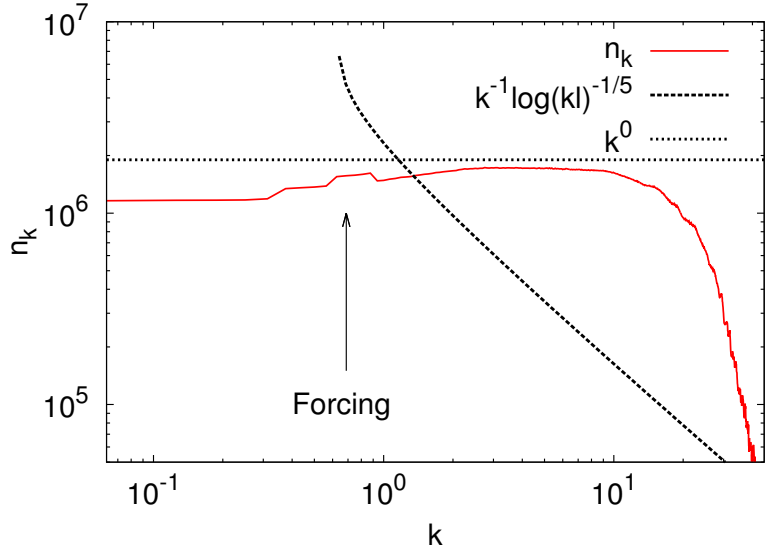}}
\caption{The wave action spectrum $n_\bk$ in a statistically non-equilibrium stationary state for the direct cascade simulation of the \ac{LWE}.  We plot two black lines to show the theoretical comparisons.  The black dashed line represent the \ac{WT} prediction of the \ac{KZ} solution $n_\bk\propto k^{-1}\ln(k\ell)$.  The black dotted line represents the thermalization of wave action.}
\label{fig:longdirectnk}
\end{figure}

For the direct cascade setup for the \ac{LWE}, we introduce dissipation profiles at the two limits of $\bk$-space that removing energy and wave action that is being constantly injected by the forcing.  We set the low wave number dissipation profile, $D_\bk^L$, to be friction, defined by (\ref{eq:Dissipation-friction}), with the friction coefficient of $\nu_{\mathrm{friction}}=5 \times 10^{-1}$ and apply a hyper-viscosity dissipation scheme at high wave numbers with a coefficient of $\nu_{\mathrm{hyper}}= 1 \times 10^{-13}$.  Forcing is situated at large scales and is defined by  (\ref{eq:Forcingdirect}), with amplitude $A=4.1\times 10^4$.  
Once the simulation has reached a statistically non-equilibrium statistical steady state, we analyze the statistics by performing  time averages over this steady state regime. We have plotted the wave action spectrum $n_\bk$ for the direct cascade regime in figure~\ref{fig:longdirectnk}.

\begin{figure}[ht!]
\centerline{\includegraphics[width=0.8\columnwidth]{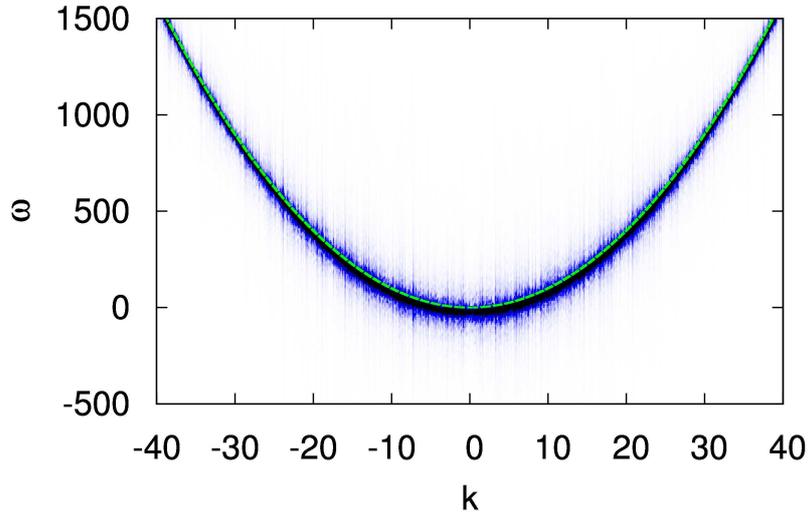}}
\caption{$(\bk,\omega)$-plot for the direct cascade simulation of the \ac{LWE}.  The green line represents the linear wave frequency $\omega_k = k^2$.}
\label{fig:disp-LWE-D}
\end{figure}

We observe that the wave action spectrum in figure~\ref{fig:longdirectnk} does not agree with the theoretical \ac{KZ} prediction \eqref{eq:directcorrected}.    There is a clear   flat spectrum of the form $n_\bk \propto k^0$.  This could correspond to several phenomena: solitons, critical balance and a thermalized state with equipartition of wave action. However, in figure~\ref{fig:disp-LWE-D}, we have plotted the $(\bk,\omega)$-plot for the same simulation at late times.  We observe an absence of any soliton structures, like the ones we saw in the decaying case (see figure~\ref{fig:wk.eps}).  Therefore, we can safely reason that the flat spectrum is not associated with the presence of solitons within the system.   The explanation for the  $k^0$ slope is more likely down to the thermalization of the system with respect to wave action around the low to the intermediate wave number range.   This behavior should be expected due to the lack of agreeing flux signs with Fj\o{}rtoft's argument for the \ac{LWE}, where a system in such a situation would tend towards a modified thermalization spectrum \cite{DNPZ92} also known as a ``warm cascade'' \cite{PNO09}. The lack of agreement with the \ac{KZ} spectrum  can be explained by the incorrect flux directions which can be most easily understood and determined using the \ac{DAM} in Subsection~\ref{sec:dam}. It remains to be understood why we observe that the wave action spectrum corresponds more closely with the thermalized \ac{KE} solution corresponding to the equipartition of wave action, and not the thermalized spectrum corresponding to the equipartition of energy.   
On the other hand, we observe that  the $(\bk,\omega)$-distribution  is not too narrow around the dispersion curve $\omega = \omega_k =  k^2$.
Thus, we cannot rule out that the \ac{CB} mechanism could be the real reason for the flat spectrum.

\begin{figure}[ht!]
\subfigure[]{\includegraphics[width=0.5\columnwidth]{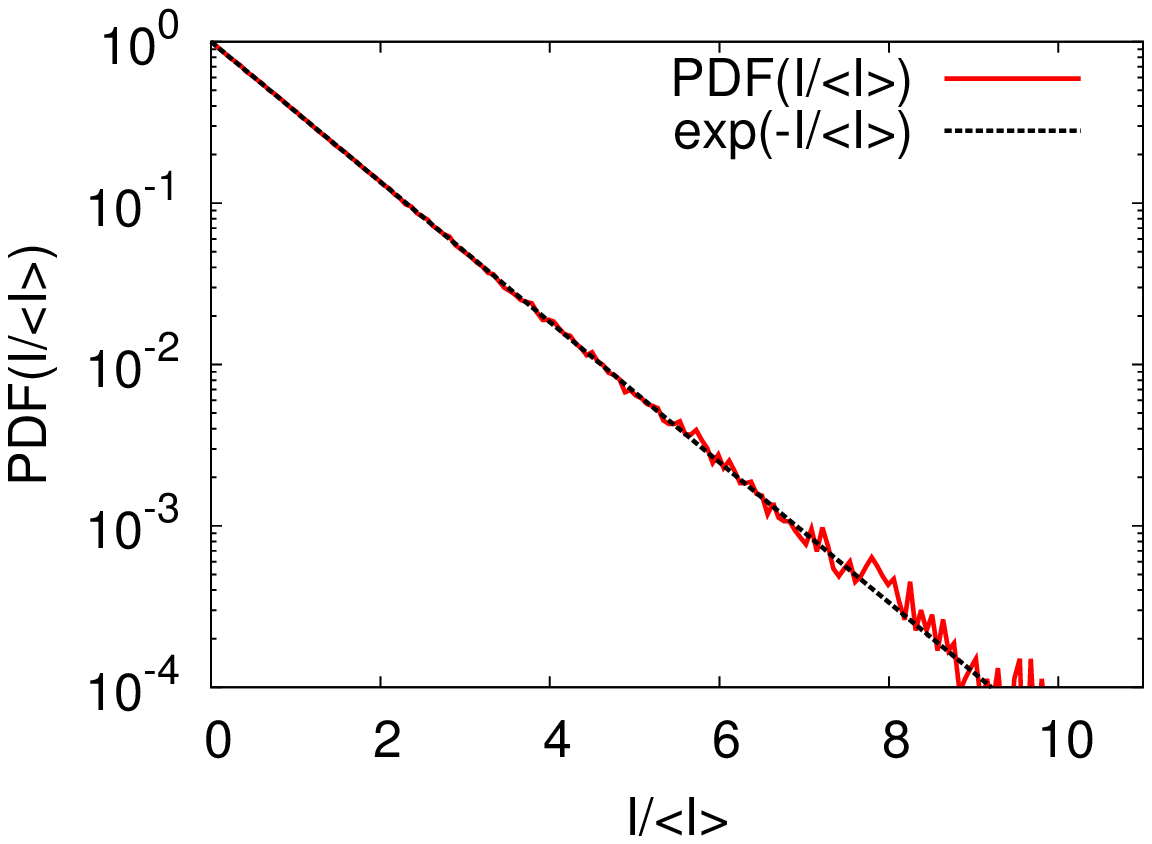}}
\hfill
\subfigure[]{\includegraphics[width=0.5\columnwidth]{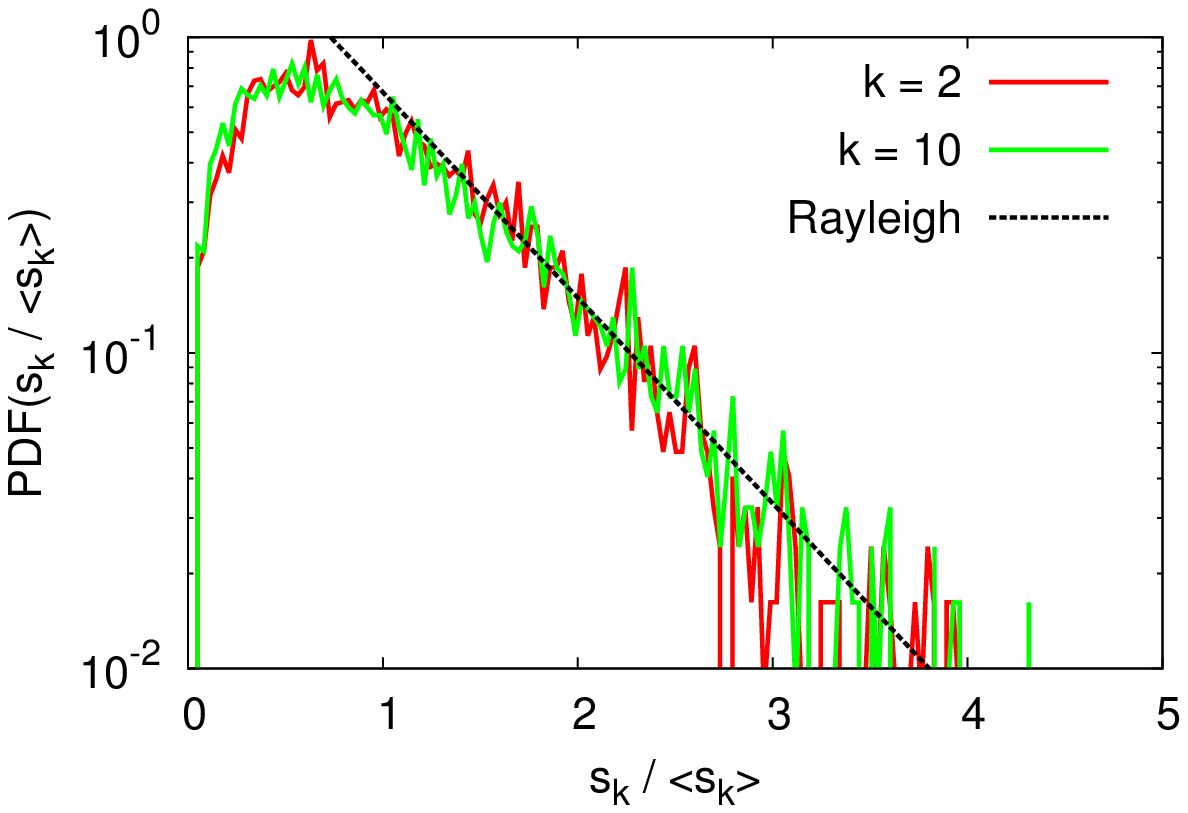}}
\caption{(a) Normalized \ac{PDF} of the wave intensity in the direct cascade simulation of the \ac{LWE}. The straight line corresponds to the Rayleigh distribution. (b) Normalized \ac{PDF} of $J_\bk$ at two wave numbers for the direct cascade in the \ac{LWE}.  The black dashed line represents the Rayleigh distribution. \label{fig:PDF-direct-lw}}
\end{figure}

In figures~\ref{fig:PDF-direct-lw} we plot \acp{PDF} of (a) the wave intensity in the physical space and (b) the Fourier-mode intensity at two different wave numbers.  In figure~\ref{fig:PDF-direct-lw}a, we observe an almost perfect agreement of the the \ac{PDF} to the Rayleigh distribution.  This indicates that the system corresponds to a Gaussian wave field - a good indication that the system is in a thermalized state of random waves.  Additionally in figure~\ref{fig:PDF-direct-lw}b we observe that both \acp{PDF} are Rayleigh distributed (here the Rayleigh distribution is fitted to the tails of the \acp{PDF}).  Again, we observe a clear agreement of the tails to the straight line of the Rayleigh distribution at large amplitudes.  Towards the core of the \acp{PDF} in figure~\ref{fig:PDF-direct-lw}b we see some depletion of the \acp{PDF} to the Rayleigh distribution the origin of which remains unclear. It is possible that
this is a signature of appearance of an embryonic condensate (the true condensate, solitons, are are not present in this system due to low wave number dissipation) - a similar effect was observed in recent simulations
of the \ac{3D} \ac{NLSE} \cite{PNO09}.

The inverse cascade displays more unusual behavior.  The inverse cascade simulation for the \ac{LWE} was setup with forcing situated at high wave numbers with a profile given by equation~\eqref{eq:Forcinginverse} and with low wave number dissipation in terms of friction and high wave number dissipation in terms of hyper-viscosity.  The parameters used for the simulation are  $A=7.7\times 10^{3}$ for the forcing amplitude, $\nu_{\mathrm{friction}}= 1 \times 10^{-1}$ and $\nu_{\mathrm{hyper}}= 1 \times 10^{-11}$ for the dissipation coefficients.  As usual, we wait for a non-equilibrium stationary state to be achieved before performing any analysis. 

\begin{figure}[ht!]
\centerline{\includegraphics[width=0.8\columnwidth]{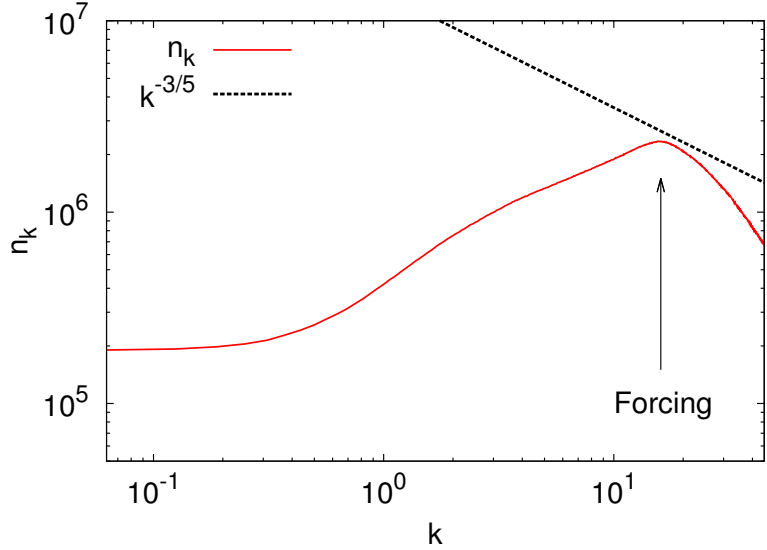}}
\caption{The wave action spectrum, $n_\bk$, in a statistically non-equilibrium stationary state for the inverse cascade simulation of the \ac{LWE}.  The straight line represents the \acs{WT} prediction of the \acs{KZ} solution of $n_\bk\propto k^{-3/5}$.}
\label{fig:longinversenk}
\end{figure}

\begin{figure}[ht!]
\centerline{\includegraphics[width=0.8\columnwidth]{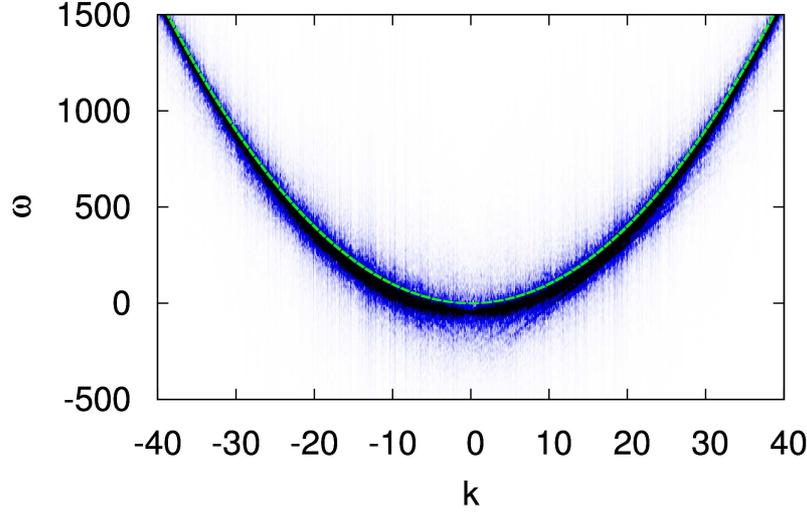}}
\caption{$(\bk,\omega)$-plot for the inverse cascade simulation of the \ac{LWE}.  The green line represents the linear wave frequency $\omega_k = k^2$.}
\label{fig:Disp-LW-inv}
\end{figure}

In figure~\ref{fig:longinversenk} we plot the wave action spectrum from the numerical simulation, time averaged once the simulation has reached a non-equilibrium stationary state. For comparison, we also show the \ac{KZ} prediction for the inverse cascade of wave action.  We observe a lack of agreement to any power-law behavior to the spectrum.  Indeed, we see a peak of the spectrum around the forcing region, while the spectrum gradually fall off at small and large $k$.  The absence of a \ac{KZ} spectrum can be explained by the incorrect flux directions, but what is really surprising is the omission of any thermalized or \ac{CB} spectrum.   The \ac{CB} in this case, is identical to the spectrum of the spectrum of the thermalization of wave action: $n_\bk \propto k^0$.  \ac{CB} has been observed in other optical systems, i.e. one described by the \ac{3D} \ac{NLSE} \cite{PNO09}.   We found that the spectrum to the left of the forcing region seemed to be determined by the level of friction used at low wave numbers.  By tuning the coefficient of friction, we adjusted the asymptotic value of the spectrum at $k=0$, and the spectrum would simply gradually decline to that level from the forcing region.  Dampening of these low wave number modes could prevent some non-local behavior that is driving the inverse cascade.   Clearly, the dampening of these modes is restricting \ac{MI} and the formation of solitons, which may have an essential role in the dynamics of the inverse cascade -  as we do not observe this behavior in the decaying simulation of the previous Subsection.  To be certain that we are not observing any effects that might be attributed to the presence of solitons, we show the $(\bk,\omega)$-plot for the simulation in figure~\ref{fig:Disp-LW-inv}.  In figure~\ref{fig:Disp-LW-inv} we observe  that no solitons are present by the lack of any coherent straight lines around dispersion curve.  However, if one looks closely, we do observe the partial formation of a soliton situated on the dispersion curve at $k=15$.  However, with the presence of dissipation this soliton does not `peel' from the wave component.  Moreover, inspection of the intensity distribution (not shown), does not provide any indication of coherent soliton like structures distinguishable from random waves.  The $(\bk, \omega)$-plot is also in good agreement to the linear dispersion relation $\omega_k = k^2$ shown by the green dashed curve, but the width of the plot is as wide as $\omega_k$ in the inverse cascade range.  This implies that the system is indeed in a strongly nonlinear regime which could point at   \ac{CB} but, as we have already mentioned, the \ac{CB} spectrum is not observed. We have also done simulations with weaker forcing amplitudes for which \ac{WT} was weak, and the result was similar to the one reported here. In fact, the reason that we increased the forcing was an attempt to strengthen the wave interactions in order to ``push" the inverse cascade through to the large scales.

\begin{figure}[ht!]
\subfigure[]{\includegraphics[width=0.5\columnwidth]{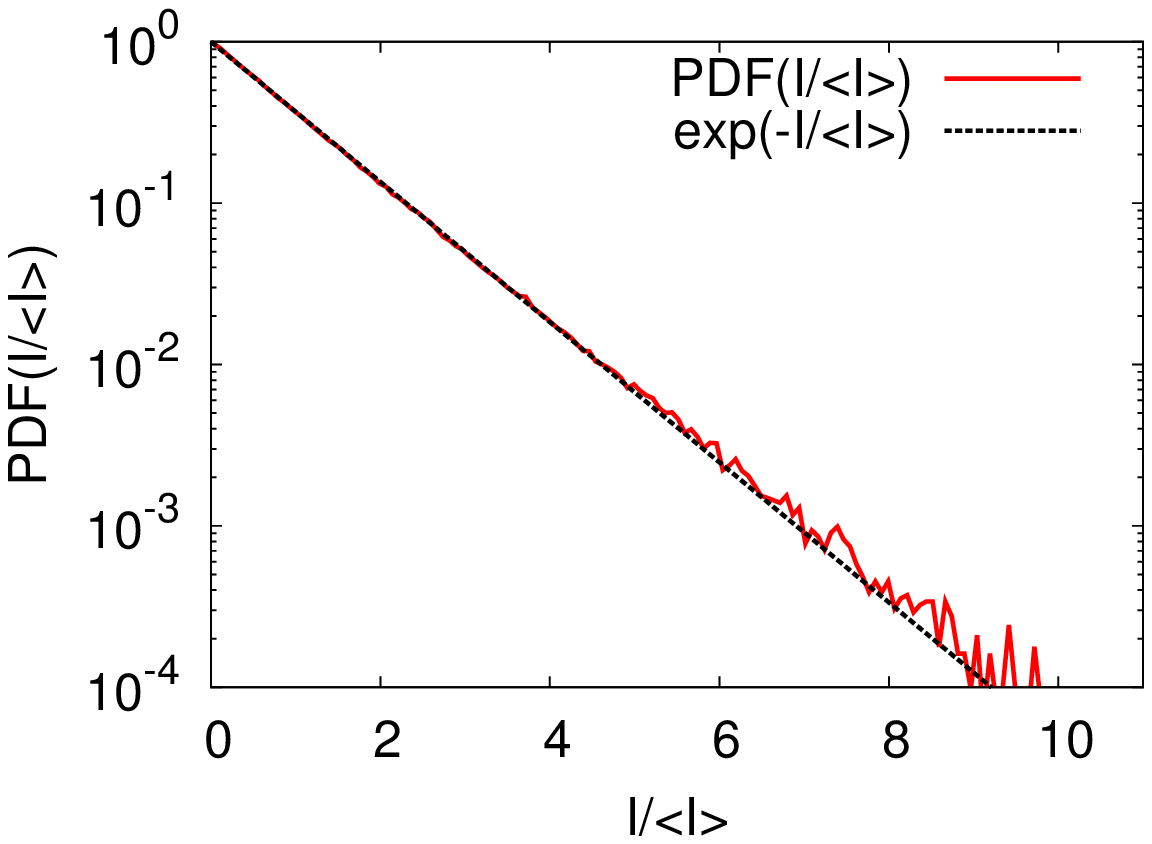}}
\hfill
\subfigure[]{\includegraphics[width=0.5\columnwidth]{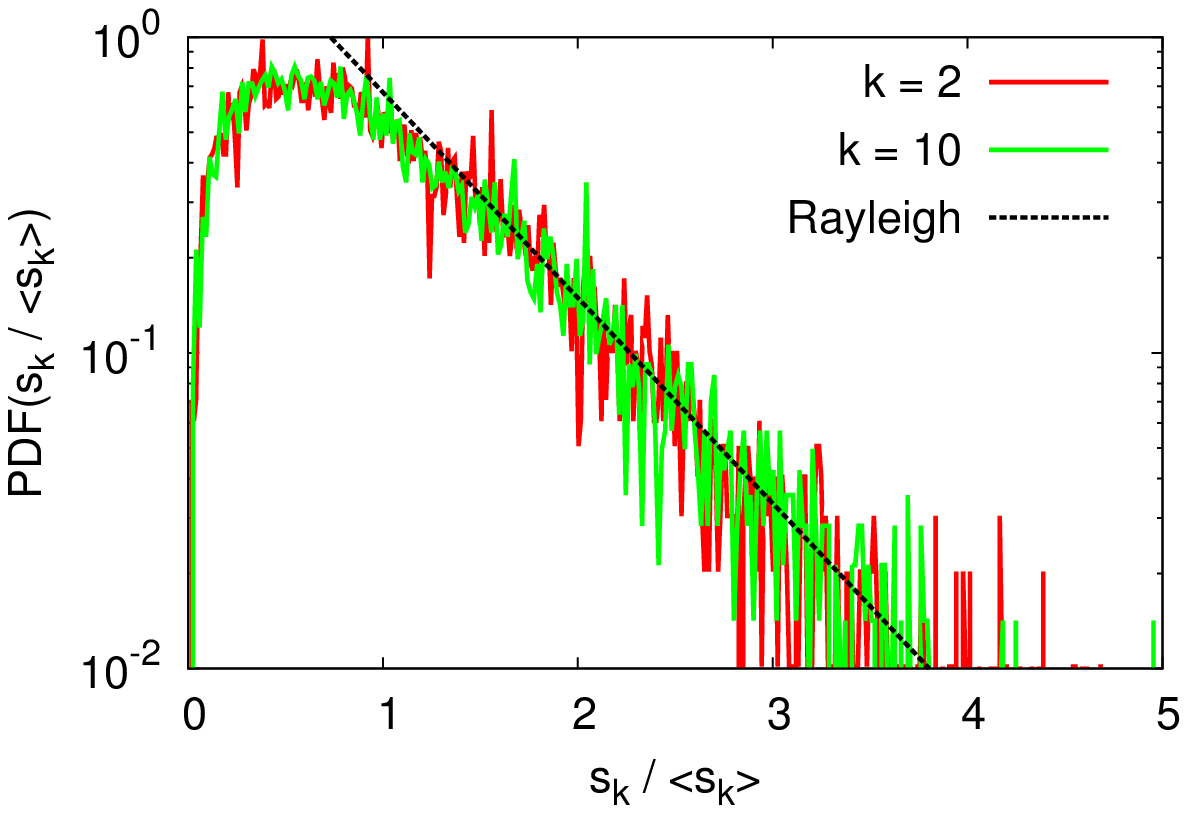}}
\caption{(a) Normalized \ac{PDF} of the wave intensity in the inverse cascade simulation of the \ac{LWE}. The straight line corresponds to the Rayleigh distribution. (b) Normalized \ac{PDF} of $J_\bk$ at two wave numbers for the inverse cascade in the \ac{LWE}.  The black dashed line represents the fitted Rayleigh distribution. \label{fig:PDF-inverse-lw}}
\end{figure}

In figures~\ref{fig:PDF-inverse-lw}  we plot \acp{PDF} of (a) the wave intensity in the physical space and (b) of the Fourier-mode intensity at two different wave numbers. Similarly to the direct cascade, we observe agreement to the Rayleigh distribution of a Gaussian wave field.  The \ac{PDF} of the wave intensity in figure~\ref{fig:PDF-inverse-lw}a is in perfect agreement to the unfitted Rayleigh distribution, whilst we observe good Rayleigh agreement of the tails of the \acp{PDF} in figure~\ref{fig:PDF-inverse-lw}b at two wave number either end of the inertial range.  Moreover, we observe a depletion of the probability at low wave mode intensities at both $k=2$ and $k=10$, the same phenomenon as seen in the direct cascade.

In the simulation of the forced-dissipated inverse cascade, we see no agreement of the \ac{KZ}, thermal or \ac{CB} spectra.  Moreover, we observe that the spectrum seems to be determined by the value of the lowest mode.  This suggests that the inverse cascade may be associated to some non-local effect established by \ac{MI}.  Indeed, we find the forced-dissipated simulation to provide more questions than answers to the behavior of \ac{1D} \ac{OWT}, and clearly more work in this area is needed to fully explain these results.

\subsection{The Short-Wave Equation}

We investigate the \ac{SWE}, (\ref{eq:eq-short}) by numerical simulations of the non-dimensionalized model of Appendix \ref{Appendix:nondim}.  We apply the same forcing and dissipation profiles that are described in Section~\ref{sec:num}.  We dissipate at low wave numbers using a hypo-viscosity profile, while using hyper-viscosity at high wave numbers. 

We begin by discussing the direct cascade simulation.   We force the system with the profile \eqref{eq:Forcingdirect}, with  amplitude $A=1.6\times 10^2$ and use the dissipation rates of $\nu_{\mathrm{hypo}}=1\times 10^{-2}$, $\nu_{\mathrm{hyper}}=1 \times 10^{-11}$ for the direct simulation. 

\begin{figure}[ht!]
\centerline{\includegraphics[width=0.7\columnwidth]{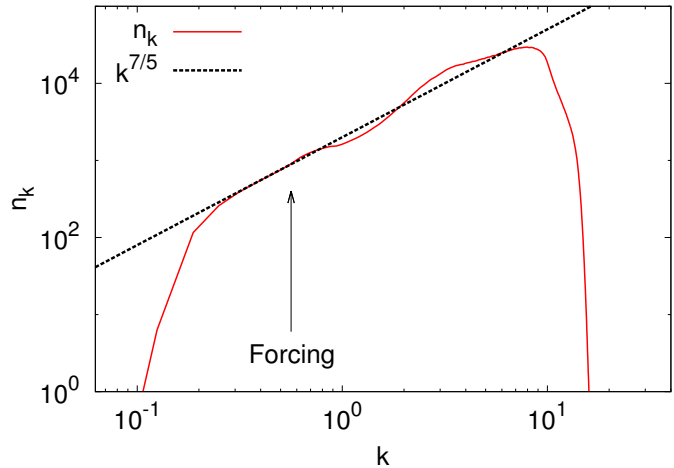}}
\caption{The wave action spectrum $n_\bk$ in a statistically non-equilibrium stationary state for the direct cascade simulation of the \ac{SWE}.  The straight line represents the \acs{WT} prediction of the \acs{KZ} solution of $n_\bk\propto k^{7/5}$.}
\label{fig:sw-n_k-dir}
\end{figure}

In figure~\ref{fig:sw-n_k-dir} we plot the wave action spectrum for the direct cascade averaged over a time window once the system has reached a statistically non-equilibrium steady state.  We observe a good agreement to the \ac{WT} prediction \eqref{eq:directshort}.  However, notice that there is some slight deviation of the spectrum in the middle of the inertial range.  This is likely to be caused by an insufficient time average of the statistics.  The agreement is surprising, as we showed that the \ac{KZ} solution, \eqref{eq:directshort}, does not produce convergence of the collision integral in the \ac{UV} limit.  However, we may be observing a non-local spectrum that is close to the local \ac{WT} prediction.  

\begin{figure}[ht!]
\subfigure[]{\includegraphics[width=0.5\columnwidth]{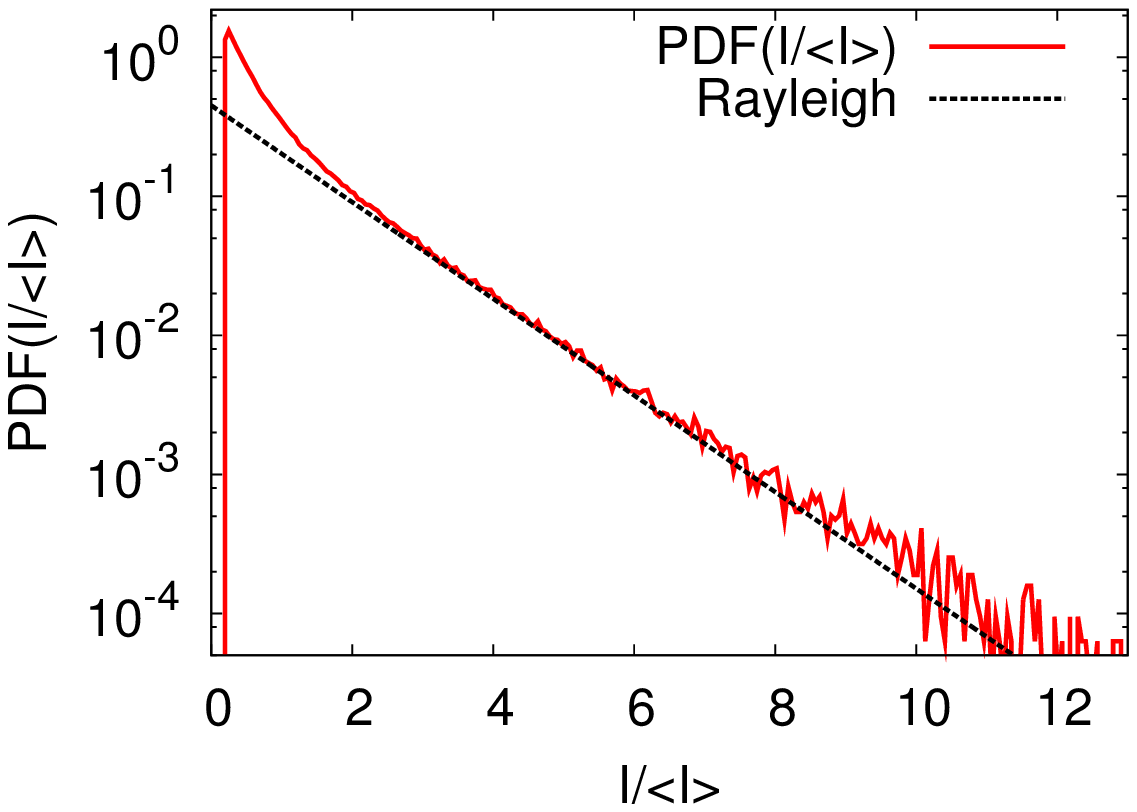}}
\hfill
\subfigure[]{\includegraphics[width=0.5\columnwidth]{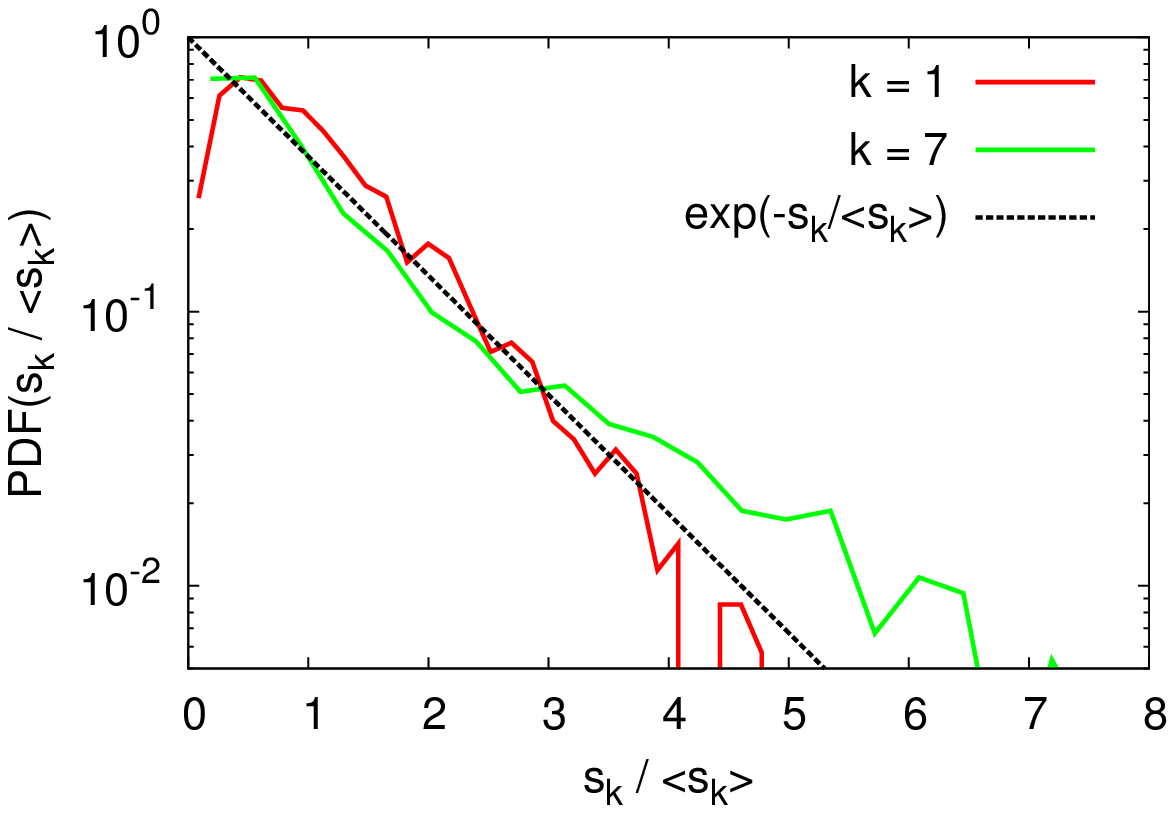}}
\caption{(a) Normalized \ac{PDF} of the wave intensity in the direct cascade simulation of the \ac{SWE}. The straight line corresponds to the fitted Rayleigh distribution. (b) Normalized \ac{PDF} of $J_\bk$ at two wave numbers for the direct cascade in the \ac{SWE}.  The black dashed line represents the Rayleigh distribution. \label{fig:PDF-direct-sw}}
\end{figure}

\begin{figure}[ht!]
\centerline{\includegraphics[width=0.8\columnwidth]{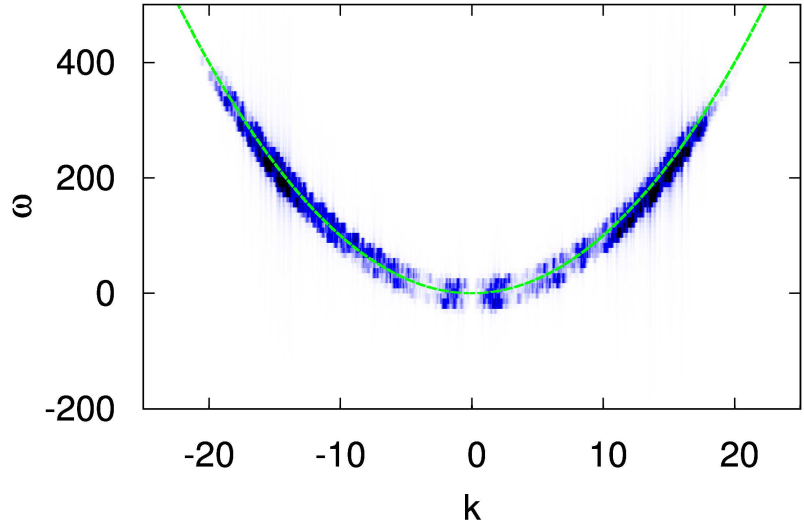}}
\caption{$(\bk,\omega)$-plot for the direct cascade simulation of the \ac{SWE}.  The dashed green line represents the linear wave frequency $\omega_k = k^2$.}
\label{fig:k-w-sw-dir}
\end{figure}

In figures~\ref{fig:PDF-direct-sw} we plot the normalized \acs{PDF} for the wave intensity in physical space and in Fourier space respectively.  We observe in figure~\ref{fig:PDF-direct-sw}a that there is a slight deviation to the Rayleigh distribution (corresponding to a Gaussian wave field) at high intensities.  This implies that there is a larger than expected occurrence of high intensity structures -  a sign of wave intermittency.  The \ac{SWE} does not seem to produce solitons, but we can numerically check this fact my investigating the $(\bk,\omega)$-plot in figure~\ref{fig:k-w-sw-dir}.  \ac{WT} theory implies that such behavior should exist as part of a wave breaking process. In addition, in figure~\ref{fig:PDF-direct-sw}a we see a large accumulation of the probability at low intensities, far more than what is predicted by the Rayleigh distribution indicating an enhanced presence of dark regions void of light.
On the other hand,  figure~\ref{fig:PDF-direct-sw}b is the \acs{PDF} of $J_\bk$ at two wave numbers at either end of the inertial range.  We see that the \ac{PDF} at $k=1$ remains relatively close to the Rayleigh distribution, with only a minor depletion in the tail. At $k=7$ we observe a large enhancement of the \ac{PDF} at large $J_\bk$, corresponding to a negative probability flux $\mathcal{F}_\bk$.  Therefore, we see behavior that is similar to that which is described by the \ac{WTLC}, figure~\ref{fig:WTLC}. Observation of the $(\bk,\omega)$-plot, figure~\ref{fig:k-w-sw-dir}, shows that actually there are no coherent structures present in the system.  This conclusion is made by the lack structures at negative $\omega$.  Moreover, we see a clear accumulation of the dispersion curve situated around the linear wave frequency $\omega_k = k^2$.  This indicates that the system is in the weak nonlinear limit.

The inverse cascade simulation of the \ac{SWE} is produced with the following parameters: The forcing amplitude is set to be $A=4.8\times 10^2$, while the dissipation rates are given by $\nu_{\mathrm{hypo}}=1 \times 10^{0}$, $\nu_{\mathrm{hyper}}= 1 \times 10^{-8}$.  Similarly to the direct simulation, we wait until a non-equilibrium stationary state is achieved, and then perform averages on the statistics.

\begin{figure}[ht!]
\centerline{\includegraphics[width=0.8\columnwidth]{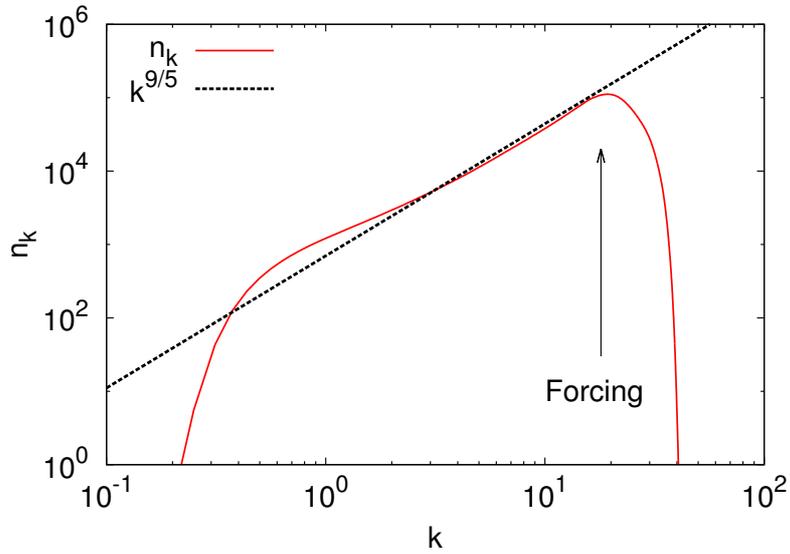}}
\caption{The wave action spectrum $n_\bk$ in a statistically non-equilibrium stationary state for the inverse cascade simulation for the \ac{SWE}.  The straight line represents the \acs{WT} prediction of the \acs{KZ} solution of $n_\bk\propto k^{-9/5}$.}
\label{fig:sw-n_k-inv}
\end{figure}

In figure~\ref{fig:sw-n_k-inv},  we observe a good agreement, for almost a decade in $\bk$-space, with the  \ac{KZ} scaling  \eqref{eq:inverseshort}. At low wave numbers, we observe a slight accumulation of wave action before the dissipation occurs. However, analysis of the locality, showed us that the inverse \ac{KZ} solution is invalid - similar puzzling behavior as in the case of the direct 
cascade simulation of the \ac{SWE}.
\begin{figure}[ht!]
\subfigure[]{\includegraphics[width=0.5\columnwidth]{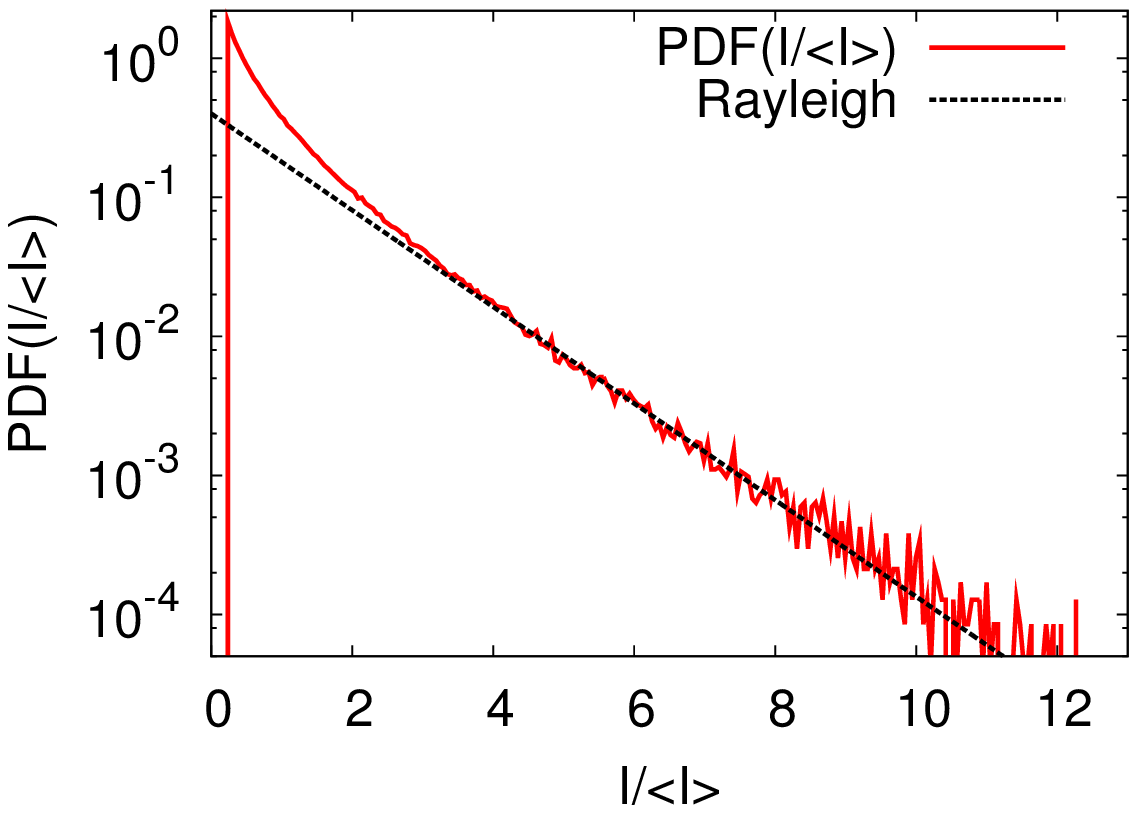}}
\hfill
\subfigure[]{\includegraphics[width=0.5\columnwidth]{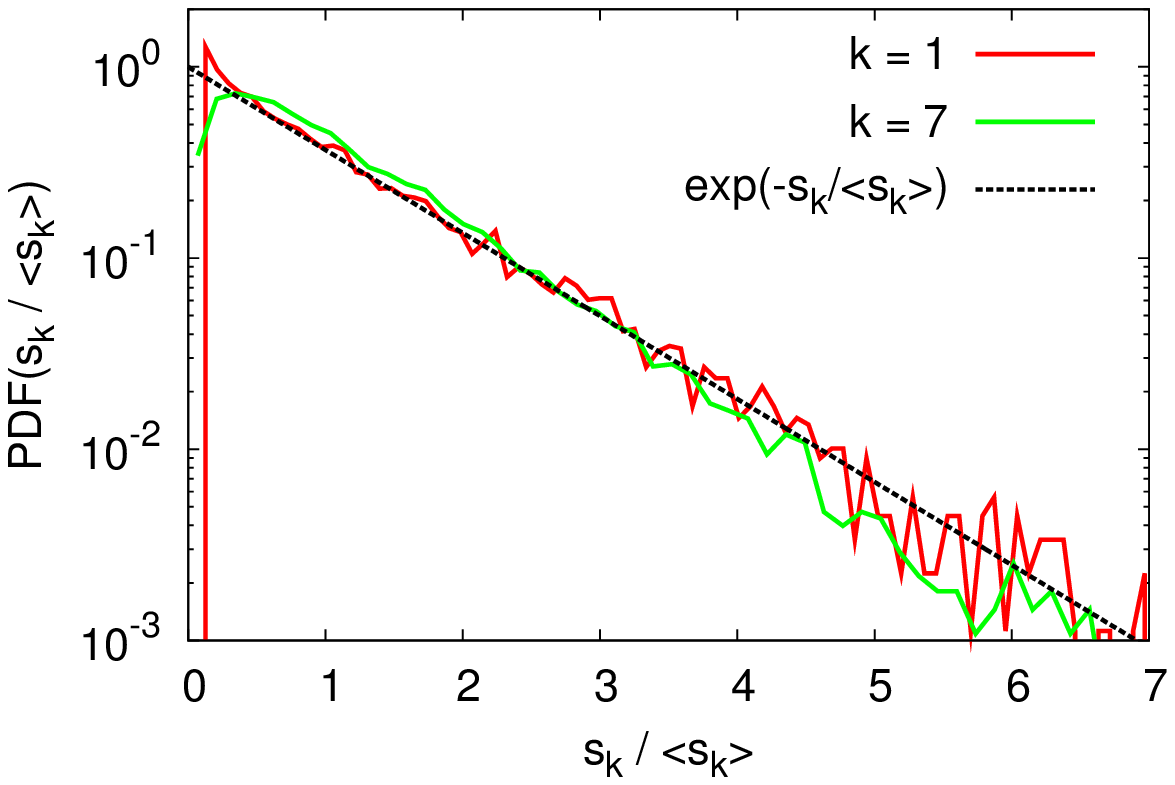}}
\caption{(a) Normalized \ac{PDF} of the wave intensity in the inverse cascade simulation of the \ac{SWE}. The straight line corresponds to a fitted Rayleigh distribution. (b) Normalized \ac{PDF} of $J_\bk$ at two wave numbers for the inverse cascade in the \ac{SWE}.  The black dashed line represents the Rayleigh distribution. \label{fig:PDF-inverse-sw}}
\end{figure}


\begin{figure}[ht!]
\subfigure[$\tau = 13.67$]{\includegraphics[width = 0.47\columnwidth]{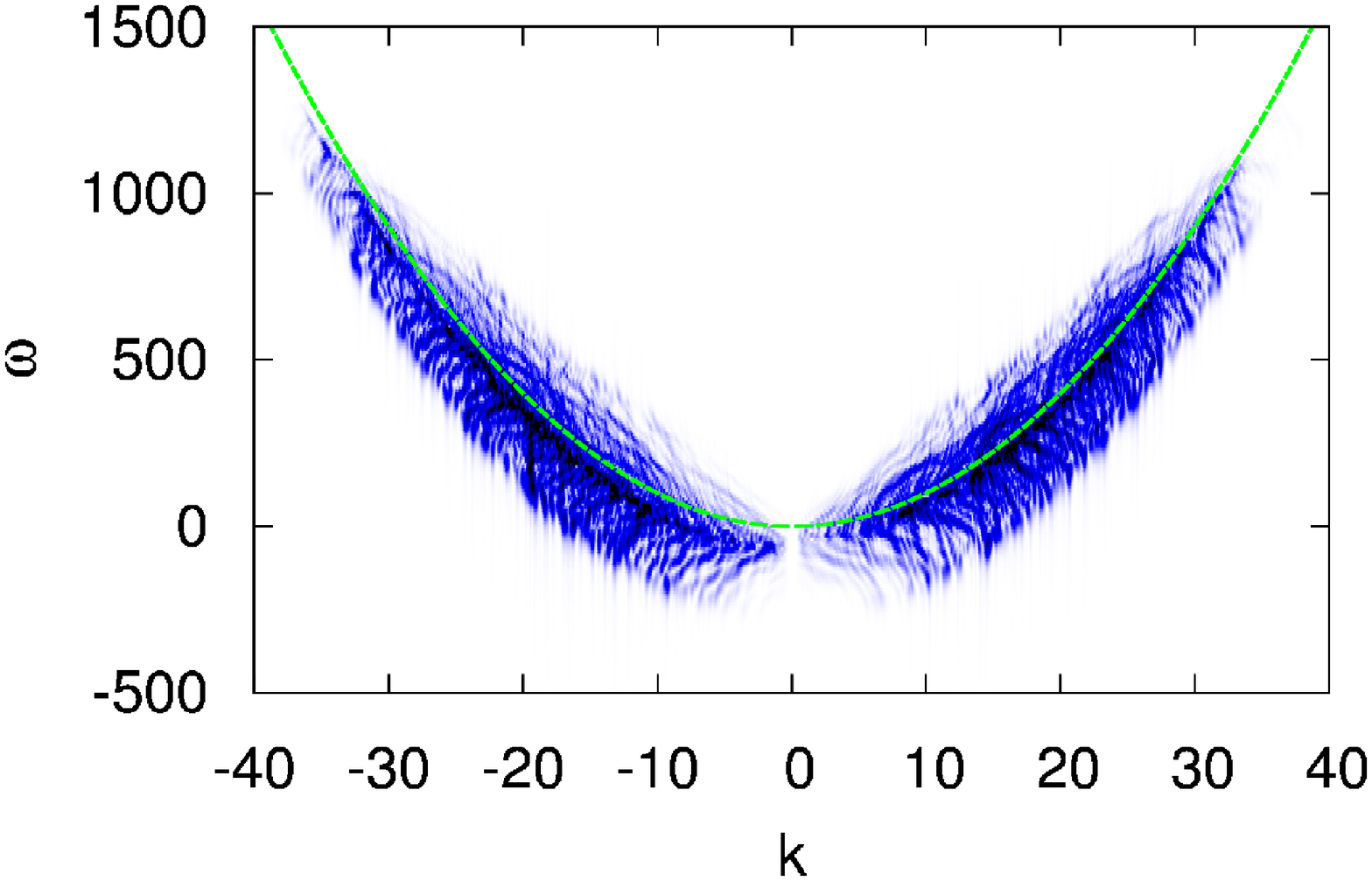}}
\hfill
\subfigure[$\tau = 18.65$]{\includegraphics[width = 0.47\columnwidth]{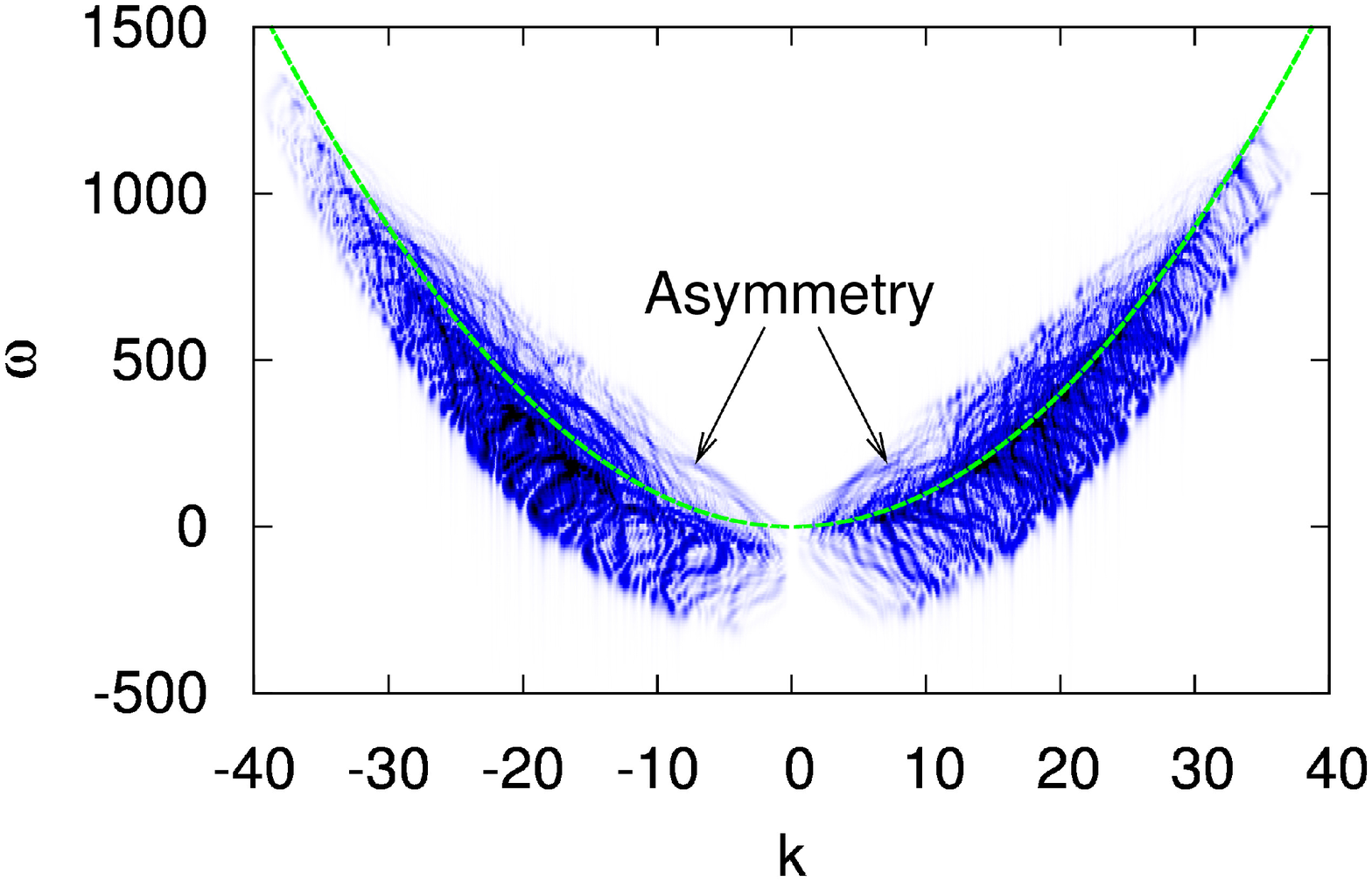}}

\subfigure[$\tau = 28.60$]{\includegraphics[width = 0.47\columnwidth]{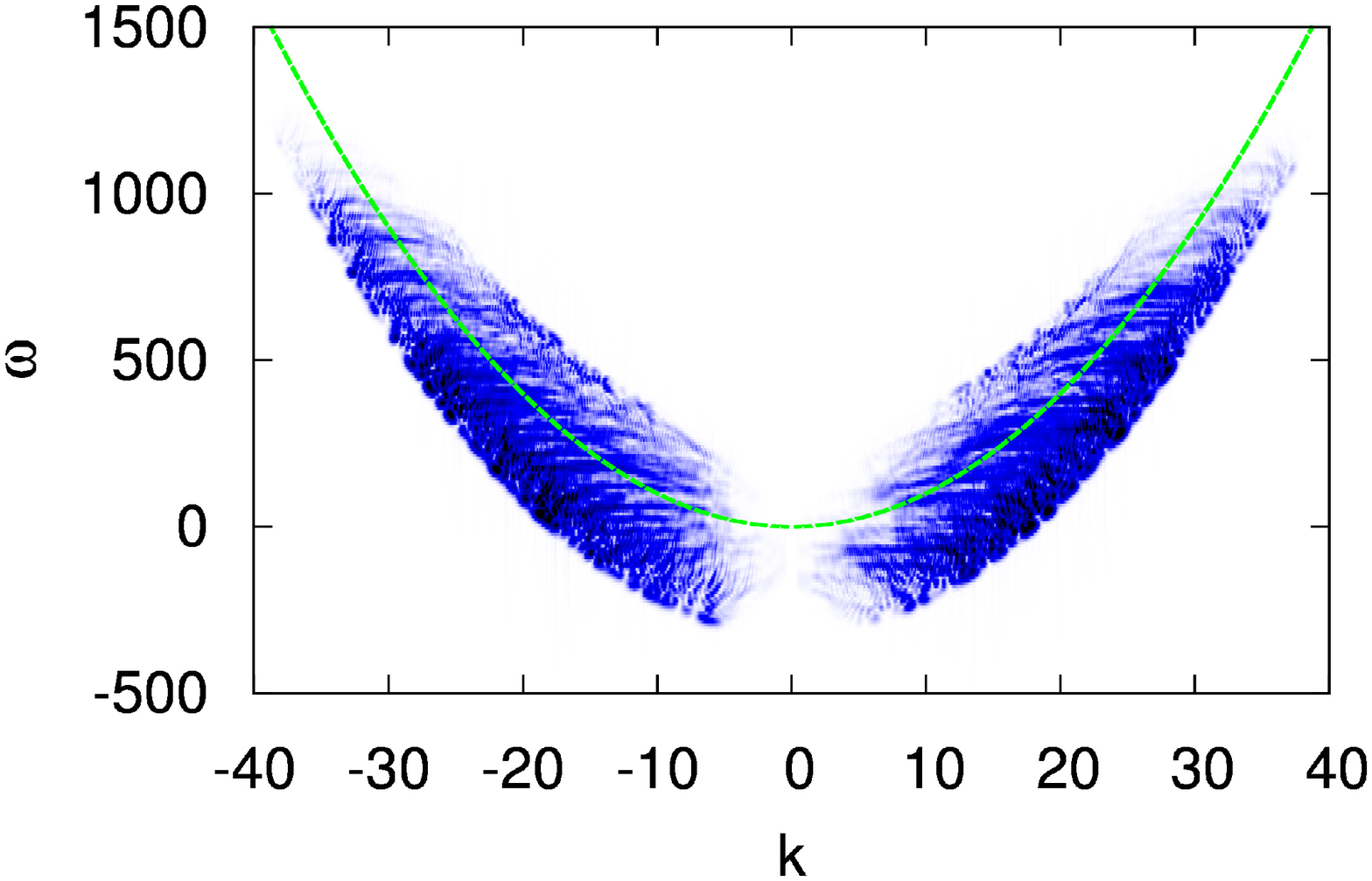}}
\hfill
\subfigure[$\tau = 33.57$]{\includegraphics[width = 0.47\columnwidth]{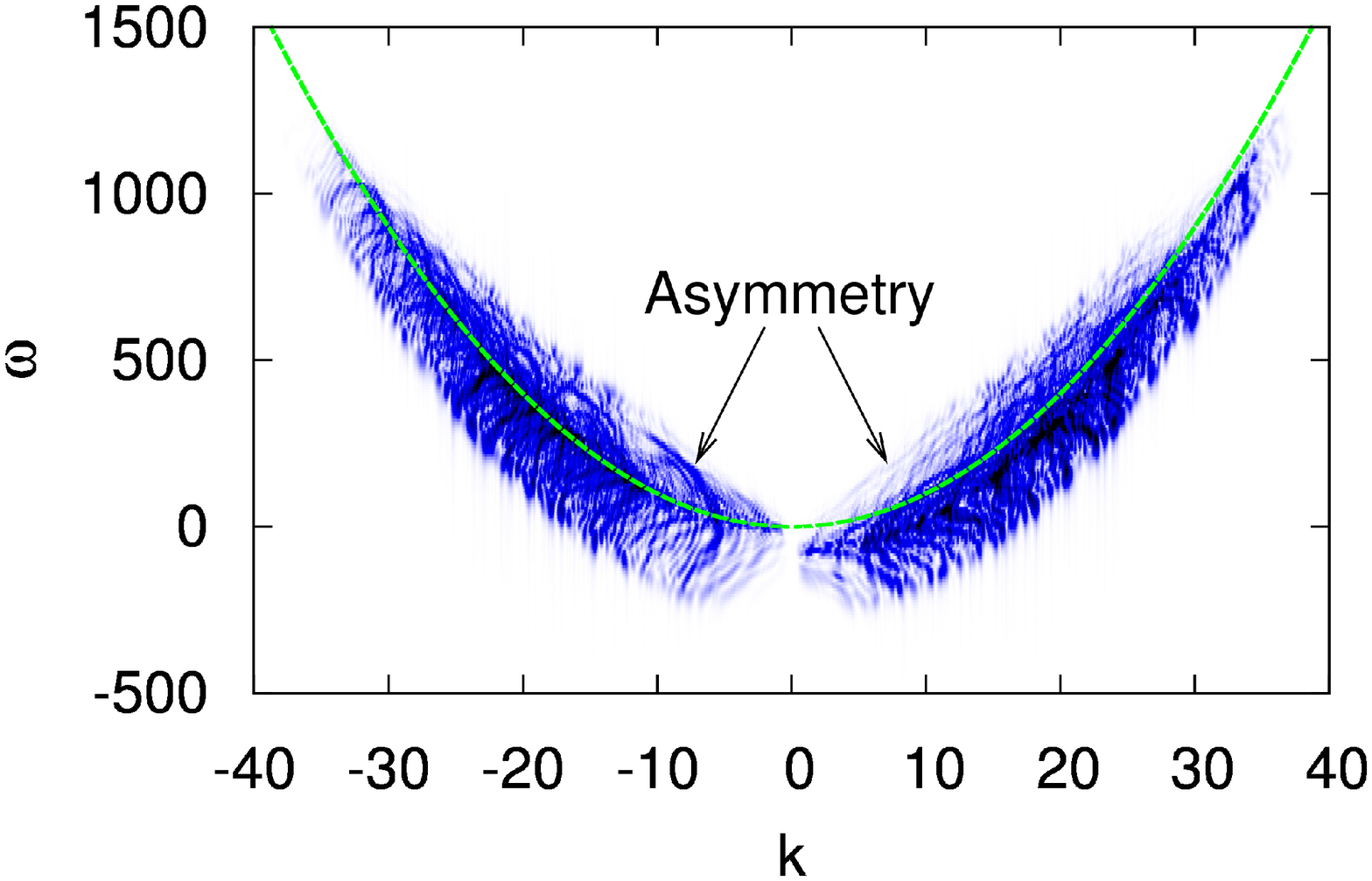}}

\caption{$(\bk,\omega)$-plots for the inverse cascade simulation of the \ac{SWE} at four different times $\tau = 13.67$, $18.65$, $28.60$ and $33.57$, where $\tau$ is in units of the linear evolution time $T_{\mathrm{L}}$ of the slowest wave mode in the system.  The dashed green line represents the linear wave frequency $\omega_k = k^2$. We observe asymmetry of the dispersion curve in figures (b) and (d) with respect to positive/negative wave numbers. \label{fig:k-w-inv-sw}}
\end{figure}

\begin{figure}[ht!]
\subfigure[]{\includegraphics[width = 0.55\columnwidth]{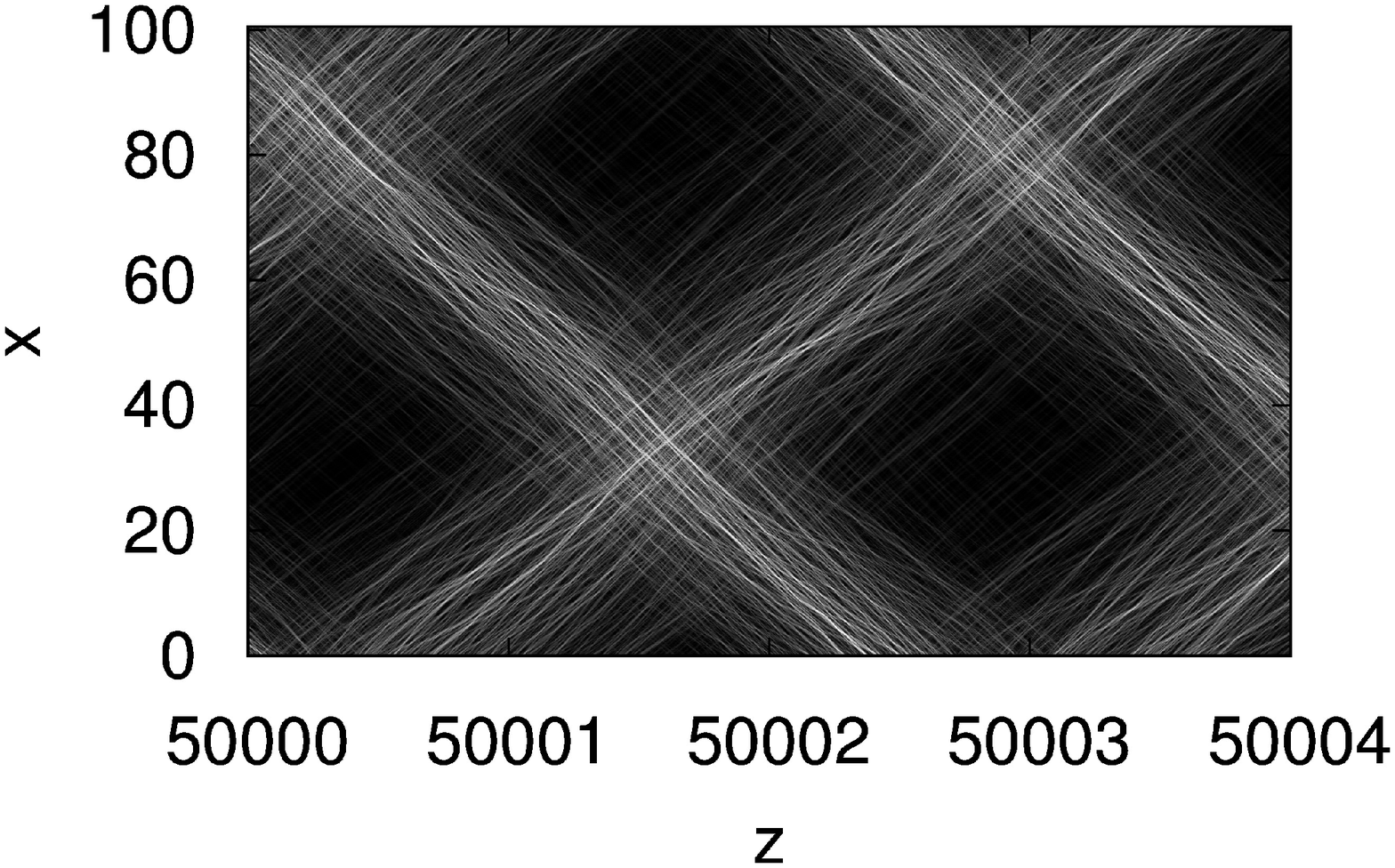}}
\hfill
\subfigure[]{\includegraphics[width = 0.45\columnwidth]{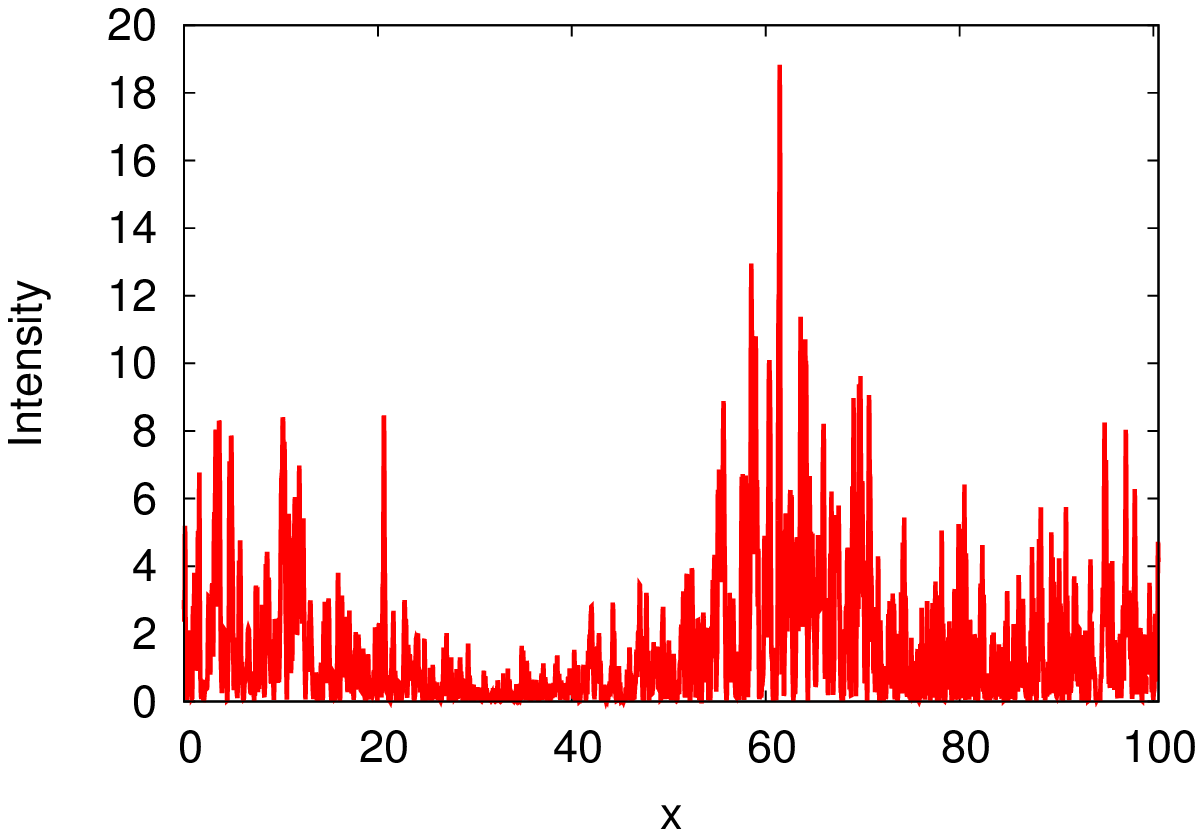}}

\caption{(a) Distribution of $I(x,z)$ at late times in the simulation of the \ac{SWE}. (b) We plot a time slice $I(x)$ at distance $z= 50001$. In (a) we observe coherent structures propagating with constant speed.  In (b) we observe a snapshot of the condensate structure. \label{fig:x-z-SWE}}
\end{figure}

In figures~\ref{fig:PDF-inverse-sw} we plot the normalized \acs{PDF} for the wave intensity in the physical space and for the Fourier-space intensities $J_\bk$.  In both plots for comparison we insert lines corresponding to a Gaussian wave field.  We observe in figure~\ref{fig:PDF-inverse-sw}a that there is a clear deviation at low intensities from Gaussianity.  This gives a clear indication of \ac{WT} intermittency, with the enhanced frequency of occurrence of dark spots  - far greater than what is predicted by a general Gaussian wave field.  For $J_\bk$ seen in figure~\ref{fig:PDF-inverse-sw}, we see a close agreement at intermediate $J_\bk$ with the Rayleigh distribution given by the black dashed line, but at large $J_\bk$ we see some slight deviation.  This is a possible sign of \ac{WT} intermittency, with the depletion of the \ac{PDF} tail from Gaussianity.  Inspection of the $(\bk,\omega)$-plots in figure~\ref{fig:k-w-inv-sw} shows  a wide spreading of the wave distribution around the linear dispersion curve $\omega_k = k^2$ indicating a nonlinearity level that is quite strong. Moreover, we observe asymmetric behavior of the wave distribution that clearly shifts from negative to positive wave numbers in figures~\ref{fig:k-w-inv-sw}b and \ref{fig:k-w-inv-sw}d.  This could indicate that the there is some symmetry breaking instability associated with the \ac{SWE}.  Furthermore, there is an interesting `hair'-like structure around the dispersion curve in figures~\ref{fig:k-w-inv-sw} which align together in a certain direction.  It consists of lines in the $(\bk,\omega)$-plot indicating a set of coherent structures propagating with  speeds that are determined by their slopes. The `hair'-like lines are not straight which indicates that the speed of such `solitons' vary with time. Also note that these structures never   `peel off' from the wave dispersion curve like it was the case in \ac{LWE} and, therefore, they never de-couple from the wave component.   In figure~\ref{fig:x-z-SWE}a we plot the distribution of intensity at late times of the simulation over the whole domain. We observe strange behavior, in which soliton like structures are seen bundled together and weakly interacting, while propagating as a bundle with constant speed.  These structures are clearly the hair-like objects seen in the $(\bk,\omega)$-plots of figures~\ref{fig:k-w-inv-sw}.  If we take a closer inspection of the coherent structures in figure~\ref{fig:x-z-SWE}, we see that they meander around the center of the bundle, but remain confined to the bundle.  We don't yet know how to explain this behavior theoretically or how it is related to the \ac{KZ} spectrum.  In figure~\ref{fig:x-z-SWE}b we show a time slice of the intensity $I$. We see a  low wave number coherent structure with high frequency peaks propagating upon it.

\section{Conclusions}

We introduced the field of \ac{1D} \ac{OWT}, where waves of light propagating through a nonlinear medium weakly interact. We presented the first ever experiment of \ac{OWT}, which is realized in a \ac{1D} setup based on
a \ac{LC} cell, and that serves us as a basic example to which we refer when developing the theoretical
framework for \ac{1D} \ac{OWT}.  We showed, by starting with an initial condition of weakly nonlinear waves, we were able, by the inverse wave action cascade described by \ac{WT} theory, to show the formation of strongly nonlinear solitons.   One intriguing aspect of \ac{OWT} is the inverse cascade, where wave action is transported to large scales.  This process is analogous to \ac{BEC} in super-cooled alkali gases.  We demonstrated this process in the experimental implementation of \ac{1D} \ac{OWT}.  We showed that by injecting high frequency waves into a nonlinear \ac{LC} cell, we observed a wave turbulent inverse cascade, followed by the condensation of incoherent photons into coherent solitons.  Moreover, we supported the experimental findings by performing numerical simulation of the \ac{LWE}. Analysis of the inverse cascade by \ac{WT} theory predicted that a pure \ac{KZ} state was not realizable.  Although the \ac{KZ} spectra are unrealizable, 
the \ac{OWT} system is still a very interesting object which is rich in physical effects and mathematical structure. We show how coherent solitons are formed by nonlinear interactions out of random incoherent waves, and analyze the importance of the soliton behavior with the random wave background.  Additionally, we determined that the inverse cascade can be described by a mixed wave action spectrum of thermal and non-equilibrium parts.  The inverse cascade is associated with the growth of nonlinearity of the system, gradually resulting in the formation of solitons by \ac{MI}.  We observed solitons in both the experiment and numerical simulations, and found that the system relaxes to a state of a single dominant soliton.  During the transition to this final state, we observed interactions between solitons and the random wave background.  With the aid of \ac{WT} theory, we developed a strong \ac{WT} description for the behavior of the system in the form of a \ac{WTLC}.  The \ac{WTLC} diagrammatically represents the behavior of the coexistence of coherent solitons  and the incoherent random wave background. 

In addition to the experimental setup, we performed numerical simulations of the \ac{LWE} and \ac{SWE} in non-equilibrium stationary regimes in forced-dissipated systems.  This allowed for the subsequent testing of the \ac{KZ} solutions. Theoretically we discovered that the \ac{LWE} emits fluxes of opposite sign to what is necessary for the \ac{KZ} states, and this was further verified by a lack of agreement of the wave action spectrum to the \ac{KZ} solutions.  Conversely, this problem does not occur for the \ac{SWE}. However, we showed that both the \ac{KZ} for the \ac{SWE} are non-local and thus should be non-realizable.  Nevertheless, we observed good agreements to the \ac{KZ} predictions for both the direct and inverse cascades. Further investigation into the fluxes yielded a non-constant energy flux for the direct cascade (not shown) - a constant flux being an essential requirement for a local \ac{KZ} solution.  This indicates that we observed a possible non-local wave action spectrum and if so, we are led to ask what non-local process determines this power law?  The answer to this can be determined with a development of a non-local theory for \ac{1D} \ac{OWT} in the short-wave limit - which is a future goal of ours.  The inverse cascade for the \ac{SWE} yielded an excellent agreement to the \ac{KZ} solution and a good indication of a constant negative flux of wave action (not shown).  This suggests that the inverse cascade is of \ac{KZ} type,  contradicting the non-locality analysis and thus raising questions in itself.  Thus further analysis of the numerical prefactor of the wave action spectrum is required and a possible explanation of the observations via a non-local theory.  Further, in the inverse cascade simulation
of the \ac{SWE}, we found a new fascinating physical phenomenon - generation of meandering solitons organized in crisscrossing bundles. 
Theoretical explanation for such a behavior or possible experimental
implementations are yet to be constructed.

In trying to tackle \ac{OWT}, we have ended with more questions than answers. We have observed very rich behavior with numerous interesting physical effects which potentially could come useful for technological applications insofar as
optical solitons are considered as potential candidates for optical interconnect and information processing. However, there remains a lack of a solid theory for explaining the observed results, and some of these results appear to be in contradiction with the classical \ac{WT}  analysis.
Thus, we consider this review as a report summarizing the first stage \ac{OWT} studies, with an open discussion of the observed results, theoretical 
advances  and remaining puzzles.  Our analysis has raised several new questions that need to be addressed with additional experiments and numerical simulations.  There is yet to be done an experimental investigation of higher dimensional \ac{OWT}, which we believe is a natural extension to the research presented in this report.  In addition, the description of the evolution of solitons, with the possible understanding of their behavior and manipulation would be valuable to any future applications in nonlinear optics. In this context OWT represents a challenging step forward, providing the occurrence of spontaneous
increase of coherence of light through weakly nonlinear wave interactions.

\section{Acknowledgements}
We acknowledge helpful discussions with V. Zakharov and with G. Assanto. 
We gratefully acknowledge the Referee for precious remarks and suggestions. 
This work has been partially supported by the Royal Society�s International Joint Project grant. 
U.B. and S.R. acknowledge support of the project ANR-07-BLAN-0246-03, {\it "turbonde"} and of the ANR international 
program, project ANR-2010-INTB-402-02	{\it "COLORS"}.

\newpage
\renewcommand{\appendixname}{}

\appendix
\addappheadtotoc
\appendixpage
\section{The Canonical Transformation}
\label{Appendix:ct}

In this Appendix, we will outline the details of applying the \ac{CT} to a six-wave system with non-resonant four-wave interactions.  This procedure is adapted from \cite{ZLF92}, where a \ac{CT} was used for converting a non-resonant three-wave system into one with resonant four-wave interactions.  The fundamental goal of the \ac{CT} is to remove the leading order non-resonant four-wave interactions by re-defining the canonical variable by a quasi-nonlinear canonical transformation. This will preserve the system's linear dynamics whilst removing the lowest order non-resonant dynamics.   The result will be a system that contains resonant wave interactions at its lowest order.

To begin, we must define a new canonical interaction variable, $c_\bk$, such that it satisfies an auxiliary Hamiltonian, $\mathcal{H}_{aux}$, in the sense that it contains arbitrary interaction coefficients at all orders\footnote{We have defined the auxiliary Hamiltonian up to six-wave interactions, which is sufficient for our systems. This will be verified {\em a posteriori}.}.  We can define a general auxiliary Hamiltonian, up to order six, for $c_\bk$ as: 
\begin{eqnarray}\label{eq:auxham}
 \mathcal{H}_{aux} &=& \frac{1}{2}\sum_{1,2,3} \tilde{V}^{1,2}_3\ \delta^{1,2}_{3}\left(c_1c_2c_3^* + \mathrm{c.c.}\right) + \frac{1}{6}\sum_{1,2,3}\tilde{U}^{1,2,3}\ \delta^{1,2,3}\left(c_1c_2c_3 + \mathrm{c.c.}\right)\nonumber\\
&&+\frac{1}{4}\sum_{1,2,3,4}\tilde{T}^{1,2}_{3,4}\ \delta^{1,2}_{3,4}\ c_1 c_2 c_3^* c_4^* + \frac{1}{6}\sum_{1,2,3,4} \tilde{X}^{1,2,3}_4\ \delta^{1,2,3}_4 \left(c_1c_2c_3c^*_4 + \mathrm{c.c.}\right)\nonumber\\
&&+\frac{1}{24}\sum_{1,2,3,4}\tilde{Y}^{1,2,3,4}\ \delta^{1,2,3,4} \left(c_1c_2c_3c_4 + \mathrm{c.c.}\right) \nonumber\\
&&+ \frac{1}{120}\sum_{1,2,3,4,5}\tilde{A}^{1,2,3,4,5}\ \delta^{1,2,3,4,5} \left(c_1c_2c_3c_4c_5 + \mathrm{c.c.}\right)\nonumber\\
&&+\frac{1}{24}\sum_{1,2,3,4,5}\tilde{B}^{1,2,3,4}_5\ \delta^{1,2,3,4}_5 \left(c_1c_2c_3c_4c^*_5 + \mathrm{c.c.}\right)\nonumber\\
&&+\frac{1}{12}\sum_{1,2,3,4,5}\tilde{C}^{1,2,3}_{4,5}\ \delta^{1,2,3}_{4,5} \left(c_1c_2c_3c^*_4c^*_5 + \mathrm{c.c.}\right)\nonumber\\
&&+\frac{1}{36}\sum_{1,2,3,4,5,6}\tilde{W}^{1,2,3}_{4,5,6}\ \delta^{1,2,3}_{4,5,6}\ c_1 c_2 c_3 c^*_4 c^*_5 c^*_6\nonumber\\
&&+\frac{1}{120}\sum_{1,2,3,4,5,6}\tilde{Q}^{1,2,3,4,5}_{6}\ \delta^{1,2,3,4,5}_{6} \left(c_1c_2c_3c_4c_5c^*_6+ \mathrm{c.c.}\right)\nonumber\\
&&+\frac{1}{48}\sum_{1,2,3,4,5,6}\tilde{R}^{1,2,3,4}_{5,6}\ \delta^{1,2,3,4}_{5,6} \left(c_1c_2c_3c_4c^*_5c^*_6+ \mathrm{c.c.}\right)\nonumber\\
&&+\frac{1}{36}\sum_{1,2,3,4,5,6}\tilde{S}^{1,2,3,4,5,6}\ \delta^{1,2,3,4,5,6} \left(c_1c_2c_3c_4c_5c_6+ \mathrm{c.c.}\right).
\end{eqnarray}
The auxiliary Hamiltonian is completely arbitrary and as such we denote all interaction coefficients with tildes to emphasize this. The idea is to construct  a new Hamiltonian, \eqref{eq:auxham}, for our system in terms of $c_\bk$ using the original Hamiltonian \eqref{eq:4wham-ot} to determine certain auxiliary interaction coefficients, whilst allowing us to explicitly choose the remaining arbitrary interaction coefficients in \eqref{eq:auxham} to eliminate the non-resonant interaction terms of \eqref{eq:4wham-ot}. 

The \ac{CT} is weakly nonlinear, thus preserving the linear dynamics of the system.  We utilize the fact that the time evolution operator is canonical \cite{ZLF92}, and then use the Taylor expansion of $a_\bk$ around $a(\bk,0) = c(\bk,0)$ giving 
\begin{equation}\label{eq:cantran}
 a(\bk,z) = c(\bk,0) + z \left.\left(\frac{\partial c(\bk,z)}{\partial z}\right)\right|_{z=0} + \frac{z^2}{2} \left.\left(\frac{\partial^2 c(\bk,z)}{\partial z^2}\right)\right|_{z=0}+\cdots.
\end{equation}
The coefficients of the \ac{CT} in \eqref{eq:cantran} can be computed using relation \eqref{eq:hamsys} applied to the auxiliary Hamiltonian, \eqref{eq:auxham}, i.e.
\begin{subequations}\label{eq:cantrancoeff}
 \begin{equation}\label{eq:cantrancoeff1}
  \left(\frac{\partial c_\bk}{\partial
z}\right)_{z=0}=-i\frac{\delta \mathcal{H}_{aux}}{\delta c^*_\bk},
 \end{equation}
\begin{equation}\label{eq:cantrancoeff2}
  \left(\frac{\partial^2c_\bk}{\partial z^2}\right)_{z=0}=-i
\frac{\partial}{\partial z}\frac{\delta \mathcal{H}_{aux}}{\delta c^*_\bk}.
\end{equation}
\end{subequations}
These coefficients will be of the form of summations, involving the canonical variable $c_\bk$ and the interaction coefficients of the auxiliary Hamiltonian.  Coefficient \eqref{eq:cantrancoeff1} is expressed below, and is simply determined by considering the variational derivative of the auxiliary Hamiltonian with respect to $c_\bk^*$:
\begin{eqnarray}
  \left(\frac{\partial c(\bk,t)}{\partial z}\right)_{t=0} &=& -i\left[    \frac{1}{2}\sum_{1,2} \tilde{V}^{1,2}_\bk\delta^{1,2}_{\bk}c_1c_2 + 2 \left( \tilde{V}^{\bk,1}_2 \right)^* \delta^{\bk,1}_{2} c_1^* c_2 \right. \nonumber \\ && \left. + \frac{1}{2} \sum_{1,2} \left( \tilde{U}^{\bk,1,2} \right)^* \delta^{\bk,1,2} c^*_1 c^*_2 + \frac{1}{2} \sum_{1,2,3} \tilde{W}^{1,2}_{\bk,3}\ \delta^{1,2}_{\bk,3} c_1 c_2 c_3^* \right. \nonumber \\ && \left. + \frac{1}{6} \sum_{1,2,3} \tilde{X}^{1,2,3}_\bk \delta^{1,2,3}_\bk c_1 c_2 c_3 + 3\left( \tilde{X}^{\bk,1,2}_3 \right)^* \delta^{\bk,1,2}_3 c^*_1 c^*_2 c_3 \right. \nonumber \\
&& + \frac{1}{6} \sum_{1,2,3} \left(\tilde{Y}^{\bk,1,2,3}\right)^*\delta^{\bk,1,2,3}c^*_1c^*_2c^*_3 \nonumber\\
&&+\frac{1}{24}\sum_{1,2,3,4}\left(\tilde{A}^{\bk,1,2,3,4}\right)^*\delta^{\bk,1,2,3,4}c^*_1c^*_2c^*_3c^*_4\nonumber\\
&&+\frac{1}{24}\sum_{1,2,3,4}\tilde{B}^{1,2,3,4}_\bk\delta^{1,2,3,4}_\bk c_1c_2c_3c_4 + 4\left(\tilde{B}^{\bk,1,2,3}_4\right)^*\delta^{\bk,1,2,3}_4c^*_1c^*_2c^*_3c_4 \nonumber\\
&&+\frac{1}{12}\sum_{1,2,3,4}2 \tilde{C}^{1,2,3}_{\bk,4}\ \delta^{1,2,3}_{\bk,4}c_1c_2c_3c^*_4 +3\left(\tilde{C}^{\bk,1,2}_{3,4}\right)^*\delta^{\bk,1,2}_{3,4}c^*_1c^*_2c_3c_4 \nonumber \\
&&+\frac{1}{12} \sum_{1,2,3,4,5} \tilde{T}^{1,2,3}_{\bk,4,5}\ \delta^{1,2,3}_{\bk,4,5}c_1c_2c_3c^*_4c^*_5 \nonumber\\
&& \left. + \frac{1}{120} \sum_{1,2,3,4,5} \tilde{Q}^{1,2,3,4,5}_{\bk} \delta^{1,2,3,4,5}_{\bk} c_1 c_2 c_3 c_4 c_5 \right. \nonumber \\ && \left. + 5\left( \tilde{Q}^{\bk,1,2,3,4}_{5} \right)^* \delta^{\bk,1,2,3,4}_{5} c^*_1 c^*_2 c^*_3 c^*_4 c_5 \right. \nonumber\\
&&+\frac{1}{48} \sum_{1,2,3,4,5} 2\tilde{R}^{1,2,3,4}_{\bk,5}\ \delta^{1,2,3,4}_{\bk,5} c_1 c_2 c_3 c_4 c^*_5 \nonumber \\ && + 4\left( \tilde{R}^{\bk,1,2,3}_{4,5} \right)^* \delta^{\bk,1,2,3}_{4,5} c^*_1 c^*_2 c^*_3 c_4 c_5 \nonumber\\
&&+\left.\frac{1}{6}\sum_{1,2,3,4,5}\left(\tilde{S}^{\bk,1,2,3,4,5}\right)^*\delta^{\bk,1,2,3,4,5}c^*_1c^*_2c^*_3c^*_4c^*_5\right].\label{eq:ct1}
\end{eqnarray}

The second coefficient, \eqref{eq:cantrancoeff2}, is derived by taking the time derivative of  \eqref{eq:ct1} and then re-substituting the expression for $\dot{c}_\bk$ from \eqref{eq:ct1}.  As one can expect, this leads to a lengthy formula which we omit from the review due to space restrictions.   With the full expression of the \ac{CT} found in terms of $c_\bk$s, we then substitute the \ac{CT} \eqref{eq:cantran} into the Hamiltonian \eqref{eq:4wham-ot}.  The result will be an expression of the Hamiltonian in terms of the new interaction representation variable $c_\bk$.  The Hamiltonian will involve the interaction coefficients from the auxiliary Hamiltonian (containing tildes) and $T^{1,2}_{3,4}$. The aim is to eliminate the non-resonant four-wave interaction contribution by selecting the values of the arbitrary interaction coefficients.  Moreover, as the system conserves wave action, we find that the \ac{CT} does not result in any additional odd $N$-wave contributions,  hence all arbitrary interaction coefficients for odd orders are automatically zero.

We find that elimination of all non-resonant four-wave contributions can be achieved by selecting 
\begin{equation}\label{eq:canelim}
 \tilde{T}^{1,2}_{3,4} = \frac{-4i\left(T^{1,2}_{3.4}\right)^*}{\omega_1 + \omega_2-\omega_3-\omega_4}.
\end{equation}
Relation \eqref{eq:canelim} is valid as the denominator does not vanish due to a lack of resonant four-wave interactions (i.e. no non-trivial solutions to the four-wave resonance condition \eqref{eq:4wrc}).  After selecting \eqref{eq:canelim}, we find that Hamiltonian \eqref{eq:4wham-ot} reduces to
\begin{eqnarray}\label{eq:6wham1}
 \mathcal{H} &=& \sum_\bk \omega_\bk c_\bk c_{\bk}^* + \frac{1}{36}\sum_{1,2,3,4,5,6} \left[\mathcal{W}^{1,2,3}_{4,5,6} - i\left(\omega_1+\omega_2+\omega_3 - \omega_4-\omega_5-\omega_6\right) \right. \nonumber \\ && \left. \times \tilde{W}^{1,2,3}_{4,5,6}\ \right]\ \delta^{1,2,3}_{4,5,6}\ c_1 c_2 c_3 c^*_4 c^*_5 c^*_6.
\end{eqnarray}
Hamiltonian \eqref{eq:6wham1} represents Hamiltonian \eqref{eq:4wham-ot} in the new canonical variable $c_\bk$, up to the leading resonant wave interaction, in this case being of order six. Notice, that within the six-wave contribution there is still an arbitrary contribution, $\tilde{W}^{1,2,3}_{4,5,6}$, that arises from the auxiliary Hamiltonian \eqref{eq:auxham}.  However, its prefactor, $\omega_1+\omega_2+\omega_3 - \omega_4-\omega_5-\omega_6$. vanishes when the six-wave resonance condition
\begin{subequations}\label{eq:6wres}
 \begin{equation}\label{eq:6wresk}
 \bk_1 + \bk_2 + \bk_3 = \bk_4 + \bk_5 + \bk_6, 
 \end{equation}
\begin{equation}\label{eq:6wresw}
 \omega(\bk_1) + \omega(\bk_2) +\omega(\bk_3) = \omega(\bk_4) + \omega(\bk_5) +\omega(\bk_6),  
\end{equation}
\end{subequations}
is satisfied. Ultimately, $\tilde{W}^{1,2,3}_{4,5,6}$ does not provide a contribution to the nonlinear wave dynamics.  Therefore, we may set $\tilde{W}^{1,2,3}_{4,5,6}$ to equal anything without altering the nonlinear wave dynamics.  For convenience  we select $\tilde{W}^{1,2,3}_{4,5,6}$ to equal minus the difference of $\mathcal{W}^{1,2,3}_{4,5,6}$ from the value taken when the resonance condition, \eqref{eq:6wres}, is satisfied. For instance, if we decompose $\mathcal{W}^{1,2,3}_{4,5,6}$ into two parts, the first being its value when the resonant condition is satisfied, say ${}^{RC}\mathcal{W}^{1,2,3}_{4,5,6}$ and the second being its residual ${}^{\perp}\mathcal{W}^{1,2,3}_{4,5,6} = \mathcal{W}^{1,2,3}_{4,5,6} - {}^{RC}\mathcal{W}^{1,2,3}_{4,5,6} $, i.e. we can then express 
\begin{equation}
 \mathcal{W}^{1,2,3}_{4,5,6} = {}^{RC}\mathcal{W}^{1,2,3}_{4,5,6} + {}^{\perp}\mathcal{W}^{1,2,3}_{4,5,6}. 
\end{equation}
By choosing
\begin{equation}
 \tilde{W}^{1,2,3}_{4,5,6} = \frac{-i{}^{\perp}\mathcal{W}^{1,2,3}_{4,5,6}}{\omega_1+\omega_2+\omega_3 - \omega_4-\omega_5-\omega_6}, 
\end{equation}
the arbitrary interaction coefficient, $\tilde{W}^{1,2,3}_{4,5,6}$, directly cancels with the residual contribution ${}^{\perp}\mathcal{W}^{1,2,3}_{4,5,6}$.   This means that the six-wave interaction coefficient will equal its value on the six-wave resonance condition over the whole of the domain. The final result yields the Hamiltonian expressed as equation \eqref{eq:6wham}.

\section{Details and Assumptions of Weak Wave Turbulence Theory}
\label{Appendix:wwt}

In this Appendix, we will outline the main technical details in deriving the evolution equations for the one-mode amplitude \ac{PDF} and for the wave action density $n_\bk$.  The means to do this is by constructing an evolution equation for a \ac{GF} that can generate evolution equations for all the statistical quantities in the wave system.

This approach can be applied to a general $N$-mode wave system\footnote{There exist several papers that consider the full $N$-mode statistics of three-wave and four-wave systems, for instance see \cite{CLN04,CLNP05}.}. However, we will only concentrate on the one-mode statistics of the system for brevity.  
Subsequently, we will only need to consider the one-mode amplitude \ac{GF} 
defined as 
\begin{equation}\label{eq:1aGen}
\mathcal{Z}_{\bk}(\lambda_\bk)= \left\langle e^{\left(\frac{L}{2\pi}\right)\lambda_\bk J_\bk}\right\rangle,
\end{equation}
where $J_\bk = |a_\bk|^2$ for wave number $\bk$.
Besides from deriving all the statistical moments of the wave field $a_\bk$ from the \ac{GF}, we can also obtain the one-mode  \ac{PDF} for wave intensities $J_\bk$ by taking the inverse Laplace transform of the one-mode amplitude \ac{GF}:
\begin{equation}\label{eq:laplace}
\mathcal{P}_\bk(s_\bk)=\frac{1}{2\pi i}\int^{i\infty+c}_{-i\infty+c}\; e^{-\left(\frac{L}{2\pi}\right)\lambda_\bk s_\bk}\;\mathcal{Z}_\bk(\lambda_\bk)\; d\lambda_\bk,
\end{equation} 
where $c$ is a constant greater than the real part of all singularities.  

Now we will derive the evolution equation for the \ac{GF} starting with the first step - the  weak nonlinearity expansion.


\subsection{The Weak Nonlinearity Expansion}
\label{sec:intvar}
The \ac{WT} strategy exploits the separation of the linear and nonlinear timescales. Namely, when considering a weakly nonlinear regime, (when wave amplitudes are small), the linear evolution time,
\begin{equation}\label{eq:lintime}
 T_{\mathrm{L}} = \frac{2\pi}{\omega_k},
\end{equation}
is much smaller than the nonlinear evolution time $T_{\mathrm{NL}}$ - the characteristic time for the nonlinear transfer of energy between waves.  This separation of timescales allows for an expansion in terms of a small parameter and then we make an average over the fast linear dynamics.  The result is a description of the nonlinear evolution of the wave system. 
First of all we note that the leading order nonlinear effects arise from the diagonal terms in the sum corresponding to the nonlinear interaction.
These terms correspond to the trivial pairings of wave numbers in equation \eqref{eq:6_wave_evol}, which do not contribute to the nonlinear exchange of energy between wave modes, but do provide an additional contribution to the frequency.  Consequently, before we can proceed, we must isolate this contribution to ensure that the weak nonlinearity expansion will be defined in a self-consistent way \cite{CLN04}.  By separating the diagonal terms from the main summation, equation~\eqref{eq:6_wave_evol} can be expressed as
\begin{equation}\label{eq:6_wave_corr}
  i\dot{c}_\bk = \left(\omega_k + \Omega_k\right)\;c_\bk + \frac{1}{12} \sum_{\mathcal{K}^{2,3}_{4,5,6}}\; \mathcal{W}^{\bk,2,3}_{4,5,6}\; c^*_2c^*_3c_4c_5c_6\; \delta^{\bk,2,3}_{4,5,6},
\end{equation}
where
\begin{equation}\label{eq:omegaNL}
\Omega_k = \frac{1}{2}\sum_{7,8}\; \mathcal{W}^{\bk,7,8}_{\bk,7,8}\; |c_7|^2\;|c_8|^2.
\end{equation}
The summation in \eqref{eq:6_wave_corr} is now taken over the set $\mathcal{K}^{2,3}_{4,5,6} = \{\bk_2,\bk_3,\bk_4,\bk_5,\bk_6: \bk, \bk_2,\bk_3 \neq \bk_4,\bk_5,\bk_6\}$.  We see that the nonlinear frequency $\Omega_k$ acts as a correction to the linear wave frequency, so that the system evolves with an effective wave frequency of $\tilde{\omega}_k = \omega_k + \Omega_k$.  
  
Let us define a scaled interaction representation variable $b_\bk$, so that it evolves on the slow nonlinear scale. This is achieved by compensating for the fast oscillating factor by
\begin{equation}\label{eq:intvar}
b_\bk = \frac{c_\bk}{\epsilon}\; e^{i\omega_kz + i\int^z_0\Omega_k dz'}.
\end{equation}
Note that we have incorporated the nonlinear frequency correction $\Omega_k$ into the oscillating factor, so that the variable $b_\bk$ varies only at the timescale of the energy transfer between the modes. We have also divided the right-hand side by a formal small parameter $\epsilon \ll 1$ for easier power counting of the nonlinearity orders (thus formally defining $b_\bk$ to be $\mathcal{O}(1)$).   
Substitution of formula \eqref{eq:intvar} into equation~\eqref{eq:6_wave_corr} gives
\begin{equation}\label{eq:b_k}
 i\dot{b}_\bk = \frac{\epsilon^4}{12}\sum_{\mathcal{K}^{2,3}_{4,5,6}}\; \mathcal{W}^{\bk,2,3}_{4,5,6}\; b^*_2b^*_3b_4b_5b_6\; \delta^{\bk,2,3}_{4,5,6}\; e^{i\omega^{k,2,3}_{4,5,6}z +i\epsilon^4\int^z_0 \Omega^{k,2,3}_{4,5,6} dz'},
\end{equation}
where $\omega^{k,2,3}_{4,5,6} = \omega_k +\omega_2 +\omega_3 -\omega_4 -\omega_5 -\omega_6$ and
$\Omega^{k,2,3}_{4,5,6} = \Omega_k +\Omega_2 +\Omega_3 -\Omega_4 -\Omega_5 -\Omega_6$.
As the nonlinear frequency correction is time-dependent, extra care should be taken when considering its contribution to \eqref{eq:b_k}.  
To separate the slow and the fast timescales, let us chose  an auxiliary intermediate time $T$, 
\begin{equation}\label{eq:timescale}
T_{\mathrm{L}} \ll T \ll T_{\mathrm{NL}}.
\end{equation}
For small wave amplitudes, $|c_\bk| \sim \epsilon \ll 1$, 
the nonlinear timescale for a six-wave process is of the order $T_{\mathrm{NL}}\sim 2\pi/\epsilon^8\omega_k$, which will be verified {\em a posteriori}.
Thus, to be specific, let us take $T =  2\pi/\epsilon^4 \omega_k$.

The main objective is to seek a solution for $b_\bk$, at the intermediate time $T$ in the form of an $\epsilon$-expansion
\begin{equation}\label{eq:b-expansion}
 b_\bk(T) = b_\bk^{(0)} + \epsilon^4 b_\bk^{(1)} + \epsilon^8 b_\bk^{(2)} + \cdots,
\end{equation}
and then solve each $\epsilon$-order of $b_\bk(T)$ by an iterative method using evolution equation \eqref{eq:b_k}. We substitute \eqref{eq:b-expansion} into \eqref{eq:b_k}, and equate $\epsilon$-orders.  The leading $\mathcal{O}(\epsilon^0)$ contribution gives
\begin{equation}\label{eq:beps0}
b_\bk^{(0)}(T) = b_\bk^{(0)}(0) =b_\bk(0).
\end{equation}
Result \eqref{eq:beps0}, implies that at leading order, the wave amplitude $b_\bk(T)$ is time-independent.  

In the following order, $ \mathcal{O}\left(\epsilon^4\right)$, we get
\begin{equation}\label{eq:beps1}
 b_\bk^{(1)}(T) = -\frac{i}{12}\sum_{\mathcal{K}^{2,3}_{4,5,6}}\; \mathcal{W}^{\bk,2,3}_{4,5,6}\; b^{(0)*}_2b^{(0)*}_3b^{(0)}_4b^{(0)}_5b^{(0)}_6\; \delta^{\bk,2,3}_{4,5,6} \; \Delta_T\left(\omega^{k,2,3}_{4,5,6} +{}^{(0)}\Omega^{k,2,3}_{4,5,6}\right).
\end{equation}
Here we have used that $b_\bk^{(1)}(0) = 0$ and have defined
\begin{equation}\label{eq:Delta_T}
\Delta_T(x)  = \int_0^T e^{i x z}dz = \frac{e^{i x T} -1 }{ix}.
\end{equation}
 and ${}^{(0)}\Omega^{k,2,3}_{4,5,6} =  \Omega_k^{(0)}+ \Omega_2^{(0)}+ \Omega_3^{(0)}- \Omega_4^{(0)}- \Omega_5^{(0)}- \Omega_6^{(0)}$
where
\begin{equation}
  {}\Omega_k^{(0)} =  \frac{\epsilon^4}{2}\sum_{7,8}\; \mathcal{W}^{\bk,7,8}_{\bk,7,8}\; |b^{(0)}_7|^2\;|b^{(0)}_8|^2.
\end{equation}
Note that, even though $\Omega_k^{(0)} \sim \mathcal{O}\left(\epsilon^4\right)$, the factor $e^{i  {}^{(0)}\Omega^{k,2,3}_{4,5,6} T} $ is $\mathcal{O}(1)$ because $T \sim \epsilon^{-4}$. Therefore, the $\epsilon$-orders of the of the left-hand side and the right-hand side are the same. In other words, it would be wrong to expand the exponential function in small $\epsilon$ here. The next order correction to $\Omega_k$ is indeed small enough ($ \sim \epsilon^{8}$) so that the exponential could be expanded, but its contribution would be nullified by the phase averaging (see below).


Finally, we use formulae \eqref{eq:beps0} and \eqref{eq:beps1} to determine $b^{(2)}_\bk(T)$:
\begin{eqnarray}\label{eq:beps2}
 b_\bk^{(2)} &=& \frac{1}{12}\sum_{\mathcal{K}^{2,3}_{4,5,6},\; \mathcal{K}^{7,8}_{9,10,11}}\left[  2\mathcal{W}^{\bk,2,3}_{4,5,6} \left(\mathcal{W}^{2,7,8}_{9,10,11}\right)^*\ b^{(0)*}_3 b^{(0)*}_9 b^{(0)*}_{10} b^{(0)*}_{11} b^{(0)}_4 b^{(0)}_5 b^{(0)}_6 b^{(0)}_7 b^{(0)}_8 \right. \nonumber \\ && \times \left. \delta^{\bk,2,3}_{4,5,6}\ \delta^{2,7,8}_{9,10,11} E_T\left(\left({{}^{(0)}\tilde\omega}^{2,7,8}_{9,10,11}\right)^*,{{}^{(0)}\tilde\omega}^{k,2,3}_{4,5,6}\right)-3\mathcal{W}^{\bk,2,3}_{4,5,6}\ \mathcal{W}^{4,7,8}_{9,10,11} \right. \\ && \left. \times b^{(0)*}_2 b^{(0)*}_3 b^{(0)*}_{7} b^{(0)*}_{8} b^{(0)}_5 b^{(0)}_6 b^{(0)}_9 b^{(0)}_{10} b^{(0)}_{11} \ \delta^{\bk,2,3}_{4,5,6}\ \delta^{4,7,8}_{9,10,11}\ E_T\left({{}^{(0)}\tilde\omega}^{4,7,8}_{9,10,11},{{}^{(0)}\tilde\omega}^{k,2,3}_{4,5,6}\right)\right],
\nonumber
\end{eqnarray}
where we have defined $E_T(x,y)$ as
\begin{equation}
 E_T(x,y) = \int^T_0 \Delta_z(x)e^{i\int^z_0 y dz'} dz,
\end{equation}
and ${}^{(0)}\tilde\omega^{k,2,3}_{4,5,6} = \tilde\omega^{k,2,3}_{4,5,6} +{}^{(0)}\Omega^{k,2,3}_{4,5,6}$.
The symmetry of $\Delta_T(x)$ implies that $E_T(x^*,y) = E_T(-x,y)$. With equations \eqref{eq:beps0}, \eqref{eq:beps1} and \eqref{eq:beps2}, we have calculated the $\epsilon$-expansion of $b_\bk$ up to $\mathcal{O}(\epsilon^8)$. This order is sufficient for the analysis contained within this review. The next objective is to substitute the above expansion into  the definition of  \ac{GF}, thereby finding the solution for \ac{GF} at the intermediate time $T$.


\subsection{Equation for the Generating Functional}


Recall that the one-mode amplitude \ac{GF} is defined in \eqref{eq:1aGen} where we can substitute
$J_\bk = |b_\bk|^2$.
 The aim is to calculate the $\epsilon$-expansion of $\mathcal{Z}_\bk$ at the intermediate time, $T$, and subsequently, derive an evolution equation, using  the results of the previous Subsection. Assume that we can represent the \ac{GF} in powers of $\epsilon$ as
\begin{equation}\label{eq:z}
 \mathcal{Z}_\bk = \mathcal{Z}_\bk^{(0)} + \epsilon^4\mathcal{Z}_\bk^{(1)} + \epsilon^8\mathcal{Z}_\bk^{(2)} + \cdots. 
\end{equation}
Then to calculate the \ac{GF},  we must average over random phases, $\langle \cdot \rangle_\phi$, and random amplitudes, $\langle \cdot \rangle_J$, in an initial \ac{RPA} field.  Once we have performed these, we can derive the evolution equation for $\mathcal{Z}_\bk$, and thus any of the one-mode statistical objects we require.

Now we will perform phase and amplitude averaging over a \ac{RPA} field for the derivation of the evolution equation for the \ac{GF}.  To achieve this, we express the \ac{GF} in powers of $\epsilon$, and represent each order in terms of the variable $b_\bk^{(0)}$.  At this stage, we can average over phases by assuming a \ac{RPA} field.  Once, phase averaged, we can average over amplitudes and compute the evolution equation.

To begin, we substitute the $\epsilon$-expansion of $b_\bk$, \eqref{eq:b-expansion}, into definition \eqref{eq:1aGen} and consider terms up to $\mathcal{O}(\epsilon^8)$.  Further simplification can be achieved by Taylor expansion of each exponential\footnote{This will not alter the the accuracy of the derivation.}, thus obtaining
 \begin{eqnarray}\label{eq:zk}
 \mathcal{Z}_\bk &= &\left\langle e^{\left(\frac{L}{2\pi}\right)\lambda_\bk |b_\bk^{(0)}|^2}\left[ 1+ \epsilon^4\left(\frac{L}{2\pi}\right)\lambda_\bk\left(b_\bk^{(0)}b_\bk^{(1)*} + b_\bk^{(0)*}b_\bk^{(1)} \right) \right.\right.\nonumber\\
&&+ \left.\left.\epsilon^8\left(\left(\frac{L}{2\pi}\right)\lambda_\bk\left(|b_\bk^{(1)}|^2 + b_\bk^{(0)}b_\bk^{(2)*} +b_\bk^{(0)*}b_\bk^{(2)}\right)\right. \right.\right.\nonumber\\
&& +\left.\left.\left.\left(\frac{L}{2\pi}\right)^2\frac{\lambda_\bk^2}{2}\left(b_\bk^{(0)}b_\bk^{(1)*}+b_\bk^{(0)*}b_\bk^{(1)}\right)^2\right)\right]\right\rangle  + \mathcal{O}(\epsilon^{12}).
\end{eqnarray}
We now proceed by averaging each term in expansion \eqref{eq:zk}.  The leading $\mathcal{O}(\epsilon^0)$ contribution of $\mathcal{Z}_\bk$ is given by
\begin{equation}\label{eq:zk0}
\mathcal{Z}_\bk^{(0)} = \left\langle e^{\left(\frac{L}{2\pi}\right)\lambda_\bk|b_\bk^{(0)}|^2}\right\rangle= \left\langle e^{\left(\frac{L}{2\pi}\right)\lambda_\bk J^{(0)}_\bk}\right\rangle = \mathcal{Z}_\bk(0),
\end{equation}
where $J^{(0)}_\bk = |b_\bk^{(0)}|^2$ is the amplitude of the interaction variable $b_\bk^{(0)}$.  For clarity, we shall omit the superscript of $J^{(0)}_\bk$ which will thus become $J_\bk$\footnote{Note that now $J_\bk$  is not the amplitude of the {\em full} interaction variable $b_\bk$, as defined in Section~\ref{sec:stat}.}.   The leading order of the $\epsilon$-expansion is labeled as $\mathcal{Z}_\bk(0)$, corresponding to the initial value of the \ac{GF}.  The next order yields
\begin{equation}\label{eq:zk1}
\mathcal{Z}_\bk^{(1)} = \left\langle \left(\frac{L}{2\pi}\right)\lambda_\bk e^{\left(\frac{L}{2\pi}\right)\lambda_\bk J_\bk}\left( b_\bk^{(0)}b_\bk^{(1)*} + b_\bk^{(0)*}b_\bk^{(1)}\right)\right\rangle,
\end{equation}
and subsequently at the final order to be considered, we obtain
\begin{eqnarray}\label{eq:zk2}
\mathcal{Z}_\bk^{(2)} &=& \left\langle \left(\frac{L}{2\pi}\right)\lambda_\bk e^{\left(\frac{L}{2\pi}\right)\lambda_\bk J_\bk}\left[|b_\bk^{(1)}|^2 + b_\bk^{(0)}b_\bk^{(2)*} + b_\bk^{(0)*}b_\bk^{(2)}\right.\right.\nonumber\\
&& \left.\left.+ \left(\frac{L}{2\pi}\right)\frac{\lambda_\bk}{2}\left(2J_\bk|b_\bk^{(1)}|^2+ \left(b_\bk^{(0)}b_\bk^{(1)*}\right)^2+\left(b_\bk^{(0)*}b_\bk^{(1)}\right)^2\right)\right] \right \rangle.
\end{eqnarray}
Having expressed $\mathcal{Z}_\bk$ in terms of $b_\bk$, we now proceed by evaluating the ensemble averages.

We must take the phase average before considering the amplitude average.  In the expansion for the \ac{GF} we can identify three distinct terms, these are $\langle b^{(0)}_\bk b^{(1)*}_\bk\rangle$, $\langle|b_\bk^{(1)}|^2\rangle$, and $\langle b^{(0)}_\bk b^{(2)*}_\bk\rangle$.  We implement Wick's rule, which states that the only contributions are those where one can pairwise match the wave numbers.   However, due to the trivial pairings leading to the nonlinear frequency correction, we are restricted to which wave numbers can be matched\footnote{See the definition of set $\mathcal{K}^{2,3}_{4,5,6}$.}.  Consequently, for $\langle b^{(0)}_\bk b^{(1)*}_\bk\rangle$ no pairings can occur and thus the average is zero.  This is also the case for $\langle \left(b^{(0)}_\bk b^{(1)*}_\bk\right)^2\rangle_\phi$ and so gives zero.

In the case of  $\langle|b_\bk^{(1)}|^2\rangle$, we are able to match the wave numbers in such a way as to not contradict the restrictions of the summations.  Applying Wick's rule, we find that
\begin{eqnarray}
  &&\left\langle b^{(0)*}_2b^{(0)*}_3b^{(0)*}_9b^{(0)*}_{10}b^{(0)*}_{11}  b^{(0)}_4b^{(0)}_5b^{(0)}_6b^{(0)}_7b^{(0)}_8   \right\rangle_\phi =\left(\delta^2_7\delta^3_8+\delta^2_8\delta^3_7\right)\nonumber\\
&&\times \left[ \delta^4_9\left(\delta^5_{10}\delta^6_{11}+\delta^5_{11}\delta^{6}_{10}\right)  +\delta^4_{10}\left(\delta^5_{9}\delta^6_{11}+\delta^5_{11}\delta^{6}_{9}\right)    +  \delta^4_{11}\left(\delta^5_{9}\delta^6_{10}+\delta^5_{10}\delta^{6}_{9}\right) \right].
\end{eqnarray}
 This gives
\begin{equation}\label{eq:b1b1}
  \left\langle |b_\bk^{(1)}|^2 \right\rangle_\phi = \frac{1}{12}\sum_{\mathcal{K}^{2,3}_{4,5,6}} |\mathcal{W}^{\bk,2,3}_{4,5,6}|^2J_2 J_3J_4J_5J_6\ \delta^{\bk,2,3}_{4,5,6}\ | \Delta_T(\tilde{\omega}^{k,2,3}_{4,5,6})|^2.
\end{equation}
Applying a similar approach to $\langle b^{(0)}_\bk b^{(2)*}_\bk\rangle_\phi$, which also does not average to zero, we arrive at
\begin{eqnarray}\label{eq:b0b2}
 \langle b^{(0)}_\bk b^{(2)*}_\bk \rangle_\phi &=& \frac{1}{12}\sum_{\mathcal{K}^{2,3}_{4,5,6}} |\mathcal{W}^{\bk,2,3}_{4,5,6}|^2 J_\bk\left(J_2+J_3\right)J_4J_5J_6\delta^{\bk,2,3}_{4,5,6}E_T\left(\left({\omega}^{k,2,3}_{4,5,6}\right)^*, {\omega}^{k,2,3}_{4,5,6}\right)\nonumber\\
&&-\frac{1}{12}\sum_{\mathcal{K}^{2,3}_{4,5,6}} |\mathcal{W}^{\bk,2,3}_{4,5,6}|^2 J_\bk J_2J_3\left(J_4J_5 +J_4J_6 + J_5J_6\right)\delta^{\bk,2,3}_{4,5,6}\nonumber\\
&&\times E_T^*\left({\omega}_{k,2,3}^{4,5,6}, {\omega}^{k,2,3}_{4,5,6}\right).
\end{eqnarray}

By defining  the wave action spectrum $n_\bk$, and the derivative of the \ac{GF} with respect to $\lambda_\bk$ as
\begin{subequations}
 \begin{equation}
  n_\bk=\left(\frac{L}{2\pi}\right)\langle J_\bk \rangle_J , 
 \end{equation}
 \begin{equation}
  \frac{\partial \mathcal{Z}_\bk}{\partial \lambda_\bk}=\frac{\partial}{\partial\lambda_\bk}\left\langle e^{\left(\frac{L}{2\pi}\right)\lambda_\bk J_\bk}\right \rangle_J = \left\langle \left(\frac{L}{2\pi}\right)J_\bk e^{\left(\frac{L}{2\pi}\right)\lambda_\bk J_\bk} \right\rangle_J,
 \end{equation}
\end{subequations}
and using our formulae  \eqref{eq:b1b1} and \eqref{eq:b0b2}, equation \eqref{eq:z} can be written in the form,
\begin{eqnarray}\label{eq:z-afterJ}
 \mathcal{Z}_\bk(T) - \mathcal{Z}_\bk(0) &=& \left(\frac{2\pi}{L}\right)^4\frac{\epsilon^8}{12}\sum_{\mathcal{K}^{2,3}_{4,5,6}}|\mathcal{W}^{\bk,2,3}_{4,5,6}|^2\delta^{\bk,2,3}_{4,5,6}\left[ \left(   \lambda_\bk\mathcal{Z}_\bk+\lambda_\bk^2\frac{\partial \mathcal{Z}_\bk}{\partial \lambda_\bk}\right)\right.\nonumber \\
&& \times \; \left.n_2n_3n_4n_5n_6 \left|\Delta_T\left({\omega}^{k,2,3}_{4,5,6}\right)\right|^2 \right.\nonumber\\  
&& +2\lambda_\bk\frac{\partial \mathcal{Z}_\bk}{\partial \lambda_\bk}\left(n_2+n_3\right)n_4n_5n_6\ \Re\left[E_T\left(\left({\omega}^{k,2,3}_{4,5,6}\right)^*, {\omega}^{k,2,3}_{4,5,6}\right)\right]\nonumber \\
&& -  2\lambda_\bk\frac{\partial \mathcal{Z}_\bk}{\partial \lambda_\bk}n_2n_3\left(n_4n_5 + n_4n_6 + n_5n_6\right)\nonumber\\
&&\left.\times \Re\left[E_T^*\left({\omega}_{k,2,3}^{4,5,6}, {\omega}^{k,2,3}_{4,5,6}\right)\right]\right],
\end{eqnarray}
up to $\mathcal{O}(\epsilon^{12})$ corrections.

In equation~\eqref{eq:z-afterJ}, we now take the large box and long evolution time (or equivalently the weak nonlinearity) limits.  The order in which the limits are taken is essential. We must take the large box limit, $N\to \infty$, before the weak nonlinearity limit, $T\sim 1/\epsilon^4\to \infty$, otherwise the width of the nonlinear resonance broadening will be smaller than the frequency grid spacing, and will therefore result in a zero contribution.  For the large box limit, the summations are replaced by integrations of the form: $\lim_{N\to \infty}\sum_{2,3,4,5,6} = \left(\frac{L}{2\pi}\right)^5\int d\bk_2\ d\bk_3\ d\bk_4\ d\bk_5\ d\bk_6$ and the Kronecker delta function is transformed into a Dirac delta function with a prefactor of $2\pi / L$. 
In the weak nonlinearity limit, we can calculate the asymptotic limits of $\Delta_T(x)$:
  \begin{equation}\label{eq:limitDelta}
 \lim_{T\to\infty} |\Delta_T(x)|^2 = 2\pi T\delta (x),
\end{equation}
and of $E_T(x,y)$:
\begin{equation}\label{eq:limitE}
 \lim_{T\to\infty}\Re[E_T(-x,x)]  = \pi T\delta (x).
\end{equation}

Moreover, as the intermediate time $T$, at which point we want to seek a solution, is smaller than the nonlinear evolution time $T_{\mathrm{NL}}$, we can make an approximation for the time derivative of $\mathcal{Z}_\bk$ by:
\begin{equation}\label{eq:limitZ}
\dot{\mathcal{Z}_\bk} \approx \frac{\mathcal{Z}_\bk(T) - \mathcal{Z}_\bk(0)}{T} . 
\end{equation}

The large box and weak nonlinearity limits, together with results \eqref{eq:limitDelta}, \eqref{eq:limitE} and \eqref{eq:limitZ}, gives the continuous description of the evolution of the \ac{GF}:
\begin{equation}\label{eq:evol_z}
\dot{\mathcal{Z}_\bk}=\lambda_\bk\eta_\bk\mathcal{Z}_\bk + \left(\lambda_\bk^2\eta_\bk - \lambda_\bk\gamma_\bk\right)\frac{\partial \mathcal{Z}_\bk}{\partial\lambda_\bk},
\end{equation}
where $\eta_\bk$ and $\gamma_\bk$ are defined in the main text as equations \eqref{eq:eta} and \eqref{eq:gamma} respectively.
By taking the inverse Laplace transform of \eqref{eq:evol_z}, we get the evolution equation for the amplitude \ac{PDF} \eqref{eq:evol_pdf}.

\section{The Zakharov Transform}
\label{Appendix:zt}

The \ac{ZT} is used to find steady state solutions to the \ac{KE}.  The transformation utilizes symmetries of the \ac{KE} to overlap six disjoint regions of the domain of the collision integral on to one another.  From the resulting integrand, one can straightforwardly derive the steady states of the \ac{KE}.   In Section~\ref{sec:zak}, we presented the transform on one of these regions.  In this Appendix, we give the four remaining transformations:
\begin{subequations}
\begin{equation}
\tilde{\bk}_1 = \frac{\tilde{\bk}\tilde{\bk}^{'}_1}{\tilde{\bk}_2^{'}},\quad \tilde{\bk}_2 = \frac{\tilde{\bk}^2}{\tilde{\bk}_2^{'}}, \quad \tilde{\bk}_3 =
\frac{\tilde{\bk}\tilde{\bk}_3^{'}}{\tilde{\bk}_2^{'}},\quad \tilde{\bk}_4 = \frac{\tilde{\bk}\tilde{\bk}_4^{'}}{\tilde{\bk}_2^{'}} \quad \mathrm{and} \quad \tilde{\bk}_5 =
\frac{\tilde{\bk}\tilde{\bk}_5^{'}}{\tilde{\bk}_2^{'}},
\end{equation}
with Jacobian $J=-\left(\tilde{\bk} / \tilde{\bk}^{'}_2 \right)^6$.
\begin{equation}
\tilde{\bk}_1 = \frac{\tilde{\bk}\tilde{\bk}_4^{'}}{\tilde{\bk}_3^{'}},\quad \tilde{\bk}_2 = \frac{\tilde{\bk}\tilde{\bk}_5^{'}}{\tilde{\bk}_3^{'}}, \quad \tilde{\bk}_3
= \frac{\tilde{\bk}^2}{\tilde{\bk}_3^{'}},\quad\tilde{\bk}_4 = \frac{\tilde{\bk}\tilde{\bk}_1^{'}}{\tilde{\bk}_3^{'}} \quad \mathrm{and} \quad \tilde{\bk}_5 =
\frac{\tilde{\bk}\tilde{\bk}_2^{'}}{\tilde{\bk}_3^{'}},
\end{equation}
with Jacobian $J=-\left(\tilde{\bk} / \tilde{\bk}^{'}_3 \right)^6$.
\begin{equation}
\tilde{\bk}_1 = \frac{\tilde{\bk}\tilde{\bk}_5^{'}}{\tilde{\bk}_4^{'}},\quad \tilde{\bk}_2 = \frac{\tilde{\bk}\tilde{\bk}_3^{'}}{\tilde{\bk}_4^{'}}, \quad \tilde{\bk}_3
= \frac{\tilde{\bk}\tilde{\bk}^{'}_{2}}{\tilde{\bk}_4^{'}},\quad\tilde{\bk}_4 = \frac{\tilde{\bk}^2}{\tilde{\bk}_4^{'}} \quad \mathrm{and} \quad \tilde{\bk}_5 =
\frac{\tilde{\bk}\tilde{\bk}_1^{'}}{\tilde{\bk}_4^{'}},
\end{equation}
with Jacobian $J=-\left(\tilde{\bk} / \tilde{\bk}^{'}_4 \right)^6$.
\begin{equation}
\tilde{\bk}_1 = \frac{\tilde{\bk}\tilde{\bk}_3^{'}}{\tilde{\bk}_5^{'}},\quad \tilde{\bk}_2 = \frac{\tilde{\bk}\tilde{\bk}_4^{'}}{\tilde{\bk}_5^{'}}, \quad \tilde{\bk}_3
= \frac{\tilde{\bk}\tilde{\bk}^{'}_{1}}{\tilde{\bk}_5^{'}},\quad\tilde{\bk}_4 = \frac{\tilde{\bk}\tilde{\bk}^{'}_2}{\tilde{\bk}_5^{'}} \quad \mathrm{and} \quad \tilde{\bk}_5 =\frac{\tilde{\bk}^2}{\tilde{\bk}_5^{'}},
\end{equation}
\end{subequations}
with Jacobian $J=-\left(\tilde{\bk} / \tilde{\bk}^{'}_5 \right)^6$.

We re-iterate that these transformations are only applicable if the collision integral is proven to be convergent upon the \ac{KZ} solutions.  If not, then the \ac{KZ} solutions are deemed non-local and the solutions found using the \ac{ZT} are unphysical.

\section{Expansion of the Long-Wave Six-Wave Interaction Coefficient}
\label{Appendix:LWE-W}

As the \ac{LWE} contains two nonlinear terms, the resulting four-wave interaction coefficient \eqref{eq:longcoefficients2}, will also contain two expressions.  \ac{WT} theory relies on that fact that the interaction coefficients are scale invariant.  However, both of these expressions  express different scalings with respect to the wave number $\bk$ (where all four wave numbers are of the same order, i.e. $k_1,k_2,k_3,k_4 \propto k$). However, for the \ac{LWE} this does not pose a problem because the long-wave limit $kl_\xi \ll 1$, provides a small parameter for us to determine a leading order, scale invariant, contribution to the four-wave interaction coefficient.  The final six-wave interaction coefficient $\mathcal{W}^{1,2,3}_{4,5,6}$, of the \ac{LWE} can be represented as a sum of three contributions:
\begin{subequations}\label{eq:longexp}
\begin{equation}
 {}^{\mathrm{L}}\mathcal{W}^{1,2,3}_{4,5,6} =  {}^1\mathcal{W}^{1,2,3}_{4,5,6}+ {}^2\mathcal{W}^{1,2,3}_{4,5,6}+ {}^3\mathcal{W}^{1,2,3}_{4,5,6},
\end{equation}
where each term is of a different scaling with $k$.  The exact expressions are given by
\begin{equation}
 {}^1\mathcal{W}^{1,2,3}_{4,5,6} = -\frac{1}{8}\displaystyle\sum^{3}_{\substack{i,j,m=1\\ i\neq j\neq
m\neq i}}\displaystyle\sum^{6}_{\substack{p,q,r=4\\ p\neq q \neq r\neq
p}}\frac{{}^1 T^{p+q-i,i}_{p,q}\;{}^1 T^{j+m-r,r}_{j,m}}{\omega^{j+m-r,r}_{j,m}}+\frac{{}^1 T^{i+j-p,p}_{i,j}\; {}^1T^{q+r-m,m}_{q,r}}{\omega^{q+r-m,m}_{q,r}},
\end{equation}
\begin{eqnarray}
{}^2\mathcal{W}^{1,2,3}_{4,5,6} &=& -\frac{1}{8}\displaystyle\sum^{3}_{\substack{i,j,m=1\\ i\neq j\neq
m\neq i}}\displaystyle\sum^{6}_{\substack{p,q,r=4\\ p\neq q \neq r\neq
p}}\frac{{}^1 T^{p+q-i,i}_{p,q}\;{}^2 T^{j+m-r,r}_{j,m}}{\omega^{j+m-r,r}_{j,m}}+\frac{{}^1 T^{i+j-p,p}_{i,j}\; {}^2T^{q+r-m,m}_{q,r}}{\omega^{q+r-m,m}_{q,r}}\nonumber\\
&&-\frac{1}{8}\displaystyle\sum^{3}_{\substack{i,j,m=1\\ i\neq j\neq
m\neq i}}\displaystyle\sum^{6}_{\substack{p,q,r=4\\ p\neq q \neq r\neq
p}}\frac{{}^2 T^{p+q-i,i}_{p,q}\;{}^1 T^{j+m-r,r}_{j,m}}{\omega^{j+m-r,r}_{j,m}}+\frac{{}^2 T^{i+j-p,p}_{i,j}\; {}^1T^{q+r-m,m}_{q,r}}{\omega^{q+r-m,m}_{q,r}}
\end{eqnarray}
and
\begin{equation}
{}^3\mathcal{W}^{1,2,3}_{4,5,6} = -\frac{1}{8}\displaystyle\sum^{3}_{\substack{i,j,m=1\\ i\neq j\neq
m\neq i}}\displaystyle\sum^{6}_{\substack{p,q,r=4\\ p\neq q \neq r\neq
p}}\frac{{}^2 T^{p+q-i,i}_{p,q}\;{}^2 T^{j+m-r,r}_{j,m}}{\omega^{j+m-r,r}_{j,m}}+\frac{{}^2 T^{i+j-p,p}_{i,j}\; {}^2T^{q+r-m,m}_{q,r}}{\omega^{q+r-m,m}_{q,r}},
\end{equation}
\end{subequations}
where ${}^1T$ and ${}^2T$ are defined in equation~\eqref{eq:longcoefficients2}.

The three contributions of ${}^{\mathrm{L}}\mathcal{W}^{1,2,3}_{4,5,6}$ given in equations~\eqref{eq:longexp}, each are of different orders in $kl_\xi$ - the small parameter arising from the long-wave limit.  Therefore, $ {}^1\mathcal{W}^{1,2,3}_{4,5,6}\sim \mathcal{O}\left((kl_\xi)^4\right)$ is the leading contribution to the six-wave dynamics, followed by ${}^2\mathcal{W}^{1,2,3}_{4,5,6}\sim \mathcal{O}\left( (kl_\xi)^6\right)$ and finally ${}^3\mathcal{W}^{1,2,3}_{4,5,6}\sim \mathcal{O}\left( (kl_\xi)^8\right)$. 

The first term in expansion (\ref{eq:longexp}), ${}^1\mathcal{W}^{1,2,3}_{4,5,6}$, is generated from the coupling of the leading cubic nonlinear term, ${}^1T$, of (\ref{eq:eq-long}) with itself. If we were to consider the \ac{1D} \ac{NLSE}, then the six-wave interaction coefficient would only consist of this contribution, ${}^1\mathcal{W}^{1,2,3}_{4,5,6}$, as ${}^2 T^{1,2}_{3,4}\equiv 0$. Integrability of the \ac{1D} \ac{NLSE} should imply that the six-wave interaction coefficient ${}^1\mathcal{W}^{1,2,3}_{4,5,6}$ is zero when the six-wave resonance condition is satisfied.  We verified that this is indeed the case by using parametrization (\ref{eq:liaparam}) and utilizing the \texttt{Mathematica} package.  

Therefore, the main contribution to the six-wave dynamics for the \ac{LWE} arise from the second (the first non-zero) term in (\ref{eq:longexp}), ${}^2\mathcal{W}^{1,2,3}_{4,5,6}$.  This contribution, with the use of (\ref{eq:liaparam}), gives a $\bk$-independent result:
\begin{equation}\label{eq:LWE-const}
  {}^{\mathrm{L}}\mathcal{W}^{1,2,3}_{4,5,6} \approx  {}^2\mathcal{W}^{1,2,3}_{4,5,6}  =  \frac{9\varepsilon_0^2n_a^8l_\xi^6k_0^4}{16K^2}. 
\end{equation}
We shall approximate the six-wave interaction coefficient of the \ac{LWE} by the constant given by \eqref{eq:LWE-const}.

\section{Locality of the Kinetic Equation Collision Integral}
\label{Appendix:local}

In this Appendix, we will derive the necessary criteria for the convergence of the collision integral in the \ac{KE}.  We will check for convergence in the limits where a combination of wave numbers diverge to infinity or tend to zero. The Dirac delta functions present in the collision integral imply that the six-wave resonance condition \eqref{eq:6wres} is always satisfied.  To make our analysis easier we will use a parametrization for the six-wave resonance condition. This parametrization is defined by making the values of two wave numbers functions of the remaining four such that the six-wave resonance condition is satisfied.  This procedure leads to the parametrization:
\begin{subequations}\label{eq:liaparam}
 \begin{eqnarray}
  \bk_2 &=& \frac{(\bk_4-\bk)(\bk_3-\bk_4)}{\bk+\bk_3-\bk_4-\bk_6} + \bk_6,\\
\bk_5 &=& \frac{(\bk_4-\bk)(\bk_3-\bk_4)}{\bk+\bk_3-\bk_4-\bk_6} + \bk + \bk_3-\bk_4. 
 \end{eqnarray}
\end{subequations}


We begin by checking for convergence in the \ac{IR} region of $\bk$-space, by determining the $\bk$-scaling of the \ac{KE} as one wave number, say $\bk_6 \to 0$.  Therefore, we must find the \ac{IR} $\bk$-scaling of each contribution in the collision integral.  We will present our analysis for a general six-wave \ac{KE}.   So we assume that the six-wave interaction coefficient has the following scaling as $k_6\to 0$:
\begin{equation}\label{eq:wscaleonesmall-ap}
\lim_{k_6 \to 0} \mathcal{W}^{\bk,1,2}_{4,5,6} \propto k_6^\xi,
\end{equation}
where $\xi \in \mathbb{R}$. Then in the same limit, the following term in the \ac{KE} behaves as 
\begin{eqnarray}
\lim_{\bk_6 \to 0}&&n_\bk n_2n_3n_4n_5n_6\left(n_\bk^{-1}+n_2^{-1}+n_3^{-1}-n_4^{-1}-n_5^{-1}-n_6^{-1}\right)\\
&\propto& n_\bk n_2n_3n_4n_5n_6\left(n_\bk^{-1}+n_2^{-1}+n_3^{-1}-n_4^{-1}-n_5^{-1}\right)\propto n_6\propto k_6^{-x},\nonumber
\end{eqnarray}
where $x$ is the wave action spectrum exponent for $n_\bk$ from \eqref{eq:powerlaw}.  Thus, we can factorize out the integral over $\bk_6$ in the \ac{KE} and write it as
\begin{equation}\label{eq:nonlocalintegral}
 \int k_6^{2\xi}n_6\ d\bk_6 \propto 2\int_{0}k_6^{2\xi-x}\ dk_6.
\end{equation}
Ultimately, convergence of the \ac{KE} in the \ac{IR} limit corresponds to convergence of integral \eqref{eq:nonlocalintegral} as $\bk_6 \to 0$. Hence, convergence of the collision integral is satisfied when 
\begin{equation}\label{eq:irconvergence-ap}
 x < 1+2\xi.
\end{equation}
Note that all other integrals over $\bk_2$, $\bk_3$, $\bk_4$, $\bk_5$ diverge in exactly the same manner as the integral over $\bk_6$. Moreover, we can check for \ac{IR} divergence if two wave numbers are simultaneously small.  When two wave numbers on the same side of the six-wave resonance sextet are small , i.e. $\bk_2$ and $\bk_3$ or $\bk_4$ and $\bk_6$, then we see an integral proportional to the square of \eqref{eq:nonlocalintegral} - doubling the rate of divergence.

If the two wave numbers are on opposite sides of the sextet, then we get convergence because of an additional vanishing contribution arising from the term
\begin{equation}
 \lim_{\bk_2,\bk_6 \to 0}\left(n_\bk^{-1} + n_2^{-1} +n_3^{-1} - n_4^{-1} - n_5^{-1} - n_6^{-1}\right) \to 0.
\end{equation}
Total convergence and thus locality of the  \ac{KZ} solutions \eqref{eq:kzspectra} is only assured if we have convergence in both \ac{IR} and \ac{UV} limits of $\bk$-space.  Therefore, we will now proceed in calculating the condition of \ac{UV} convergence of the collision integral. 


To check for \ac{UV} convergence, we will consider the scaling of the collision integral as one wave number diverges to infinity. The conservation of momentum, implied by the Dirac delta function of wave numbers, means that if one wave number diverges, so must a second from the opposite side of the sextet\footnote{Divergence of two wave numbers on the same side of the sextet would violate the Dirac delta function involving frequencies.}.  This can be verified by parametrization \eqref{eq:liaparam}, i.e.  if we force $\bk_6$ to diverge, then so must $\bk_2$.  Indeed, due to this second divergence, we must integrate over the second diverging wave number and additionally compute the $\bk$-scaling of the resulting Jacobian.  

To begin, let us assume that the six-wave interaction coefficient scales in the following way with the diverging wave number $\bk_6$:
\begin{equation}
\lim_{\bk_6\to \infty} \mathcal{W}^{\bk,2,3}_{4,5,6} \propto k_6^\eta,\label{eq:scalingk_6large-ap}
\end{equation}
where $\eta \in \mathbb{R}$.  Due to the integration of an additional diverging wave number (in this case $\bk_2$), we must consider the $\bk$-scaling of the Jacobian corresponding to the transformation between $\bk_6$ and $\bk_2$.  The Jacobian is expressed as
\begin{equation}\label{eq:jacob}
 \left|\frac{\partial \omega^{k,2,3}_{4,5,6}}{\partial \bk_2}  \right|^{-1} = \frac{1}{2|\bk + \bk_3 - \bk_4 - \bk_6|},
\end{equation}
where we have used the fact that $\bk_5 = \bk + \bk_2 + \bk_3 - \bk_4-\bk_6$.  From equation~\eqref{eq:jacob}, we observe that the Jacobian produces a contribution to the collision integral $\propto k_6^{-1}$, and moreover, we find that the following expression in the collision integral scales as
\begin{eqnarray}
\lim_{\bk_6\to\infty}&& \left(n_\bk^{-1}+n_2^{-1}+n_3^{-1}-n_4^{-1}-n_5^{-1}-n_6^{-1}\right)n_\bk n_2n_3n_4n_5n_6 \nonumber\\
&&\propto k_6^{-1-2x}\left(k_6^0 + k_6^{-2}\right).
\end{eqnarray}
The first term on the right-hand side is a contribution arising from the Jacobian and the product of $n_\bk$s.  The second term stems from the difference of $n_\bk$s (shown in the brackets in equation~\eqref{eq:kinetic}) giving rise to two contributions.  The $k_6^0$ scaling results from the difference of the four non-divergent $n_\bk$s, while the $k_6^{-2}$ factor comes from the leading order Taylor expansion of the two divergent $n_\bk$s. Therefore, collecting all the contributions together, we find that the the condition for the \ac{UV} convergence of the collision integral is given by
\begin{equation}\label{eq:uvconvergence-ap}
\eta < x.
\end{equation}
Subsequently, the overall locality region for the \ac{KZ} solutions for the six-wave \ac{KE} is given by
\begin{equation}
\eta < x < 1+ 2\xi.
\end{equation}

\section{Derivation of the Differential Approximation Model}
\label{Appendix:dam}
In this Appendix we will outline the derivation of the \ac{DAM} from the \ac{KE}.  If one assumes that only super-local wave interactions occur, then the \ac{KE} can be simplified into a differential model.  The \ac{DAM} will describe the evolution of the wave action density $n_\bk$ in $\omega$-space - the space determined by angle averaging over wave vectors. This results in a dimensional-independent description for $n_\bk$.  Consequently, the wave action density in $\omega$-space is defined as 
\begin{equation}
N_\omega = n_\omega\; \frac{d\bk}{d\omega_k},
\end{equation}
so that the total wave action is equal to
\begin{equation}
\int N_\omega\; d\omega = \int n_\bk\; d\bk.
\end{equation}
The \ac{KE} \eqref{eq:kinetic}, can then be re-expressed in terms of $\omega$:
\begin{eqnarray}\label{eq:kineticw}
\frac{\partial N_\omega}{\partial z} &=& \int S^{\; \omega, 2,3}_{\; 4,5,6} \ n_\omega n_2n_3n_4n_5n_6 \ \left(\frac{1}{n_\omega}+\frac{1}{n_2}+\frac{1}{n_3}-\frac{1}{n_4}-\frac{1}{n_5}-\frac{1}{n_6}\right)\nonumber\\
&&\times\ d\omega_{2} d\omega_{3}d\omega_{4}d\omega_{5}d\omega_{6},
\end{eqnarray}
where we have defined a new interaction coefficient given by
\begin{equation}
 S^{\;\omega, 2,3}_{\;4,5,6} = \left\langle \frac{\epsilon^8\pi}{6}|\mathcal{W}^{\bk,2,3}_{4,5,6}|^2\delta^{\bk,2,3}_{4,5,6}\right\rangle\frac{d\bk}{d\omega_k}\frac{d\bk_2}{d\omega_2}\frac{d\bk_3}{d\omega_3}\frac{d\bk_4}{d\omega_4}\frac{d\bk_5}{d\omega_5}\frac{d\bk_6}{d\omega_6}.
\end{equation}
$S^{\; \omega, 2,3}_{\; 4,5,6}$ has the same symmetry properties as $\mathcal{W}^{\bk,2,3}_{4,5,6}$ given in \eqref{eq:sym6}. The strategy in deriving the \ac{DAM} from the \ac{KE} is outlined in \cite{DNPZ92}, and involves multiplying \eqref{eq:kineticw} by some arbitrary smooth function, $f(\omega)=f_\omega$, and then integrating with respect to $d\omega$. Using the symmetry of the interaction coefficient $S^{\; \omega, 2,3}_{\; 4,5,6}$, this procedure gives
\begin{eqnarray}
\int \dot{n}_\omega f(\omega) \omega^{-1/2}d\omega &=& \frac{1}{6}\int S^{\;\omega, 2,3}_{\; 4,5,6}\ n_\omega n_2n_3n_4n_5n_6\nonumber \\
&&\times\ \left(\frac{1}{n_\omega}+\frac{1}{n_2}+\frac{1}{n_3}-\frac{1}{n_4}-\frac{1}{n_5}-\frac{1}{n_6}\right) \\
&&\times\left(f_\omega+f_2+f_3-f_4-f_5-f_6\right)d\omega_{2}d\omega_{3}d\omega_{4}d\omega_{5}d\omega_{6},\nonumber
\end{eqnarray}
where $f_i = f(\omega_i)$, with $i=k,2,3,4,5,6$.  The super-locality assumption of wave interactions implies that each $\omega_i$ with $i=2,3,4,5,6$ can be considered to be close to $\omega_k$, such that each $\omega_i$ is within a small deviation $p_i$ of $\omega_k$, i.e. $\omega_i=\omega_k(1+p_i)$ for $i=2,3,4,5,6$. This permits the Taylor expansion of the two brackets involving $n_\omega^{-1}$s and $f$s around the deviations up to $\mathcal{O}(p^3)$.  Furthermore, by approximating $n_\omega n_2n_3n_4n_5n_6 \approx n_\omega^6$ and using the scale invariance  property of $S^{\; \omega, 2,3}_{\; 4,5,6}$, we gain the following equation:
\begin{eqnarray}
\int \dot{n}_\omega f(\omega) \omega^{-1/2}\ d\omega &=& S_0\int \omega^{9/2}\ n_\omega^6\frac{\partial^2}{\partial \omega^2}\left(\frac{1}{n_\omega}\right)\frac{\partial^2f}{\partial \omega^2}\ d\omega,
\end{eqnarray}
where
\begin{eqnarray}
 S_0 &=& \frac{1}{24}\int S(1,1+p_2,1+p_3,1+p_4,1+p_5,1+p_6) \nonumber \\
&& \times \left(p_2^2+p_3^2-p_4^2-p_5^2p_6^2\right)^2 \delta^{p_2,p_3}_{p_4,p_5,p_6}\ dp_{2}dp_3dp_4dp_5dp_6.
\end{eqnarray}
Then from applying integration by parts we get
\begin{equation}
 \int \dot{n}_\omega f(\omega) \omega^{-1/2}\ d\omega = S_0\int \left(\frac{\partial^2}{\partial \omega^2}\left[\omega^{9/2}\ n_\omega^6\frac{\partial^2}{\partial \omega^2}\left(\frac{1}{n_\omega}\right)\right]\right)f(\omega)\ d\omega.
\end{equation}
As $f(\omega)$ is an arbitrary function, the two integrals must be satisfied for all choices of $f(\omega)$.  Therefore, their integrands must equal one another.  This results in the formulation of the \ac{DAM}:
\begin{equation}
\dot{n}_\omega =  S_0\omega^{1/2}\frac{\partial^2}{\partial \omega^2}\left[\omega^{\frac{9}{2}}n_\omega^6\frac{\partial^2}{\partial \omega^2}\left(\frac{1}{n_\omega}\right)\right].
\end{equation}

\section{The Bogoliubov Dispersion Relation}
\label{Appendix:bogoliubov}

We will derive the expression for the Bogoliubov dispersion relation for the description of waves upon a condensate background.  The dynamics of waves propagating in the presence of a condensate differ from the pure linear waves of the system.  The Bogoliubov dispersion relation comprises of a nonlinear correction to the linear frequency that emerges due to the existence of a strong condensate.
The strategy in deriving the formula for the Bogoliubov dispersion relation is to consider the expansion the wave function $\psi(x,z)$ around a homogeneous condensate background.  In the most ideal situation, the condensate is represented as the dynamics of the zeroth Fourier mode.  However, in reality the condensate may be described by a range of Fourier modes situated at low $\bk$.
For simplicity, we shall consider a perturbative expansion of small disturbances on a uniform (zeroth mode) condensate.

The condensate is described by an $x$-independent solution to the \ac{LWE}, (\ref{eq:eq-long}), i.e. a solution of the form $\psi(x,z)=\psi_c(z)$, where $\psi_c(z)$ is given by
\begin{equation}\label{eq:condfreq}
\psi_c(z)=\psi_0\exp(-i\omega_{c}z/2q),
\end{equation}
with $\omega_{c}=-I_0/2\tilde{\psi}^2l_\xi^2$ and where $I_0=|\psi_0|^2$ is the intensity of the condensate (zeroth mode). Solution $\psi_c(z)$
describes the background rotation of a uniform condensate in (\ref{eq:eq-long}) with a rotation frequency of $\omega_c$.

We expand the \ac{LWE} is powers of $\phi(x,z)$, the wave function for disturbance upon the condensate, and linearize with respect to this perturbation, where we have defined
\begin{equation}\label{eq:perturbationMI}
\psi(x,z) = \psi_c(z)\left[1+ \phi(x,z)\right],
\end{equation}
with $|\phi(x,z)|\ll 1$.  Substituting relation (\ref{eq:perturbationMI}) into the \ac{LWE}, (\ref{eq:eq-long}), and linearizing to the first order in $\phi(x,z)$, gives a linear evolution equation for $\phi(x,z)$.  Assuming the disturbance $\phi(x,z)$ takes the form of a single monochromatic plane wave:
\begin{equation}
\phi(x,z) = A\exp\left(i\bk x - \frac{i\Omega_kz}{2q}\right) + A^*\exp\left(-i\bk x + \frac{i\Omega_kz}{2q}\right),
\end{equation}
where $A$ is a complex amplitude of the wave and $\Omega_k$ is the frequency of the waves (disturbances) upon the condensate.   Then by equating both types of exponentials, we can derive the dispersion relation for the propagation of weakly nonlinear waves upon the condensate: 
\begin{equation}\label{eq:bogdis}
\Omega_k = \sqrt{\left(1+\frac{I_0}{\tilde{\psi}^2}\right)k^4 - \frac{I_0}{\tilde{\psi}^2l_\xi^2}k^2}.
\end{equation}
To obtain the Bogoliubov frequency of the original wave function $\psi(x,z)$, we must include the frequency in which the condensate is rotating, $\omega_c$. Therefore, the Bogoliubov dispersion relation for a weakly nonlinear wave in the presence of a condensate is given by
\begin{eqnarray}\label{eq:bogoliubov-ap}
\omega_k &=& \omega_c + \Omega_k,\nonumber\\
&=& -\frac{I_0}{2\tilde{\psi}^2l_\xi^2} +\sqrt{\left(1+\frac{I_0}{\tilde{\psi}^2}\right)k^4 - \frac{I_0}{\tilde{\psi}^2l_\xi^2}k^2}.
\end{eqnarray}

\section{Non-Dimensionalization}
\label{Appendix:nondim}

The dimensional models for \ac{1D} \ac{OWT} contain several physical constants and parameters.  For convenience and clarity, it is preferred to consider the dimensionless forms of the equations.  Therefore in this Appendix, we present our non-dimensional descriptions for the \ac{LWE}, \eqref{eq:eq-long} and the \ac{SWE}, \eqref{eq:eq-short}.  We follow the non-dimensionalization that was performed in \cite{CPA06}:
\begin{equation}\label{eq:nondim}
\psi= (\tilde{\psi}/\sqrt{\alpha})\psi^*(x^*,
z^*),\quad x^* = x/x_c\sqrt{\alpha}, \quad z^*=z/z_c\alpha
\end{equation}
with $\tilde{\psi}^2=2K/\varepsilon_0k^2_0n_a^4l_\xi^4$, $z_c=2ql_\xi^2$ and $x_c=l_\xi$. The electrical coherence length of the \ac{LC} is defined as $l_\xi=\sqrt {\pi K / 2 \Delta
\varepsilon} (d/ V_0)$.

Non-dimensionalization \eqref{eq:nondim}, expresses the \ac{LWE} in the form:
\begin{equation}\label{eq:LWEnondim}
i\frac{\partial \psi}{\partial z} = - \frac{\partial^2 \psi }{\partial x^2} -
\frac{1}{2}\psi|\psi|^2 - \frac{1}{2\alpha}\psi \frac{\partial^2
|\psi|^2}{\partial x^2},
\end{equation}
where $\alpha$ is a tunable parameter that adjusts the strength of the second nonlinear term and for clarity we have dropped the superscripts on the non-dimensional variables.  The nonlinearity of the system can be adjusted by increasing or decreasing the magnitude of the wave function $\psi(x,z)$, however, parameter $\alpha$ provides additional control of the ratio of the the two nonlinear terms. Indeed, the \ac{LWE} was derived in a regime of a long-wave limit that implies that $\left(k^*\right)^2/\alpha \ll 1$ (where $k^*$ is the dimensionless wave number), and therefore we must ensure that the long-wave limit it satisfied.

Similarly, we can derive a dimensionless expression for the \ac{SWE} using the same dimensionless variables given in \eqref{eq:nondim}.  This gives the dimensionless version of the \ac{SWE} as
\begin{equation}\label{eq:SWEnondim}
 i\frac{\partial \psi}{\partial z} = - \frac{\partial^2 \psi }{\partial x^2} + \frac{1}{2}\psi \frac{\partial^{-2}
|\psi|^2}{\partial x^{-2}}.
\end{equation}
Notice, that in \eqref{eq:SWEnondim} $\alpha$ does not appear.

\section{The Intensity Spectrum}
\label{Appendix:Intspec}

Experimentally, due to the difficulty in measuring the phase of the wave function $\psi(x,z)$, we cannot determine $n_\bk$ easily. However, we can measure the spectrum of wave Intensity
$N_\bk=|\left(|\psi|^2\right)_\bk|^2$.  Theoretically, we are able to relate the
$k$-scaling for $N_\bk$ with that of the \ac{KZ} solution derived from the \ac{KE}, (\ref{eq:kinetic}).  In this Appendix, we present the derivation of this relationship.  First, we must consider the expression for the light intensity in $\bk$-space.  We begin by considering the usual definition for the Fourier transform in $\mathbb{R}$, this implies
\begin{subequations}
\begin{eqnarray}
I_\bk=(|\psi|^2)_\bk &=&\int \psi(x)\psi^*(x)e^{-i\bk x} \ dx, \\
&=& \int a_{1}a^*_{2}\ \delta^{\bk,2}_1 \ d \bk_1 \ d\bk_2.
\end{eqnarray}
\end{subequations}
Hence, for the intensity spectrum $N_\bk = \langle |I_\bk|^2 \rangle$ this gives
\begin{equation}
\langle |I_\bk|^2 \rangle = \int \langle a_1 a_2 a_3^* a_4^* \rangle\ 
\delta^{\bk,4}_1 \ \delta^{\bk,2}_{3}\ d \bk_1\ d \bk_2\ d \bk_3\ d \bk_4.
\end{equation}
The next step is to average over phases, in a \ac{RPA} wave field. This implies that only wave number pairings $\bk_1=\bk_4$,
$\bk_2=\bk_3$ and $\bk_1=\bk_3$, $\bk_2=\bk_4$ will contribute to the intensity spectrum. Therefore, the intensity spectrum can be expressed as
\begin{subequations}
\begin{eqnarray}
N_\bk &=& \int \left\langle |a_1|^2\right\rangle\left\langle|a_2|^2\right\rangle \delta^\bk\delta^\bk d
\bk_1 \;d\bk_2 + \int \left\langle |a_1|^2 |a_2|^2 \right\rangle \delta^{\bk,2}_1 \delta^{\bk,2}_1\ d
\bk_1\; d \bk_2 \\
&=&\left(\int n_1 \delta^\bk\ d \bk_1\right)^2 + \int n_1 n_2\ \delta^{\bk,2}_1\ d\bk_1\;  d\bk_2.\label{eq:intspectrum}
\end{eqnarray}
\end{subequations}
We will examine the intensity spectrum, when the system is in a statistically non-equilibrium stationary state and when the \ac{KZ} solution is realized.  Therefore, we can assume that the wave action spectrum is of the \ac{KZ} form, i.e. $n_\bk = Ck^{-x}$, where $C$ is a constant determining the amplitude of the spectrum and $x$ is the spectrum exponent.  The first term of equation (\ref{eq:intspectrum}) contains a Dirac delta function, centered around $\bk=0$. This implies that the contribution from this term will only appear at the zeroth mode\footnote{Numerically and experimentally this may be seen as a contribution at low wave numbers around $\bk=0$.}. On the other hand, the second term in equation (\ref{eq:intspectrum}) will contribute on the  whole of $\bk$-space, and will determine the $k$-scaling for the intensity spectrum. The $k$-scaling of the intensity spectrum will only be observed if the integral of the second term is finite.  We proceed by determining a sufficient condition for its convergence.  We can re-express the second term by using the Dirac delta function to eliminate one of the integration variables, i.e.
\begin{eqnarray}
\int n_1 n_2\ \delta^{\bk,2}_1\ d \bk_1\ d  \bk_2 &=& \int n_1 n_{1-\bk}\ d \bk_1 \nonumber\\
&=& C^2 \int k_1^{-x} |\bk_1-\bk|^{-x}\ d \bk_1.\label{eq:intscale}
\end{eqnarray}
If expression (\ref{eq:intscale}) converges on the \ac{KZ} solution, then the integral  will yield the intensity spectrum power-law scaling.  To check for convergence, we change the integration variable to a non-dimensional variable $s=\bk_1/\bk$. Then, the intensity spectrum can be approximated by the integral:
\begin{eqnarray}\label{eq:ints}
N_\bk&\approx& C^2k^{-2x+1}\int^{\infty}_{-\infty} s^{-x}|s-1|^{-x}\ d s.\nonumber \\
&\approx& C^2k^{-2x+1}\int^{\infty}_{0} s^{-x}\left(|s-1|^{-x}+
|s+1|^{-x}\right) d s.
\end{eqnarray}
Convergence must be checked in the regions where $s\to 0$ and $s \to \infty$.  As $s\to 0$, relation \eqref{eq:ints} behaves as
\begin{equation}
\int_0 s^{-x}\left(|s-1|^{-x}+
|s+1|^{-x}\right) d s \propto \int_0 s^{-x}\ ds,
\end{equation}
and therefore, the integral converges for $x<1$.

In the limit when $s\to \infty$, integral \eqref{eq:ints} can be written as
\begin{equation}\label{eq:intsinfty}
\int^\infty s^{-x}\left(|s-1|^{-x}+
|s+1|^{-x}\right) d s \propto \int^\infty s^{-2x}\ ds.
\end{equation}
The right-hand side of expression \eqref{eq:intsinfty} converges for $x > 1/2$.  Therefore, the intensity spectrum integral, (\ref{eq:ints}), is convergent in the region:
\begin{equation}\label{eq:intspecconvergence}
1/2 < x < 1, 
\end{equation}
where $x$ is exponent of the power-law for the \ac{KZ} solution of the \ac{KE}.  Experimentally at present, we can only produce the inverse cascade scenario and subsequently, we are only interested in measuring the intensity spectrum in the inverse cascade regime.  The exponent of the \ac{KZ} solution for the inverse cascade is given in equation~\eqref{eq:inverse} ($x=3/5$), which lies inside the region of convergence \eqref{eq:intspecconvergence}, (and so the intensity spectrum is observable) and corresponds to an intensity spectrum of
\begin{equation}
N_\bk\propto k^{-1/5}. 
\end{equation}

\newpage
\addcontentsline{toc}{section}{Bibliography}

\bibliographystyle{IEEEtran}
\bibliography{bibliography}   

\end{document}